\documentclass[final,3p,times,twocolumn]{elsarticle}

\usepackage{graphicx}
\usepackage{amssymb}

\usepackage{amsmath}
\usepackage{amsfonts}
\usepackage{bm}
\usepackage{verbatim}
\usepackage{multirow}
\usepackage{booktabs} 
\usepackage[normalem]{ulem} 

\usepackage{color}
\definecolor{gray}{rgb}{0.5,0.5,0.5}
\definecolor{dred}{rgb}{0.5,0.0,0.0}
\definecolor{dgreen}{rgb}{0.0,0.5,0.0}
\definecolor{dblue}{rgb}{0.0,0.0,0.5}
\definecolor{violet}{rgb}{0.7,0.0,0.5}

\definecolor{lred}{rgb}{1.0,0.5,0.5}
\definecolor{lgreen}{rgb}{0.5,1.0,0.5}
\definecolor{lblue}{rgb}{0.5,0.5,1.0}

\def\nablab{{\bm \nabla}}

\journal{arXiv} 

\begin{document}

\begin{frontmatter}
\title{Similarity of magnetized plasma wake channels behind \\ relativistic laser pulses with different wavelengths}

\author[rokk,naka]{A.~Bierwage\corref{cor1}}
\ead{bierwage.andreas@qst.go.jp}
\author[kpsi]{T.Zh.~Esirkepov}
\ead{esirkepov.timur@qst.go.jp}
\author[kpsi]{J.K.~Koga}
\author[kpsi]{A.S.~Pirozhkov}

\cortext[cor1]{Corresponding author}

\address[rokk]{National Institutes for Quantum and Radiological Science and Technology, Rokkasho Fusion Institute, Rokkasho, Aomori 039-3212, Japan}
\address[naka]{National Institutes for Quantum and Radiological Science and Technology, Naka Fusion Institute, Naka, Ibaraki 311-0193, Japan}
\address[kpsi]{National Institutes for Quantum and Radiological Science and Technology, Kansai Photon Science Institute, Kizugawa, Kyoto 619-0215, Japan}

\begin{abstract}
Using particle-in-cell simulations of relativistic laser plasma wakes in the presence of an external magnetic field, we demonstrate that there exists a parameter window where the dynamics of the magnetized wake channel are largely independent of the laser wavelength $\lambda_{\rm las}$. One condition for this manifestation of ``limited similarity'' is that the electron density $n_{\rm e}$ is highly subcritical, so that the plasma does not affect the laser. The freedom to choose a convenient laser wavelength can be useful in experiments and simulations. In simulations, an up-scaled wavelength (and, thus, a coarser mesh and larger time steps) reduces the computational effort, while limited similarity ensures that the overall structure and evolutionary phases of the wake channel are preserved. In our demonstrative example, we begin with a terrawatt$\cdot$picosecond pulse from a ${\rm CO}_2$ laser with $\lambda_{\rm las} = 10\,\mu{\rm m}$, whose field reaches a relativistic amplitude at the center of a sub-millimeter-sized focal spot. The laser is shot into a sparse deuterium gas ($n_{\rm e} \sim 10^{13}\,{\rm cm}^{-3}$) in the presence of a tesla-scale magnetic field. Limited similarity is demonstrated in 2D for $4\,\mu{\rm m} \leq \lambda_{\rm las} \leq 40\,\mu{\rm m}$ and is expected to extend to shorter wavelengths. Assuming that this limited similarity also holds in 3D, increasing the wavelength to $40\,\mu{\rm m}$ enables us to simulate the after-glow dynamics of the wake channel all the way into the nanosecond regime.
\end{abstract}

\end{frontmatter}




\section{Introduction}
\label{sec:intro}

Laser-induced wake waves \cite{Akhiezer56} have been studied extensively during the last decades, in particular in the context of particle acceleration \cite{Tajima79} and for the generation of compact sources of energetic photons \cite{ELIwhitebook2011, Corde13, Pirozhkov17b}. Many interesting and useful applications have emerged, and there still exists much potential for expansion \cite{BulanovSV16}.

Beginning with the present paper, we investigate the dynamics of laser-induced wakes in a parameter regime that, to our knowledge, has not been examined before: we consider a laser pulse with relativistic intensity \cite{Mourou06} and study the formation and long-term evolution of the wake channel that this laser produces in a magnetically confined plasma as used in current mainstream thermonuclear fusion reactor designs \cite{IterPlan2018, Tobita17}. Here, we begin with relativistic laser wakes in the cold plasma limit, which is valid up to the nanosecond scale, laying the basis for future work that will include thermal motion.

Present-day laser systems that are able to accelerate electrons to relativistic speeds operate with principal wavelengths in the near infrared: $\lambda_{\rm las} \sim 0.8$...$10\,\mu{\rm m}$ \cite{Danson15, Polyanskiy11}. Since the electron densities in magnetically confined fusion (MCF) plasmas produced in tokamak and stellarator devices rarely exceed $10^{14}\,{\rm cm}^{-3}$ \cite{Greenwald02}, they are effectively transparent to this kind of electromagnetic radiation, so that the laser pulse passes nearly unchanged through the plasma, while plowing away all electrons in its path. The accelerated electrons will be deflected by the $-e{\bm v}\times{\bm B}_{\rm ext}$ force, where $-e$ is the electron's charge, ${\bm v}$ its velocity vector, and ${\bm B}_{\rm ext}$ is the externally imposed magnetic field. The magnetic field in present-day MCF devices is typically a few tesla strong, which can be regarded as an intermediate strength in the sense that 
$\omega_{\rm Be}/\omega_{\rm pe} \sim 1$; i.e., the Larmor frequency
\begin{equation}
\omega_{\rm Be} = e B_{\rm ext}/(\gamma m_{\rm e}) \approx 2\pi\times 30\,{\rm GHz}\times B\,[{\rm T}]/\gamma
\label{eq:wBe}
\end{equation}

\noindent is comparable to the Langmuir frequency
\begin{equation}
\omega_{\rm pe} = \left(n_{\rm e} e^2/(\epsilon_0 m_{\rm e})\right)^{1/2} \lesssim 2\pi\times 90\,{\rm GHz};
\label{eq:wpe}
\end{equation}

\noindent where $m_{\rm e}$ is the electron's rest mass, $\gamma = (1 - v^2/c^2)^{-1/2}$ the Lorentz factor, and $\epsilon_0$ the vacuum permittivity. (We employ SI units in this work.)

The effect of an externally imposed magnetic field ${\bm B}_{\rm ext}$ on laser plasmas has received increasing attention during the last decade \cite{Hosokai06, Hur08, Hosokai10, Bourdier12, Schmit12, BulanovSV13, Ghotra15, Wilson17}, but these works focused primarily on magnetization effects in the vicinity of the laser pulse. Here, our motivation is also to learn how the magnetized plasma channel evolves long after the laser pulse has gone.

Even without an external magnetic field, long-lived plasma channels can be routinely seen in experiments where intense laser pulses are shot through subcritical gas. Simulations have revealed the formation of a quasi-static azimuthal magnetic field and magnetic vortices that are maintained by electron currents flowing along the channel axis \cite{BulanovSV96, Kuznetsov01, BulanovSV05, Nakamura08, Nakamura10}.

In order to study the after-glow dynamics numerically in our parameter regime of interest, the simulations need to cover a relatively long time interval (nanoseconds), include the ion response, and cover a relatively large spatial volume (centimeters). The simulation box should encompass the entire plasma channel around the focal point, with sufficiently large buffer zones in each direction to minimize boundary effects. At the same time, a high spatio-temporal resolution (sub-micron, sub-femtosecond) is needed in order to capture the evolution of the laser's electromagnetic field and the motion of electrons accelerated to relativistic speeds. This constitutes a multi-scale problem and poses many challenges, including unphysical boundary effects, numerical heating, and a large computational cost.

The need for large computational resources is connected with the necessity to perform these simulations in three dimensions (3D), so as to capture the intrinsically 3D motion of charged particles in an external magnetic field ${\bm B}_{\rm ext}$. Although 2D simulations can be used to identify interesting parameter regimes and trends, the reduced degrees of freedom and enforced symmetries imply that great care is needed when trying to extract physical insight from 2D results. The ability to perform 3D simulations is important for identifying verifiable aspects of the 2D results and obtaining accurate qualitative and quantitative information that can be validated against experiments.

If the dynamical structures of interest are larger than the laser wavelength $\lambda_{\rm las}$, the computational cost (memory, time) is determined by the value of $\lambda_{\rm las}$, which determines the spatial resolution, the time step and (via its focal length) the required box size. It can be shown (\ref{apdx:resource}) that even a single long-time 3D simulation (not to speak of parameter scans and convergence tests) would be prohibitively expensive if it was to be performed with realistic laser wavelengths in the range $\lambda_{\rm las} \sim 0.8$...$10\,\mu{\rm m}$. Under some conditions, however, this problem can be circumvented by using an artificially increased wavelength.

In this paper, we identify a parameter regime, where such an artificial scaling of the laser wavelength $\lambda_{\rm las}$ can be justified on the basis of so-called {\it limited similarity} or {\it limited scaling} \cite{Block67, Esirkepov12}. In Section~\ref{sec:setup} we describe the simulation setup, where a relativistically strong laser pulse is shot into a sparse deuterium gas with highly subcritical electron density and a moderate external magnetic field. In Section~\ref{sec:channel}, the properties of the magnetized plasma wake channel produced by a relativistic laser pulse with wavelength $\lambda_{\rm las} = 10\,\mu{\rm m}$ are illustrated in 2D. In Section~\ref{sec:limsim}, 2D simulations are used to demonstrate the limited similarity of the magnetized wake channel for laser wavelengths in the range $4$...$40\,\mu{\rm m}$. Using a scaled wavelength $\lambda_{\rm las} = 40\,\mu{\rm m}$, we are able to perform long-time 3D simulations, and first results are reported in Section~\ref{sec:3d}. A summary and conclusions are given in Section~\ref{sec:conclusion}.

\section{Simulation scenario, model and codes}
\label{sec:setup}

\subsection{Reference case}
\label{sec:setup_ref}

As a starting point and reference case, we consider a short picosecond-scale pulse from a terrawatt-scale ${\rm CO}_2$ gas laser with wavelength $\lambda_{\rm las} = 10\,\mu{\rm m}$. The laser pulse travels through a quasi-neutral deuterium plasma with number density $3\times 10^{13}\,{\rm cm}^{-3}$ in the presence of a $2\,{\rm T}$ magnetic field, as is typical for present-day tokamaks.

We follow the dynamics of the laser-induced wake channel for up to 1 nanosecond. This means that the plasma can be safely assumed to be ideal (collisionless). Thermal motion will be ignored for simplicity, although it should be noted that, under typical tokamak conditions with thermal energies around $3\,{\rm keV}$, electrons travel about $30\,{\rm mm}$ and deuterons about $0.5\,{\rm mm}$ per nanosecond. The effect of thermal motion is briefly discussed in Section~\ref{sec:conclusion} and will be examined in a separate paper, which will be dedicated to the physics of the laser-induced wake channels. The present paper lays the conceptual and computational foundations.

A right-handed Cartesian coordinate system $(x,y,z)$ in the laboratory frame of reference is used and the laser travels in the positive $x$ direction. The simulation domain is
\begin{gather}
0 \leq x - x_{\rm win}(t) \leq L_x, \\
-L_y/2 \leq y \leq L_y/2, \\
-L_z/2 \leq z \leq L_z/2;
\end{gather}

\noindent where $x_{\rm win}(t)$ is the trailing edge of the simulation window along the laser path and the time is denoted by $t$. All simulations begin with $x_{\rm win}(t=0) = 0$. In simulations with a moving window, $x_{\rm win}(t)$ advances with the speed of light $c$ after a suitable delay.

The initial electron density is uniform, except for a narrow boundary layer, where it is ramped up or down. In mathematical form, it may be written as
\begin{equation}
n_{\rm e}(x,y,z|t=0) = n_{\rm e0}\times H(x|s_x) H(y|s_y) H(z|s_z),
\end{equation}

\noindent with $n_{\rm e0} = 3\times 10^{13}\, {\rm cm}^{-3}$ and
\begin{equation}
H(\xi|s) = \left\{ \begin{array}{ccc}
1 & : & \xi_{\rm min} \leq \xi \leq \xi_{\rm max}, \\
{\rm ramp\,up} & : & \xi_{\rm min}-s \leq \xi \leq \xi_{\rm min}, \\
{\rm ramp\,down} & : & \xi_{\rm max} \leq \xi \leq \xi_{\rm max}+s, \\
0\,{\rm (vacuum)} & : & {\rm else}.
\end{array}\right.
\end{equation}

\noindent The thickness of the boundary layer where $n_{\rm e}$ is ramped up or down linearly is $s_x = s_y = s_z = 150\,\mu{\rm m}$. The thickness of the surrounding vacuum region is also $s_{x,y,z}$, so that $y_{\rm max} = -y_{\rm min} = L_y/2 - 2s_y$ (same for $z$) in the transverse direction. Along the laser path we use $x_{\rm min} = 2 s_x$ and $x_{\rm max} = \infty$ in the case of a moving window, or $x_{\rm max} = L_x - 2 s_x$ for a stationary window.

\begin{figure}[tbp]
\centering
\includegraphics[width=0.48\textwidth]{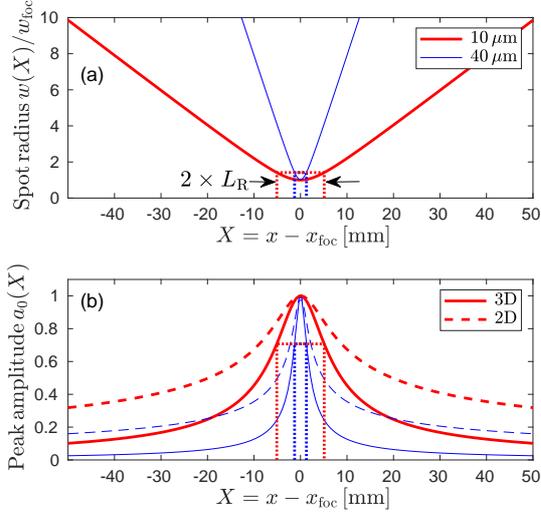}
\caption{Spatial dependence of (a) the $1/{\rm e}$ amplitude radius $w(X)$ and (b) the normalized amplitude $a_0(X)$ of a Gaussian pulse propagating through vacuum in the paraxial wave approximation. The red curves are for a laser wavelength of $\lambda_{\rm las} = 10\,\mu{\rm m}$ and blue for $40\,\mu{\rm m}$. In panel (b), solid lines are used for the 3D case and dashed lines for 2D.}
\label{fig:01_paraxial}%
\end{figure}

\subsection{Laser model}
\label{sec:setup_laser}

The Gaussian laser pulse is linearly polarized, with its electric and magnetic field vectors ${\bm E}_{\rm las}$ and ${\bm B}_{\rm las} = \nablab\times{\bm A}_{\rm las}$ oscillating along the $y$ and the $z$ axis, respectively. The laser wavelength, wavenumber and angular frequency are related as
\begin{equation}
\lambda_{\rm las} = 2\pi/k_{\rm las} = 2\pi c/\omega_{\rm las}.
\end{equation}

\noindent We are interested in laser-plasma interactions near the threshold of the relativistic regime \cite{Mourou06}, so we focus the laser such that its normalized amplitude
\begin{equation}
a_0 = \frac{e A_{\rm las}}{m_{\rm e}c} = \frac{e E_{\rm las}}{m_{\rm e}\omega_{\rm las} c} = \frac{e E_{\rm las} \lambda_{\rm las}}{2\pi m_{\rm e} c^2};
\label{eq:anrm}
\end{equation}

\noindent reaches $a_{0,{\rm foc}} = 1$ at the center of the focal spot, which is located at $(x,y,z) = (x_{\rm foc},0,0)$.

Since the electron density in a tokamak is highly subcritical (transparent) for near-infrared light, the laser pulse propagates as if in vacuum, where its electromagnetic field can be well approximated by the solution of the paraxial wave equation \cite{Boyd}. The complex-valued amplitude of the electric field can then be written as
\begin{align}
\tilde{E}(X,y,z,t) \; =& \; E_0 \exp\left(i\omega_{\rm las} (t - t_{\rm D}) - ik(X + x_{\rm foc})\right) \; \times \nonumber
\\
& \; \exp\left(-\frac{[t - t_0(X,y,z)]^2}{w_t^2}\right) \; \times \nonumber
\\
& \; \sqrt{\frac{w_{\rm foc}}{w(X)}} {\rm e}^{-\frac{y^2}{w(X)^2}} {\rm e}^{i\left(\frac{\omega_{\rm las} y^2}{2 c R(X)} + \frac{\phi_0(X)}{2}\right)} \; \times \nonumber
\\
& \; \left[\sqrt{\frac{w_{\rm foc}}{w(X)}} {\rm e}^{-\frac{z^2}{w(X)^2}} {\rm e}^{i\left(\frac{\omega_{\rm las} z^2}{2 c R(X)} + \frac{\phi_0(X)}{2}\right)}\right]^{N-2};
\label{eq:paraxial_E}
\end{align}

\noindent where $N=2$ ($z=0$) for 2D and $N=3$ for 3D. The auxiliary coordinate
\begin{equation}
X = x - x_{\rm foc}
\label{eq:x}
\end{equation}

\noindent has its origin at the focal point, where the laser field takes the form of a plane wave. The first line of Eq.~(\ref{eq:paraxial_E}) describes a harmonic wave. The second line is the Gaussian temporal envelope of the laser pulse with $1/{\rm e}$ amplitude width parameter $w_t$. The functions
\begin{align}
t_0(X,y,z) &= t_{\rm D} - \left(y^2 + z^2\right)/\left(2 c R(X)\right),
\label{eq:paraxial_t0}
\\
R(X) &= \left(X^2 + L_{\rm R}^2\right)/X,
\label{eq:paraxial_R}
\end{align}

\noindent describe the shape of the pulse envelope with curvature radius $R(X)$. The corresponding deformation of the optical wave front with respect to the $y$-$z$ plane is captured by the complex exponents on the last two lines of Eq.~(\ref{eq:paraxial_E}), where
\begin{equation}
\phi_0(X) = \arctan\left(X/L_{\rm R}\right)
\label{eq:paraxial_p}
\end{equation}

\noindent measures the phase shift of the wave front relative to the plane wave at $X=0$. Also on the last two lines, there are factors describing the transverse profile of the laser spot and the variation of the field amplitude with the spot size, where
\begin{equation}
w(X) = w_{\rm foc} \left(1 + X^2/L_{\rm R}^2\right)^{1/2}
\label{eq:paraxial_w}
\end{equation}

\noindent is the $1/{\rm e}$ amplitude radius of the Gaussian spot in the $y$-$z$ plane. Note that the last line of Eq.~(\ref{eq:paraxial_E}) is present only in the 3D case ($N=3$), where $\tilde{E} \propto w^{-1}(X)$. In the 2D case ($N=2$, $z = 0$) we have $\tilde{E} \propto w^{-1/2}(X)$ and the phase shift is $\phi_0(X)/2$.

\begin{table}[tbp]
\centering
\begin{tabular}{lcc}
\hline\hline
Laser wavelength & $\lambda_{\rm las}$ & $10\,\mu{\rm m}\times \sigma$ \\
Norm.\ amplitude & $a_{\rm 0,foc}$ & $1$ \\
Spot size & $d_{\rm foc}$ ($w_{\rm foc}$) & $0.15$ ($0.13$) ${\rm mm}$ \\
Pulse length & $\tau_{\rm pulse}$ ($w_t$) & $2$ ($1.7$) ${\rm ps}$ \\
Delay & $t_{\rm D}$ & $5\times\tau_{\rm pulse}$ \\
\hline\hline
\end{tabular}
\caption{Laser parameters. The wavelength is scanned with respect to reference case (10 microns) as $\lambda_{\rm las} = \sigma \times 10\,\mu{\rm m}$ with scaling factors $\sigma = 0.4,1,2,4,8$. The pulse has a Gaussian profile in time and space, and it is linearly polarized, consisting of $E_y$ and $B_z$ field components.}
\label{tab:parm_laser}
\end{table}

The Rayleigh length $L_{\rm R}$ appearing in Eqs.~(\ref{eq:paraxial_R})--(\ref{eq:paraxial_w}) can be written in terms of the $1/{\rm e}$ amplitude waist radius $w_{\rm foc}$ as
\begin{equation}
L_{\rm R} \equiv \pi w_{\rm foc}^2/\lambda_{\rm las}.
\label{eq:rayleigh}
\end{equation}

\noindent At a distance $L_{\rm R}$ from the focal point $x_{\rm foc}$, the mean laser intensity $|\tilde{E}|^2$ in 3D is 2 times lower than at the focus ($\sqrt{2}$ times lower in 2D).

As an example, Fig.~\ref{fig:01_paraxial} shows the $X$-dependence of the pulse radius $w(X)$ and the normalized amplitude $a_0(X)$ for two wavelengths, $\lambda_{\rm las} = 10\,\mu{\rm m}$ and $40\,\mu{\rm m}$. The Rayleigh length is indicated by dotted lines at $X = \pm L_{\rm R}$.

In the simulation, the laser is injected by applying an oscillating electromagnetic source on the left-hand boundary ($x = 0$) of the simulation domain. The electric component of the source is chosen to be the sine (imaginary) part of $\tilde{E}$ in Eq.~(\ref{eq:paraxial_E}) evaluated at $X=-x_{\rm foc}$,
\begin{align}
E_{\rm src}(r,t) \; = \; & \Im\left\{\tilde{E}(x=0)\right\} \nonumber
\\
= \; & E_0 \mathcal{E}(r,t) \sin(\omega t - \phi(r)).
\end{align}

\noindent The amplitude factor $\mathcal{E}$, the phase $\phi$ and the phase front delay $t_0$ are given by
\begin{align}
\mathcal{E}(r,t) \; = \; & \left(\frac{w_{\rm foc}}{w(x_{\rm foc})}\right)^{\frac{N-1}{2}} \exp\left(\frac{-r^2}{w^2(x_{\rm foc})}\right) \exp\left(-\frac{(t - t_0)^2}{w_t^2}\right),
\label{eq:src_e}
\\
\phi(r) \; = \; & -\omega t_0(r) - \frac{N-1}{2}\arctan\left(\frac{x_{\rm foc}}L_{\rm R}\right),
\label{eq:src_phi}
\\
t_0(r) \; = \; & t_{\rm D} + \frac{x_{\rm foc} r^2}{2 c(x_{\rm foc}^2 + L_{\rm R}^2)};
\label{eq:src_t0}
\end{align}

\noindent where $r^2 = y^2 + z^2$ in 3D ($N=3$) and $r^2 = y^2$ in 2D ($N=2$).

\begin{table}[tbp]
\centering
\begin{tabular}{cc|c@{$\;\;\;$}c@{$\;\;\;$}c} 
\hline\hline
\multicolumn{2}{c|}{Box size [mm]} & \multicolumn{3}{c}{Resolution} \\
$L_x$ & $L_y$ & $\lambda_{\rm las}/\Delta x$ & $\lambda_{\rm las}/\Delta y$ & $N_{\rm PPC}$ \\
\hline
$10.2$ & $\sigma \times 6$ & $32$ & $8$ & $1$ \\
& & ($8,16,32,64$) & ($8,16$) & ($1,4,10$) \\
\hline\hline
\end{tabular}
\caption{Numerical parameters for 2D EPOCH simulations with moving window and weakly reflecting ``open'' or ``simple outflow'' boundaries. The length $L_x$ of the window is fixed, while its height $L_y$ is varied with the wavelength scaling factor $\sigma$, where $\sigma = 1$ corresponds to the reference case with $\lambda_{\rm las} = 10\,\mu{\rm m}$. $N_{\rm PPC}$ is the number of simulation particles (electrons) per cell. Ions are immobile. The values in parentheses on the bottom row were used in convergence tests.}
\label{tab:parm_num}
\end{table}

\begin{table*}[tbp]
\centering
\begin{tabular}{c|c|cc|cc|c@{$\;\;\;$}c@{$\;\;\;$}c} 
\hline\hline
Case & Dims. & Laser wavelength & Focusing time & \multicolumn{2}{c|}{Box size} & \multicolumn{3}{c}{Resolution} \\
& & $\lambda_{\rm las}$ $[\mu{\rm m}]$ & $t_{\rm foc}$ $[{\rm ps}]$ & $L_x$ $[{\rm mm}]$ & $L_{y,z}$ $[{\rm mm}]$ & $\lambda_{\rm las}/\Delta x$ & $\lambda_{\rm las}/\Delta (y,z)$ & $N_{\rm PPC}$ \\
\hline
(i) & 2D & $10$ ($\sigma=1$) & $180$ & $101.5$ & $6$ & $16$ & $8$ & $1$ \\
(ii) & 2D & $20$ ($\sigma=2$) & $96$ & $51.5$ & $6$ & $32$ & $8$ & $1$ \\
(iii) & 2D & $40$ ($\sigma=4$) & $50$ & $26.5$ & $6$ & $32$ & $8$ & $1$ ($10$) \\
(iv) & 3D & $40$ ($\sigma=4$) & $50$ & $26.5$ & $6$ & $16$ & $8$ & $1$ \\
\hline\hline
\end{tabular}
\caption{Numerical parameters for the 2D and 3D simulations with stationary window and ``open'' (EPOCH) or absorbing (REMP) boundaries. 2D simulations are performed for three different wavelengths, $\lambda_{\rm las} = \sigma \times 10\,\mu{\rm m}$ with scaling factors $\sigma = 1,2,4$. The transverse box size $L_y = L_z$ is fixed while varying its length as $L_x = 2 x_{\rm foc} + 1.5\,{\rm mm} = 100\,{\rm mm}/\sigma + 1.5\,{\rm mm}$. The rationale behind the chosen box size is discussed in the first part of \protect\ref{apdx:resource}. The results shown in this paper were obtained using 1 PPC (simulation particles per cell) per species (electrons and deuterons). The $40\,\mu{\rm m}$ case in 2D was also simulated with 10 PPC per species (in parentheses for case (iii)), which gave very similar result.}
\label{tab:parm_num_fix}
\end{table*}

\subsection{Parameters}
\label{sec:setup_param}

The laser parameters for the 10 micron reference case are given in Table~\ref{tab:parm_laser}, where the focal spot size and pulse length are specified in terms of the respective full-width-half-maximum (FWHM) diameters of the intensity,
\begin{align}
d_{\rm foc}({\rm FWHM})  \; = \; & w_{\rm foc} \times \sqrt{2\ln 2},
\\
\tau_{\rm pulse}({\rm FWHM}) \; = \;& w_t \times \sqrt{2\ln 2},
\end{align}

\noindent whose relation to $w_{\rm foc}$ and $w_t$ follows from
\begin{equation}
|\hat{E}|^2 \propto e^{-2r^2/w_{\rm foc}^2} \equiv 2^{-r^2/(d_{\rm foc}/2)^2} = e^{-4\ln 2/d_{\rm foc}^2}.
\end{equation}

\noindent The values in Table~\ref{tab:parm_laser} are chosen such that, even for the largest wavelength ($80\,\mu{\rm m}$) in our parameter scans, it is ensured that $d_{\rm foc}/{\rm max}\{\lambda_{\rm las}\} \gtrsim 2$ and that the pulse still contains $c\tau_{\rm pulse}/{\rm max}\{\lambda_{\rm las}\} \gtrsim 7.5$ wave cycles. The constant time delay $t_{\rm D}$ in Eq.~(\ref{eq:src_t0}) is used to ensure that the jump of the laser amplitude at the head of the laser pulse is sufficiently small, such that the associated numerical artifacts are negligible. Here, we choose $t_{\rm D} = (5,5,5,7.5,12.5)\times \tau_{\rm pulse}$ for $\lambda_{\rm las} = (0.4,1,2,4,8)\times 10\,\mu{\rm m}$, where the increase of $t_{\rm D}$ for larger $\lambda_{\rm las}$ is needed to compensate for the larger curvature of the wave front (second term in Eq.~(\ref{eq:src_t0})).

Note that the FWHM pulse length $\tau_{\rm pulse} = 2\,{\rm ps}$ of our Gaussian laser pulse corresponds to the $1/{\rm e}$ amplitude length $\ell_{\rm pulse} = 2c \tau_{\rm pulse}/\sqrt{2\ln 2} \approx 1.0\,{\rm mm}$. This is shorter than the theoretically predicted optimum for the generation of a large-amplitude wakefield with a flat-top pulse: $\ell_{\rm opt}^{\rm flat} = 3.8202\times c\omega_{\rm pe}^{-1} \approx 0.6 \times \lambda_{\rm pe}$ for $a_0 = 1$ \cite{BulanovSV16}, which is $3.7\,{\rm mm}$ for our parameters. Our choice of a shorter pulse length has a practical motivation as it reduces the energy required to reach the relativistic regime. A lower pulse energy, in turn, allows to reduce the cost of the laser or to increase its repetition rate.

\begin{figure}
[tb]
\centering
\includegraphics[width=0.48\textwidth]{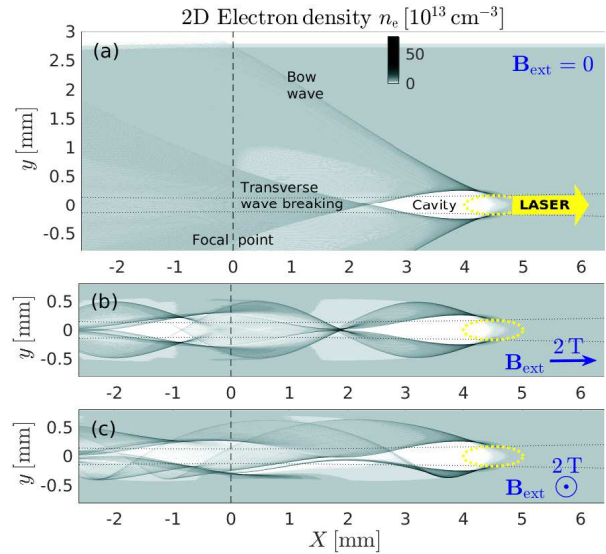}
\caption{Snapshots of the electron density $n_{\rm e}(X,y)$ in the wake of a laser pulse with $\lambda_{\rm las} = 10\,\mu{\rm m}$ and $\tau_{\rm pulse} = 2\,{\rm ps}$ in 2D EPOCH simulations with moving window and immobile ions (parameters in Tables~\protect\ref{tab:parm_laser} and \protect\ref{tab:parm_num} with scaling factor $\sigma = 1$). Results are shown for cases (a) without external magnetic field, (b) with an axial magnetic field $B_x = 2\,{\rm T}$, and (c) with a perpendicular out-of-plane magnetic field $B_z = 2\,{\rm T}$. The laser has entered from the left and propagates to the right. The location and width ($1/{\rm e}$ radius of the amplitude) of the laser pulse is indicated by a dotted yellow ellipse. The black dotted curves indicate the local spot diameter $2w(X)$ of the laser (cf.~Fig.~\protect\ref{fig:01_paraxial}(a)), and the vertical dashed lines indicate the location of the focal point ($X = 0$). Note that these snapshots show only a portion of the simulation domain and the contour plots were averaged over $4\times4$ grid points, so that the density spikes are smoothed out on that scale.}
\label{fig:02_win10mm_10um_b0-bx2-bz2_ne}%
\end{figure}

We will focus primarily on the case where an external magnetic field with strength $B_{\rm ext} = 2\,{\rm T}$ is applied in the $x$-direction (${\bm B}_{\rm ext} = B_{\rm ext} \hat{\bm e}_x$); i.e., along the laser axis. Only a few results for the unmagnetized case $B_{\rm ext} = 0$ and with a perpendicular field (${\bm B}_{\rm ext} = B_{\rm ext} \hat{\bm e}_z$) are presented for comparison (Section~\ref{sec:channel_laser}, Fig.~\ref{fig:02_win10mm_10um_b0-bx2-bz2_ne}).

Short-time simulations with a moving window, which cover less than $100\,{\rm ps}$ are performed with immobile ions. The ion response becomes important after a few $100\,{\rm ps}$, and we take it into account in simulations with a stationary window, where we follow the after-glow dynamics of the plasma wake channel for up to $1\,{\rm ns}$. In those cases, the ions are chosen to have the mass and charge of deuterium, so that $n_{\rm D} = n_{\rm e}$ initially.

The numerical parameters are summarized in Tables~\ref{tab:parm_num} and \ref{tab:parm_num_fix} for simulations with moving and stationary simulation windows, respectively. In the 2D simulations with a moving window, we fix the focal distance $x_{\rm foc} = 50\,{\rm mm}$ and the window length $L_x = 10.2\,{\rm mm}$, while varying its height $L_y = \sigma\times 6\,{\rm mm}$ in proportion to the wavelength-dependent initial spot radius $w(x=0)$ in Eq.~(\ref{eq:paraxial_w}) via the wavelength scaling factor $\sigma$. In the 2D and 3D simulations with stationary windows, we fix $w(x=0)$ and, hence, use the same transverse box size $L_y = L_z = 6\,{\rm mm}$ in all cases, while varying the box length as $L_x = 2 x_{\rm foc} + 1.5\,{\rm mm} = 100\,{\rm mm}/\sigma + 1.5\,{\rm mm}$ in proportion to the wavelength-dependent focal distance, $x_{\rm foc} = 50\,{\rm mm}/\sigma$, which decreases with increasing wavelength $\lambda_{\rm las} = \sigma\times 10\,\mu{\rm m}$.

For ease of comparison, snapshot times will be given in terms of
\begin{equation}
t' = t - t_{\rm foc};
\label{eq:tfoc}
\end{equation}

\noindent i.e., relative to the instant where the laser reaches its (theoretical) focal point $x_{\rm foc}$. In the simulations with stationary windows, $t_{\rm foc}$ varies with $L_x$ and the values can be found in Table~\ref{tab:parm_num_fix}.

All results reported in this paper were obtained using only 1 particle per cell (PPC). Convergence tests in 2D indicate that this is sufficient for the scenario and phenomena discussed in this paper. The physical reason is that the plasma is collisionless and has a highly subcritical density, so individual particles do not interact with each other and the laser is not altered by the plasma's presence. Moreover, we are presently concerned only with the overall large-scale structure of the plasma wake and its associated electromagnetic field.\footnote{Singular structures and associated phenomena are not well-resolved, as will be briefly discussed in Section~\protect\ref{sec:limsim_demo} using Fig.~\protect\ref{fig:10_win10mm_singularity}.}

We have performed convergence tests with respect to the spatial resolution and the number of simulation particles in 2D. The range of values scanned is shown in parentheses in Table~\ref{tab:parm_num} (bottom row) and Table~\ref{tab:parm_num_fix} ($N_{\rm PPC}$ column). The results of these convergence tests are not presented here, since they are similar and readily reproducible using the EPOCH code \cite{Arber15}. The spatial resolution for the 3D simulation (case (iv) in Table~\ref{tab:parm_num_fix}) was shown to be sufficient in 2D, so the 3D results are considered to be reliable for the low-density plasma at hand.

\subsection{Codes}
\label{sec:model_codes}

For the purpose of verification, we have performed our simulations with two relativistic particle-in-cell (PIC) codes, EPOCH \cite{Arber15} and REMP \cite{Esirkepov01}. Details about the equations solved and numerical methods used can be found in the respective references for each code.

The 2D results reported in Sections~\ref{sec:channel} and \ref{sec:limsim} were obtained with EPOCH using ``open'' or ``simple outflow'' boundaries, which gave similar results: the radially expanding (defocusing) laser pulse was partially reflected, but with a negligible effect as its amplitude was attenuated by a factor $\sim 10^{-3}$ when it passed again through $X = 0$.\footnote{For reflected pulses, see Fig.~\protect\ref{fig:12_fix_evol-maxEB_1ns} in Section~\protect\ref{sec:limsim_discuss}.}
Preliminary studies were performed using EPOCH versions 4.9.2 and 4.11.0, which gave similar results to version 4.14.4 that was used here. EPOCH is freely available and all simulations should be easily reproducible with the {\tt input.deck} files provided in this paper's meta data. Most of the 2D simulations, especially with laser wavelengths $> 10\,\mu{\rm m}$, can be completed within a few hours using a few 100 CPU cores.

The 3D results presented in Section~\ref{sec:3d} were obtained with REMP using open boundaries supplemented with an absorbing layer of thickness $2 s_{x,y,z}$ in all directions, which damps all electromagnetic fluctuations with a super-Gaussian factor after the laser has entered the simulation domain.

REMP and EPOCH were benchmarked against each other and gave similar results in long-time 2D simulations (up to $1\,{\rm ns}$) and in short-time 3D simulations ($\lesssim 0.1\,{\rm ns}$). At later times (after a few $100\,{\rm ps}$), we encountered discrepancies between the 3D results of REMP and our adaptation of EPOCH. The reason for these discrepancies has not been found yet, so we chose to report only the 3D results of our in-house code REMP in this paper. Note that the standard EPOCH code loads the quasi-particles in a uniform but random fashion. For the present study, we have modified EPOCH such that the quasi-particles are loaded ``uniformly'' (i.e., at equidistant positions in regions where the density is constant), as we usually do with REMP. This makes some discretization effects clearly visible and was helpful when making code-to-code comparisons. EPOCH was run with the default 1st-order b-spline (triangular) shape function,\footnote{In the present scenario, 2D EPOCH simulations using a 3rd-order b-spline shape function took 30\% longer and gave results that were very similar to those obtained with the default 1st-order b-spline.}
whereas REMP uses a quadratic form.

\section{Evolution of a magnetized laser plasma wake channel at low density in 2D}
\label{sec:channel}

\subsection{Effect of magnetization in the vicinity of the laser}
\label{sec:channel_laser}

\begin{figure}
[tbp]
\centering
\includegraphics[width=0.48\textwidth]{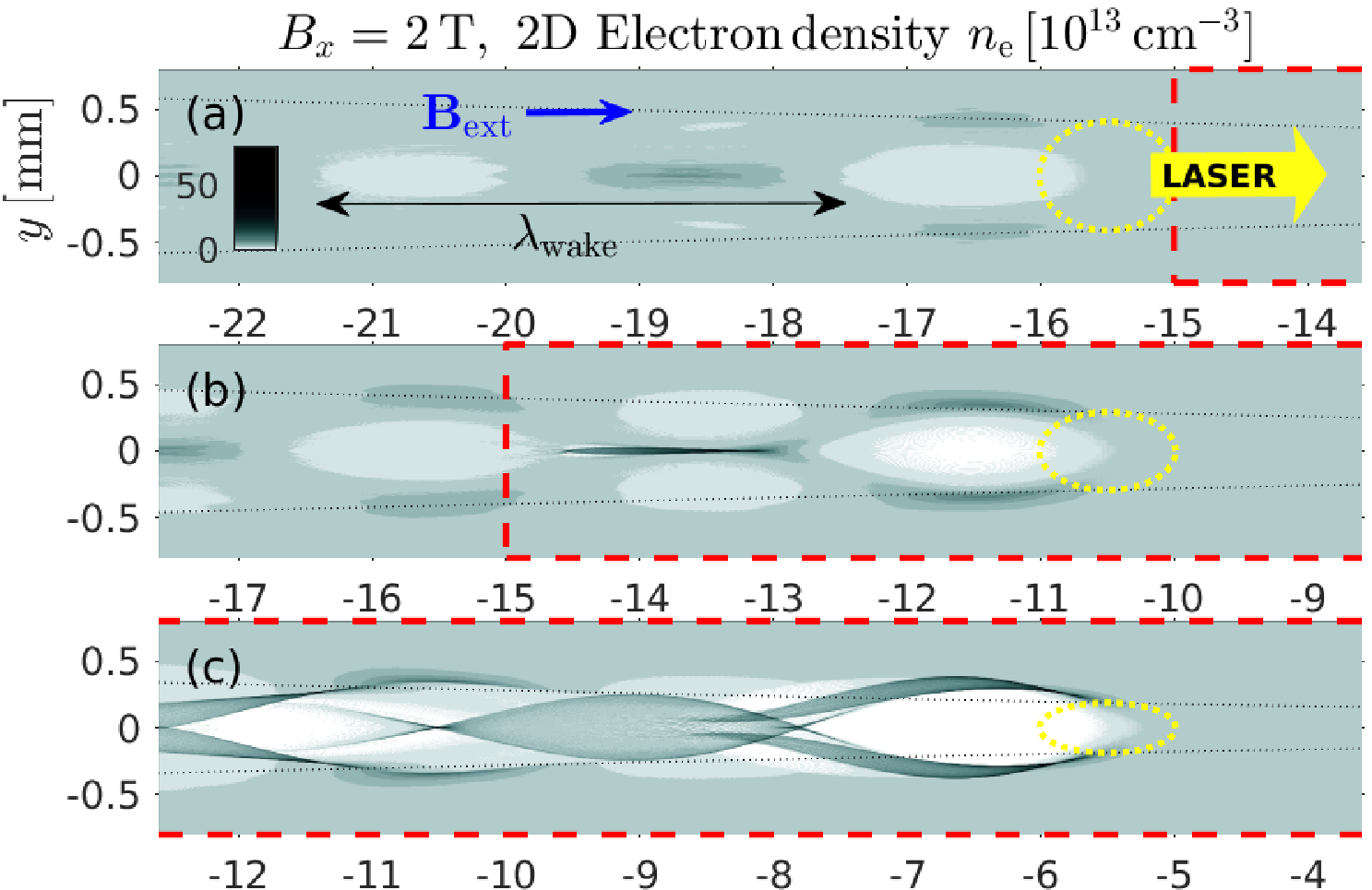}
\includegraphics[width=0.48\textwidth]{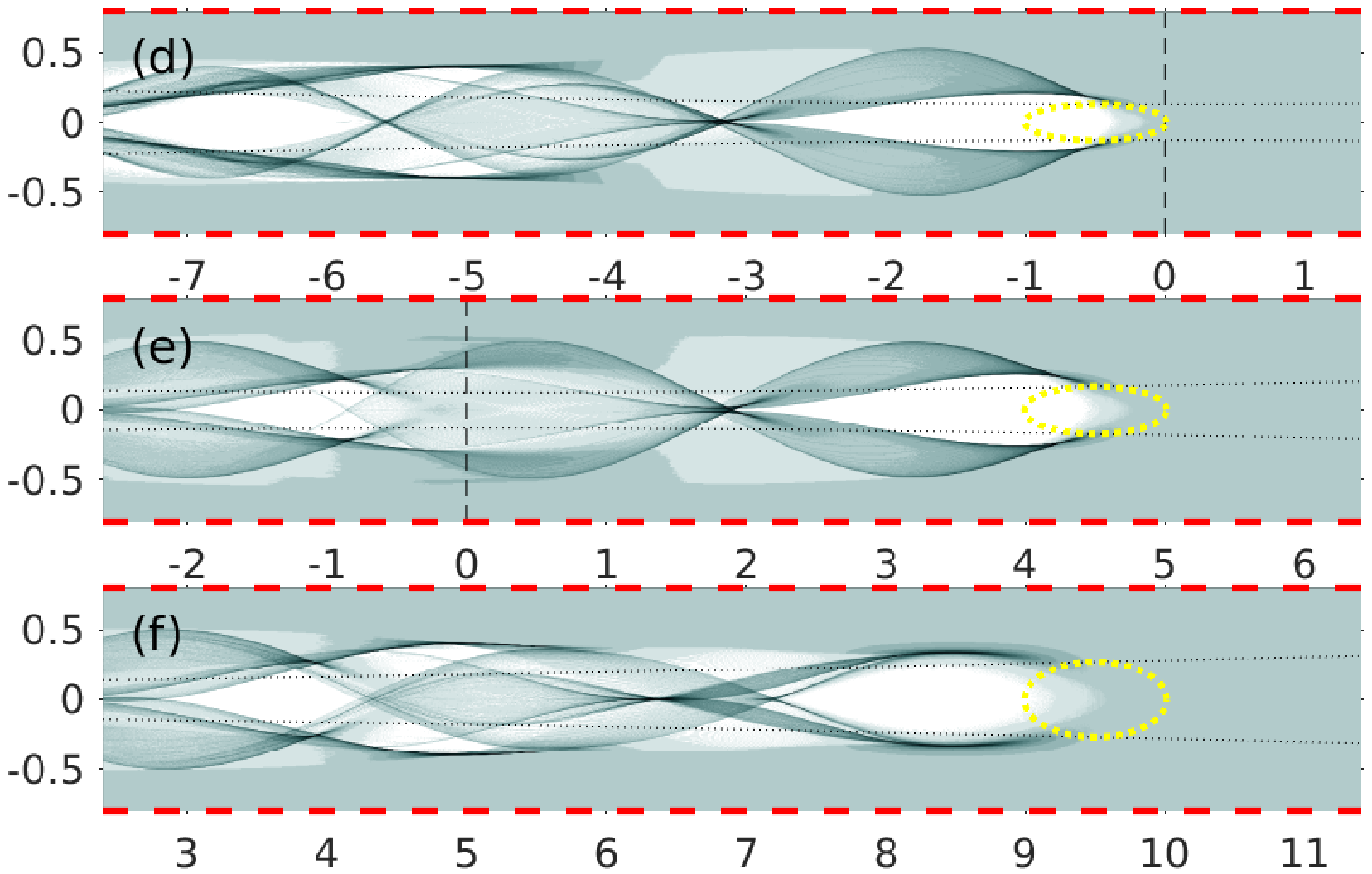} \includegraphics[width=0.48\textwidth]{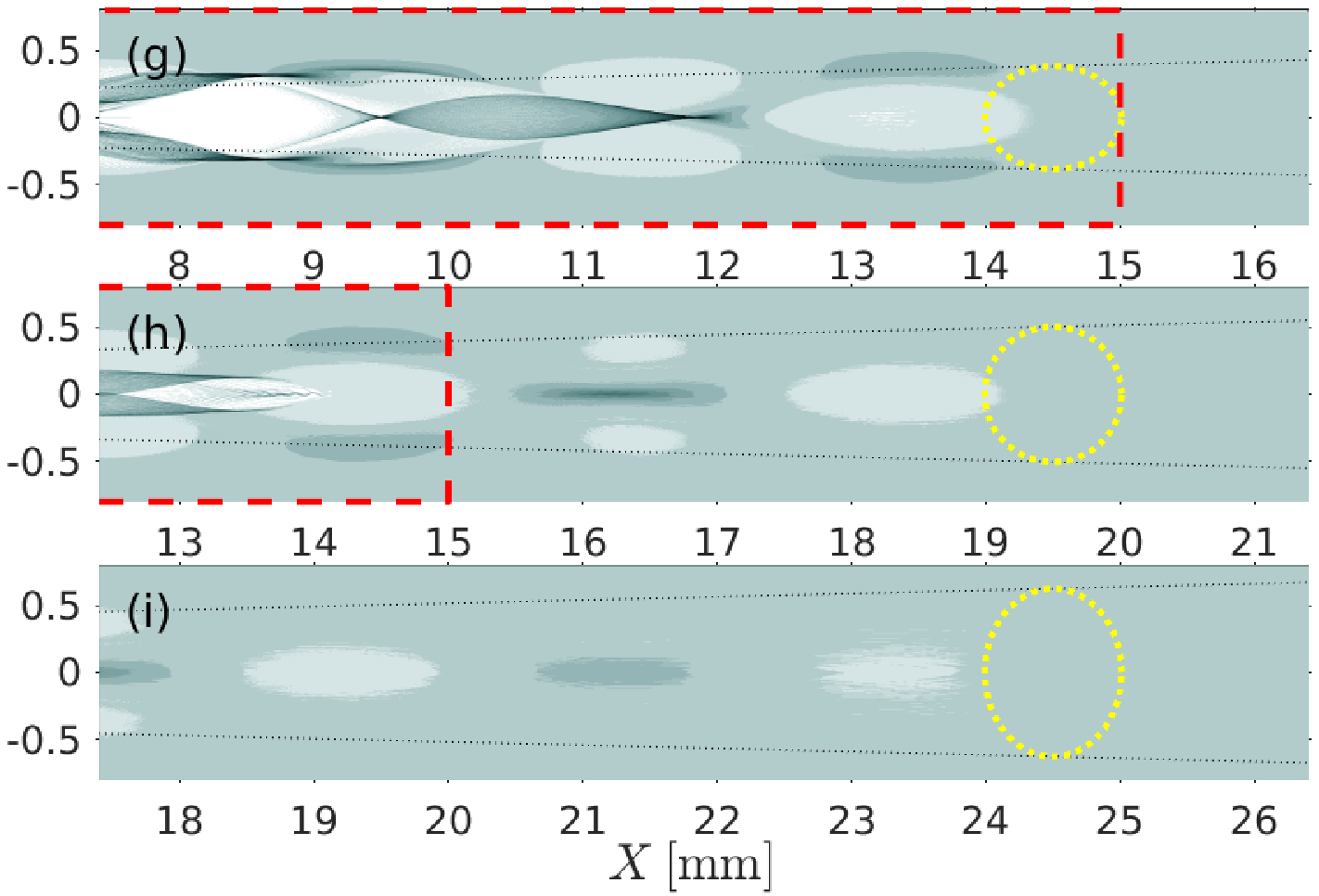}
\caption{Evolution of the electron density $n_{\rm e}(X,y)$ in the wake of a laser pulse in the same 2D case as in Fig.~\ref{fig:02_win10mm_10um_b0-bx2-bz2_ne}(b) above, with $\lambda_{\rm las} = 10\,\mu{\rm m}$ and an axial magnetic field, $B_x = 2\,{\rm T}$. Each snapshot is arranged as in Fig.~\protect\ref{fig:02_win10mm_10um_b0-bx2-bz2_ne}, with the addition of a red dashed rectangle indicating the estimated length $L_{\rm cav}$ of the region, where the laser is sufficiently intense to carve out an electron-free cavity (cf.~Eq.~(\protect\ref{eq:len_cav_ref})). In the present case, we have $L_{\rm cav}(10\,\mu{\rm m}) \approx 30\,{\rm mm}$, which is only partially visible in each panel. Snapshot (e) is the same as in Fig.~\ref{fig:02_win10mm_10um_b0-bx2-bz2_ne}(b).}
\label{fig:03_win10mm_10um_bx2-evol_ne}%
\end{figure}

We begin with a short introduction of the laser-driven electron dynamics in the parameter regime of interest. For illustration, we use the 2D simulation results in Figs.~\ref{fig:02_win10mm_10um_b0-bx2-bz2_ne} and \ref{fig:03_win10mm_10um_bx2-evol_ne}, which show snapshots of the perturbed electron density $n_{\rm e}(x,y)$ behind a 2 picosecond laser pulse with a wavelength of 10 microns that propagates through a sparse gas with low electron density $n_{\rm e0} = 3\times 10^{13}\,{\rm cm}^{-3}$. The laser parameters are given in Table~\ref{tab:parm_laser} (with scaling factor $\sigma = 1$) and the numerical parameters are listed in Table~\protect\ref{tab:parm_num}. The laser was followed for about $0.33\,{\rm ns}$ with a moving window covering $10.2\,{\rm mm}$ ($34\,{\rm ps}\times c$) along the $x$ direction. Parallelized over 1,280 CPU cores, each simulation took about 13 hours on the supercomputer JFRS-1 \cite{jfrs1}.

Figure~\ref{fig:02_win10mm_10um_b0-bx2-bz2_ne} shows the structure of the laser wake in three cases: (a) without external magnetic field, $B_{\rm ext} = 0$, (b) with an axial field $B_x = 2\,{\rm T}$ directed along the laser path, and (c) with a perpendicular out-of-plane field $B_z = 2\,{\rm T}$. The snapshots were taken at a time where the laser had traveled about $4.5\,{\rm mm}$ past its focal point.

In Fig.~\ref{fig:02_win10mm_10um_b0-bx2-bz2_ne}(a) one can see the structure of the unmagnetized laser wake in the relativistic regime, which consists of an electron-free positively charged cavity and a detached bow wave \cite{Esirkepov08}. Immediately after the laser has passed, the weakly perturbed electrons near the cavity boundary perform approximately radial plasma oscillations with angular frequency $\omega_{\rm pe}$. In agreement with theoretical predictions for $a_0 \sim 1$ \cite{Akhiezer56, Tajima79, BulanovSV16}, this produces a wake with characteristic wavelength
\begin{equation}
\lambda_{\rm wake}(B_{\rm ext}=0) \approx 2\pi c/\omega_{\rm pe} = \lambda_{\rm pe} \approx 6\,{\rm mm}
\label{eq:lWake_b0}
\end{equation}

\noindent behind the laser pulse moving away with velocity $c$. However, for the present combination of parameters, only the first half cycle is clearly visible in the form of a cavity because transverse wave breaking \cite{BulanovSV97} dominates and overshadows the subsequent wakes, which have a D-shaped form but are not clearly visible in Fig.~\ref{fig:02_win10mm_10um_b0-bx2-bz2_ne}(a).

Panels (b) and (c) of Fig.~\ref{fig:02_win10mm_10um_b0-bx2-bz2_ne} show that adding an external magnetic field with a strength of $2\,{\rm T}$ (and otherwise identical parameters) gives rise to a $-e{\bm v}\times{\bm B}_{\rm ext}$ force that is sufficiently strong to confine the bow waves near the wake channel and cause repeated transverse wave breaking, so that the laser pulse leaves behind a trail of multiple wakes and cavities. In the case with an axial magnetic field, Fig.~\ref{fig:02_win10mm_10um_b0-bx2-bz2_ne}(b) shows that the oscillations possess nearly perfect up-down symmetry. The oscillation frequency is consistent with that predicted theoretically for an extraordinary electromagnetic wave near the upper hybrid (UH) resonance \cite{BulanovSV13},
\begin{equation}
\omega_{\rm UH} = \sqrt{\omega_{\rm pe}^2 + \omega_{\rm Be}^2} \approx 2\pi \times 75\,{\rm GHz}.
\label{eq:wUH}
\end{equation}

\noindent which produces a magnetized wake with characteristic wavelength
\begin{equation}
\lambda_{\rm wake} \equiv 2\pi c/\omega_{\rm UH} \approx 4\,{\rm mm}
\label{eq:lWake}
\end{equation}

\noindent behind the laser pulse moving away with velocity $c$. When the magnetic field is perpendicular to the laser axis, the two modes of oscillation --- $\omega_{\rm pe}$ and $\omega_{\rm Be}$ --- exist independently, which produces asymmetric flow patterns, as can be seen in Fig.~\ref{fig:02_win10mm_10um_b0-bx2-bz2_ne}(c) for an out-of-plane magnetic field that causes counter-clockwise electron gyration. Since the snapshots in Figs.~\ref{fig:02_win10mm_10um_b0-bx2-bz2_ne}(b) and \ref{fig:02_win10mm_10um_b0-bx2-bz2_ne}(c) were taken during the relativistic phase, the oscillation period is somewhat obscured by the multi-flow structures associated with the repeated transverse breaking of wake and bow waves. The $4\,{\rm mm}$ wave cycle can be seen more clearly in weakly perturbed (nonrelativistic) regions away from the focal point, as will be briefly discussed in Section~\protect\ref{sec:channel_len} below for the case with axial magnetic field (for instance, see Fig.~\protect\ref{fig:03_win10mm_10um_bx2-evol_ne}(a)).

One physical consequence of the smaller radial excursion of the magnetized bow wave is that the induced electric field is weaker, while the induced current density and associated magnetic perturbations are stronger than in the unmagnetized case. From the viewpoint of numerical modeling, the magnetic confinement of the transversely accelerated electrons near the wake channel is advantageous, because it can prevent the bow wave from reaching the unphysical boundaries of the simulation box in the transverse direction, as happened in the unmagnetized case in Fig.~\ref{fig:02_win10mm_10um_b0-bx2-bz2_ne}(a). The minimal transverse size of the simulation box is then determined only by the initial spot size of the unfocused pulse at $x = 0$ ($X = -x_{\rm foc}$) (cf.~Fig.~\ref{fig:01_paraxial}(a)).

\begin{figure*}
[tb]
\centering
\includegraphics[width=0.48\textwidth]{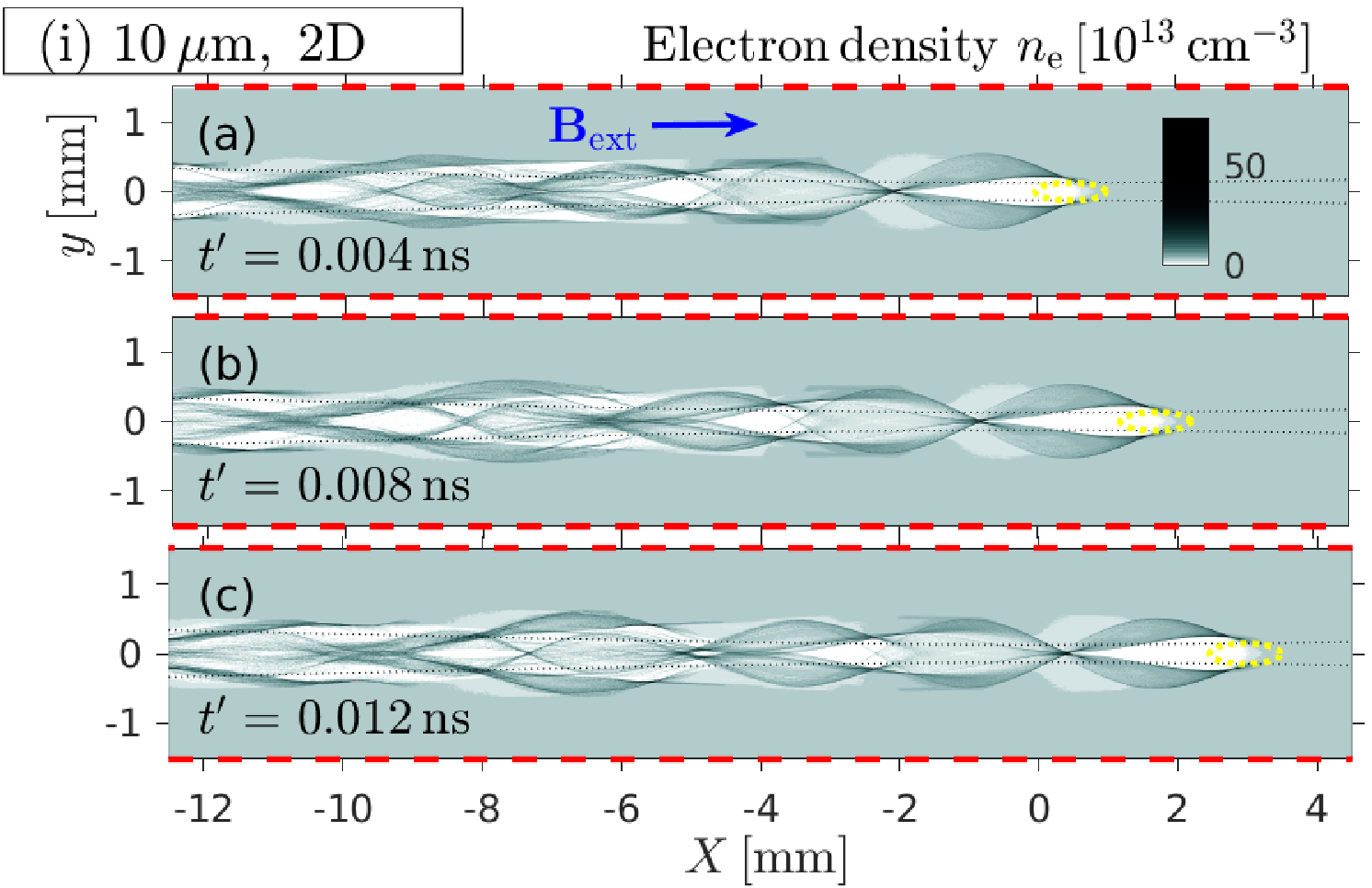} \includegraphics[width=0.48\textwidth]{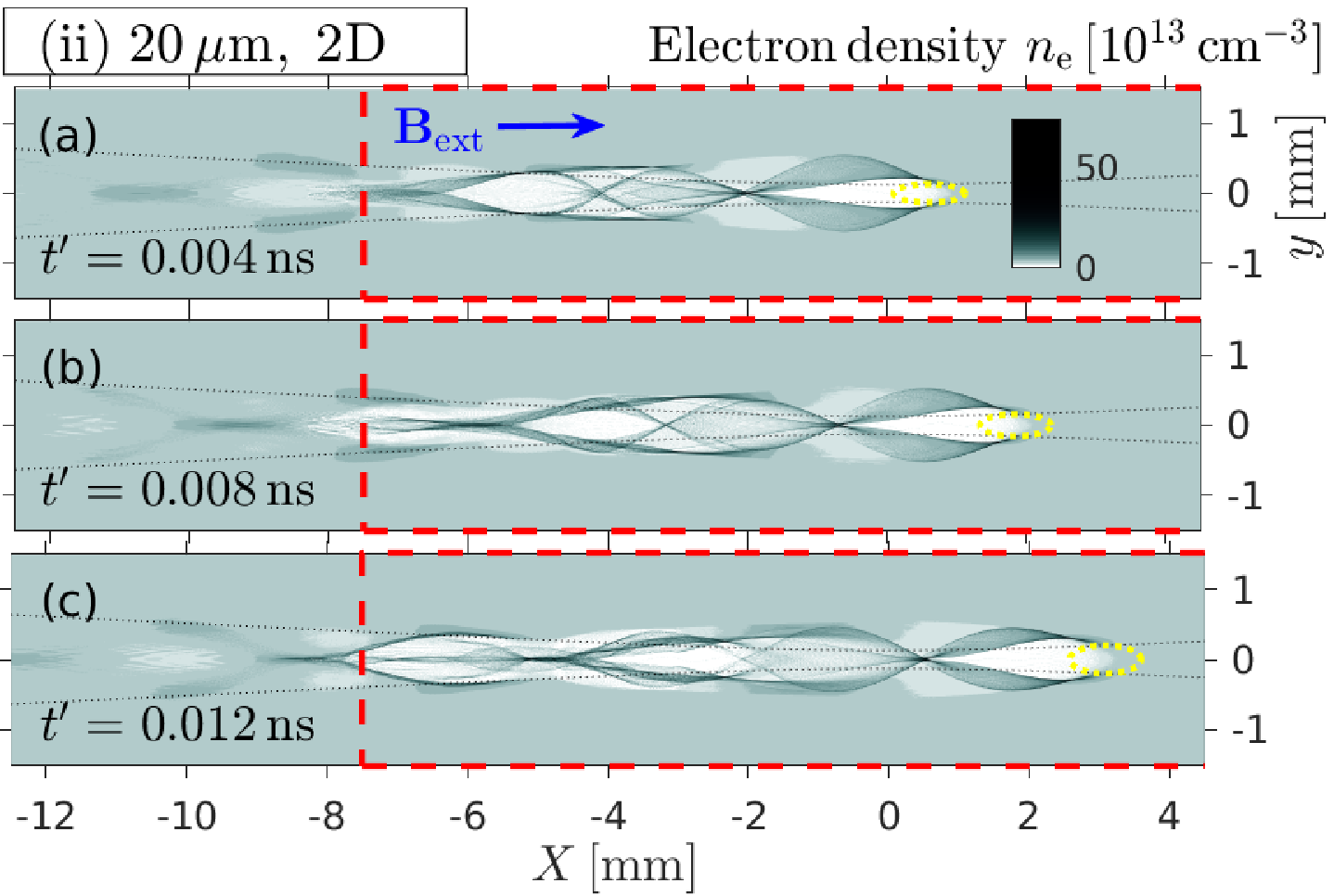} \\
\includegraphics[width=0.48\textwidth]{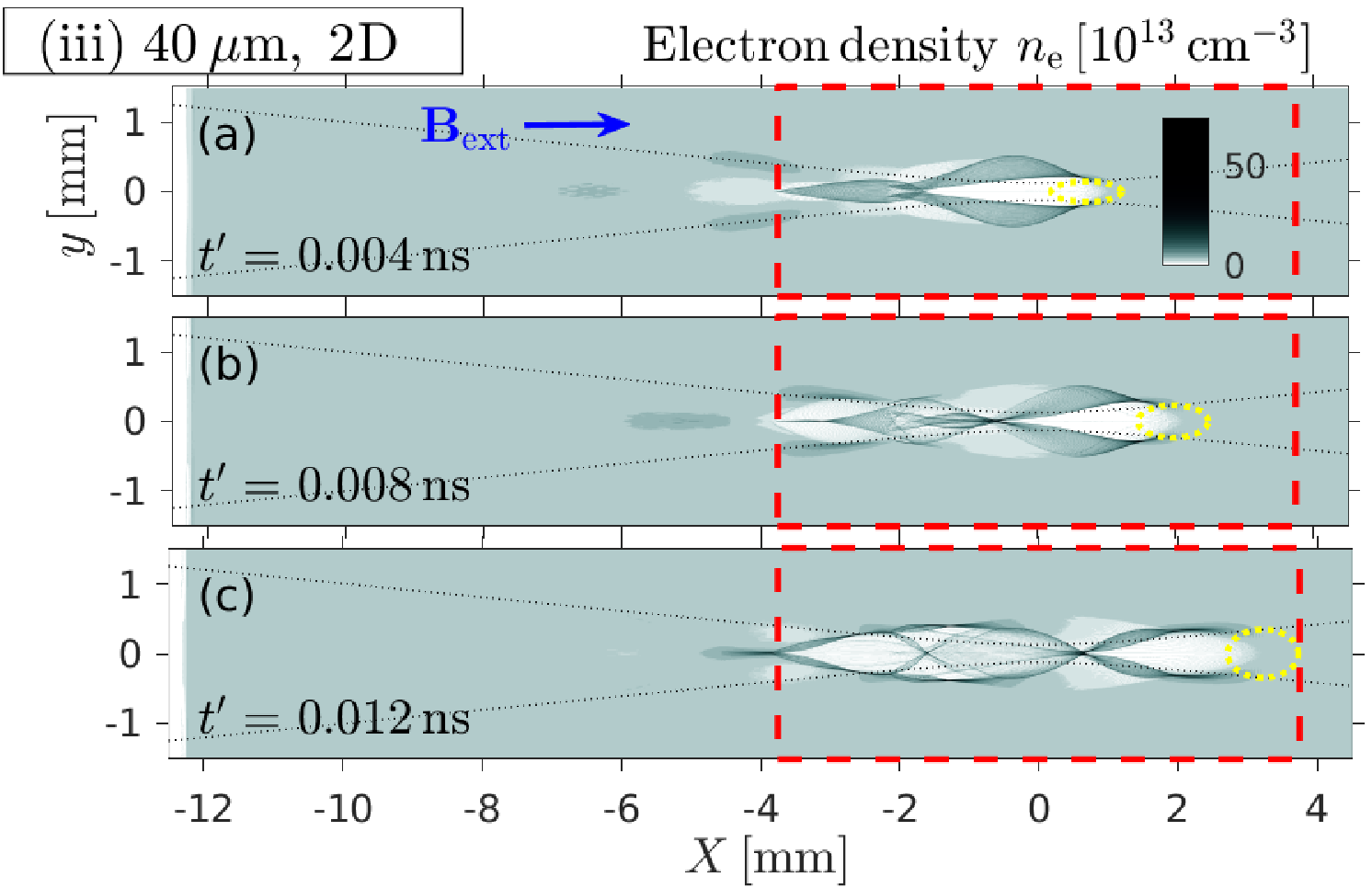}
\includegraphics[width=0.48\textwidth]{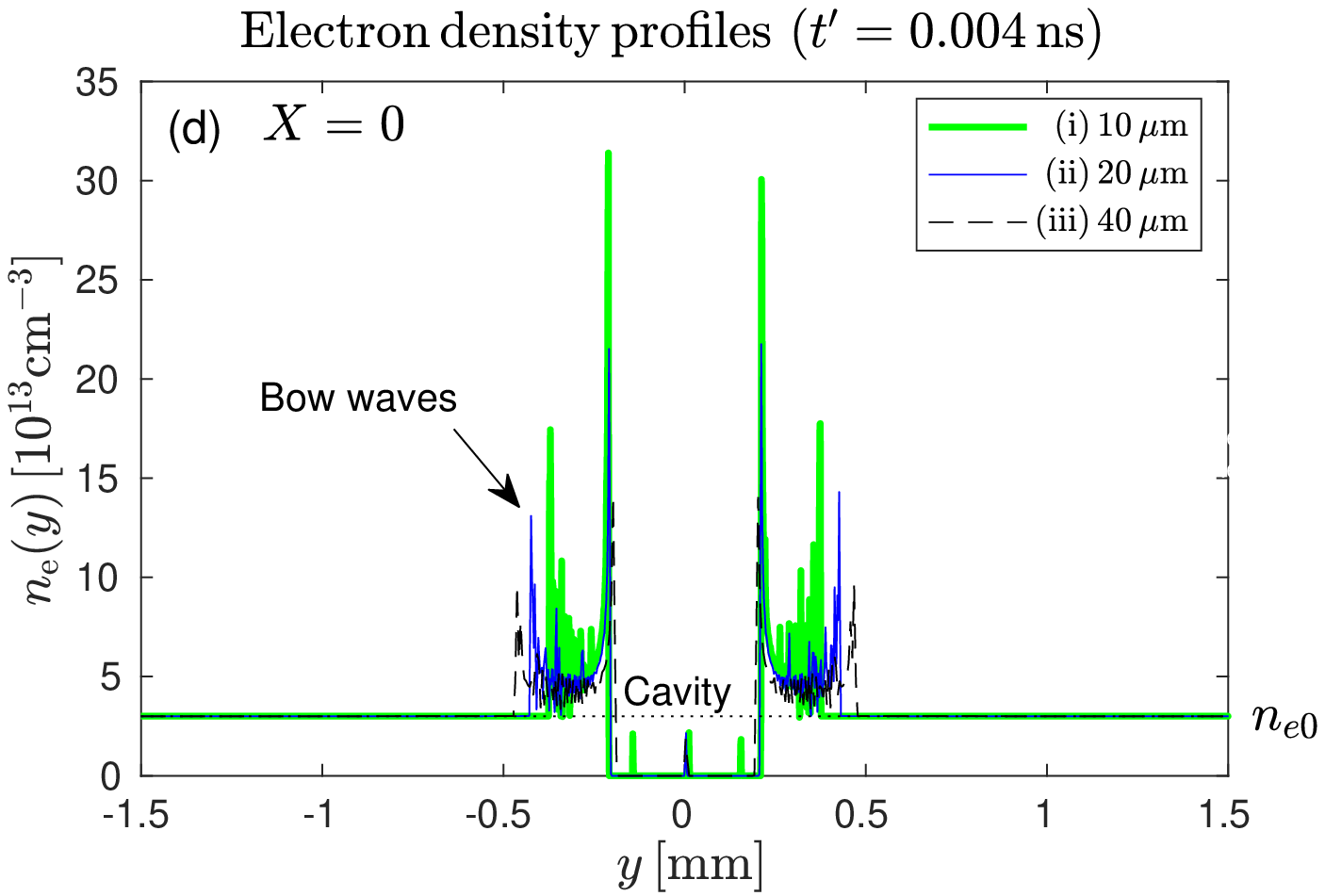}
\caption{Early phase ($t' = t - t_{\rm foc} \lesssim 12\,{\rm ps}$) of the evolution of the magnetized plasma wake in three cases: (i) $\lambda_{\rm las} = 10\,\mu{\rm m}$, (ii) $\lambda_{\rm las} = 20\,\mu{\rm m}$, and (iii) $\lambda_{\rm las} = 40\,\mu{\rm m}$ (i.e., $\sigma = 1,2,4$). For each case, three snapshots (a)--(c) of the electron density $n_{\rm e}(X,y)$ are shown in close succession, $4\,{\rm ps}$ apart, while the laser pulse (yellow ellipse) propagates away from its focal point ($X=0$). The black dotted curves indicate the local spot radius $w(X)$ of the laser pulse, and the red dashed rectangle indicates the cavitation length $L_{\rm cav} \approx 30\,{\rm mm}/\sigma$. For snapshot (a) of each case, panel (d) shows the transverse profile $n_{\rm e}(y)$ of the electron density measured at $X=0$. (2D EPOCH simulations with stationary window and mobile ions, using the parameters in Tables~\protect\ref{tab:parm_laser} and \protect\ref{tab:parm_num_fix}.)}
\label{fig:04_fix_2d_10-40um_ne}%
\end{figure*}

As we have already noted in the introduction, 2D simulations cannot fully capture the true dynamics of laser plasma wakes in the presence of an external magnetic field. At first glance, the choice of an external field along the $z$ direction as in Fig.~\ref{fig:02_win10mm_10um_b0-bx2-bz2_ne}(c) may seem to be more easily justified for 2D simulations than the use of an in-plane field, such as $B_{\rm x} = 2\,{\rm T}$ in Fig.~\ref{fig:02_win10mm_10um_b0-bx2-bz2_ne}(b). However, with an out-of-plane field, the particle motion is unmagnetized in the third dimension, so that, in reality, electrons would be pulled into the wake channel along the $z$-axis and perform plasma oscillations when the laser pulse has a finite extent in that direction. Since this effect is absent in 2D simulations with an out-of-plane magnetic field, such a setup is not useful for our purpose of investigating the long-time evolution of the plasma wake channel, where parallel streaming of charged particles along the magnetic field is expected to be a key factor that determines the channel's evolution and life time.

\begin{figure*}
[tb]
\centering
\includegraphics[width=0.48\textwidth]{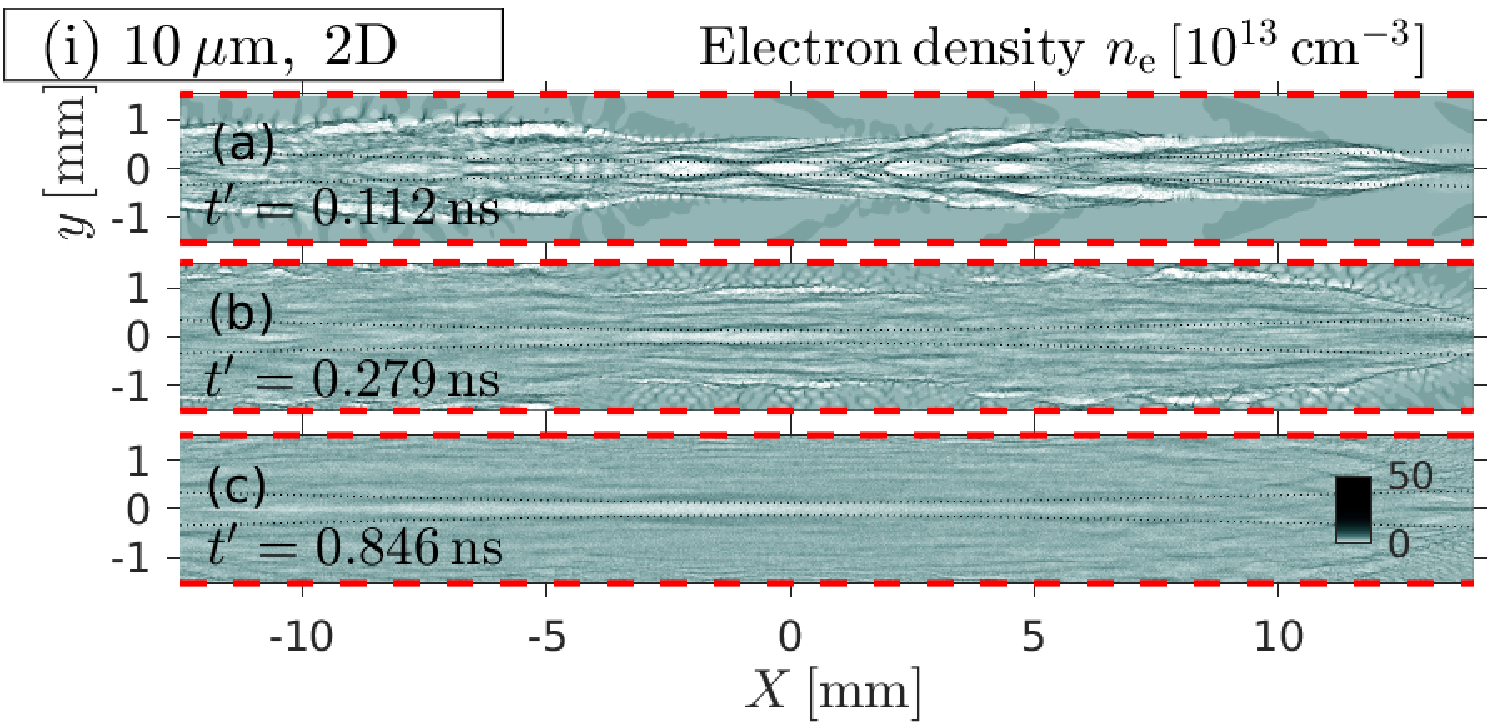} \includegraphics[width=0.48\textwidth]{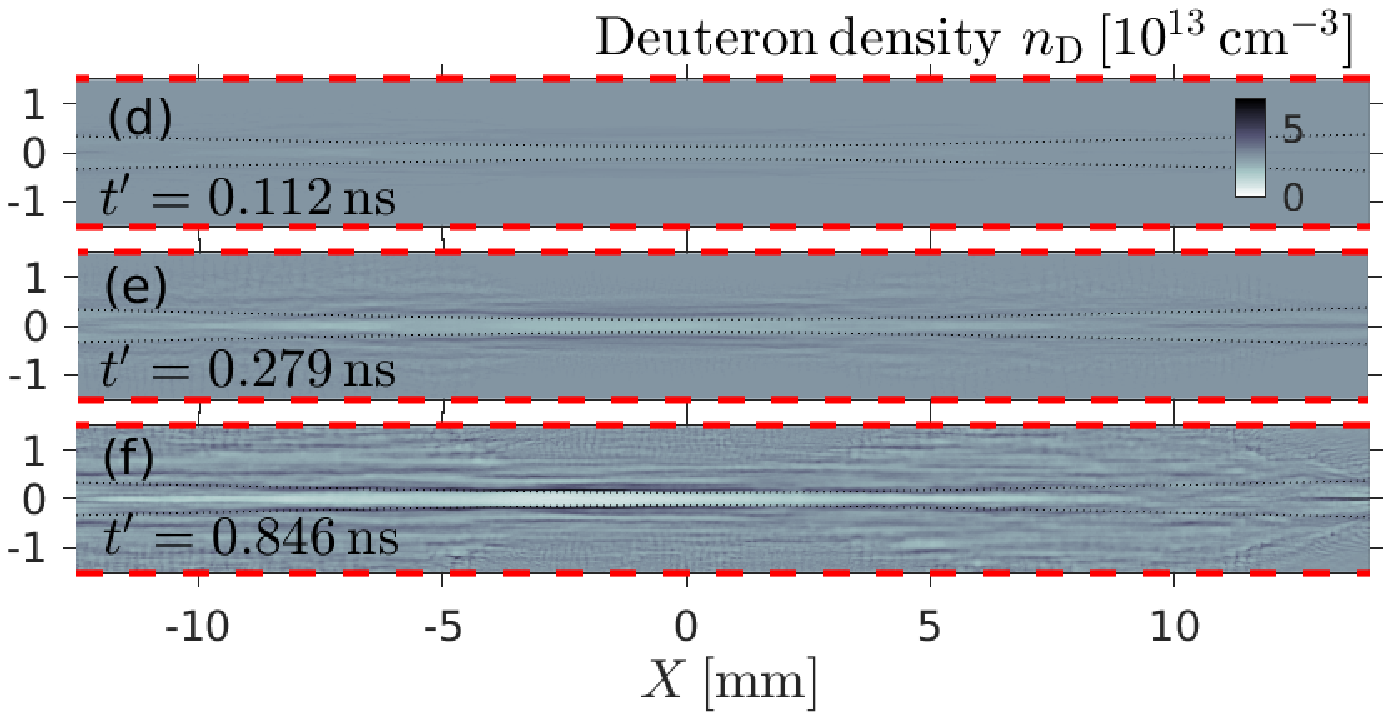} \\
\includegraphics[width=0.48\textwidth]{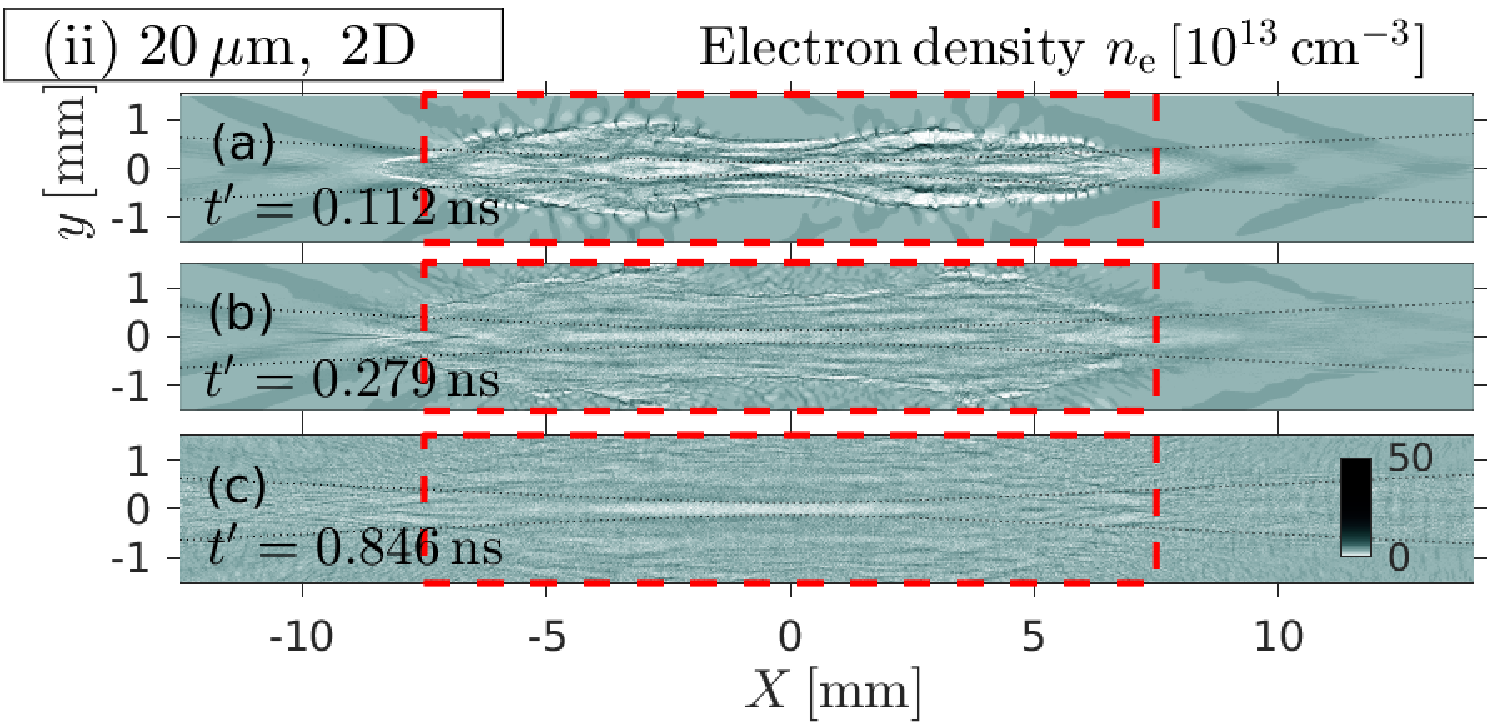}
\includegraphics[width=0.48\textwidth]{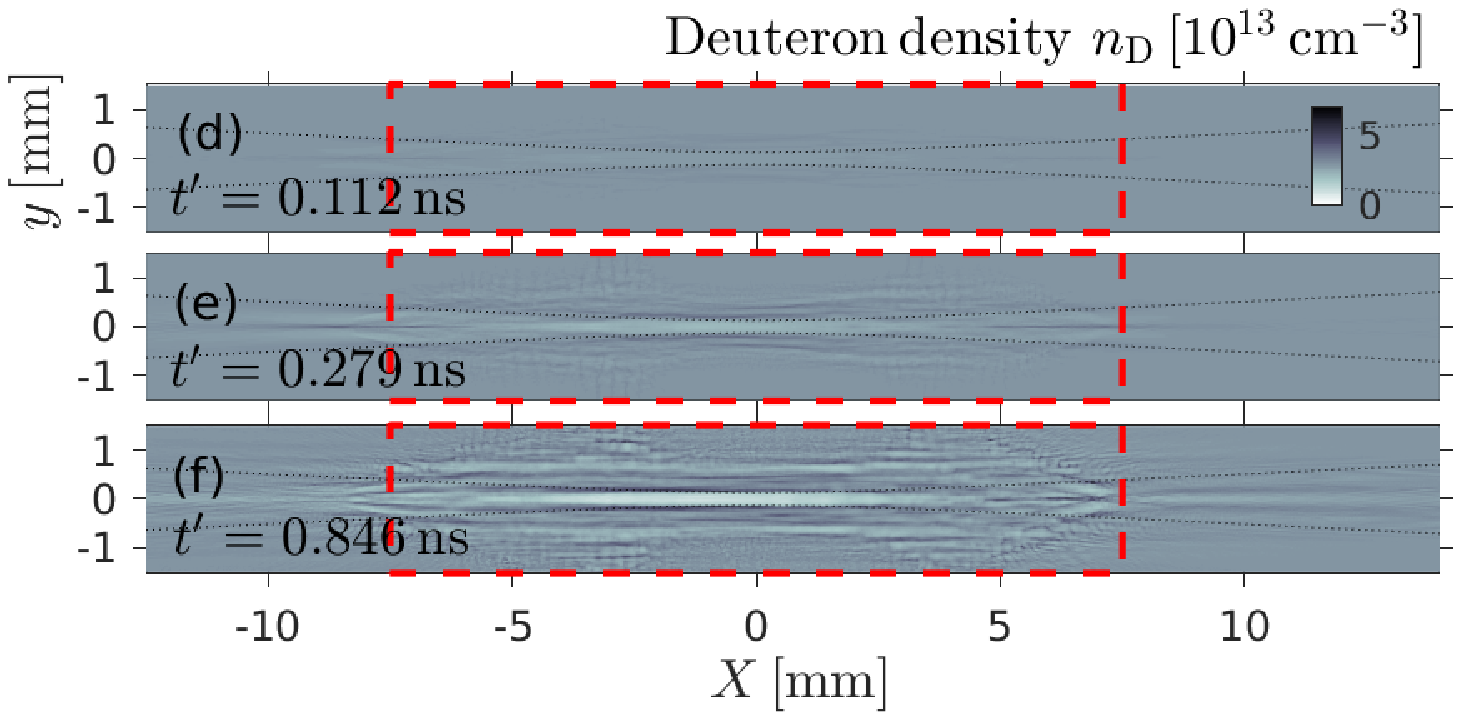} \\
\includegraphics[width=0.48\textwidth]{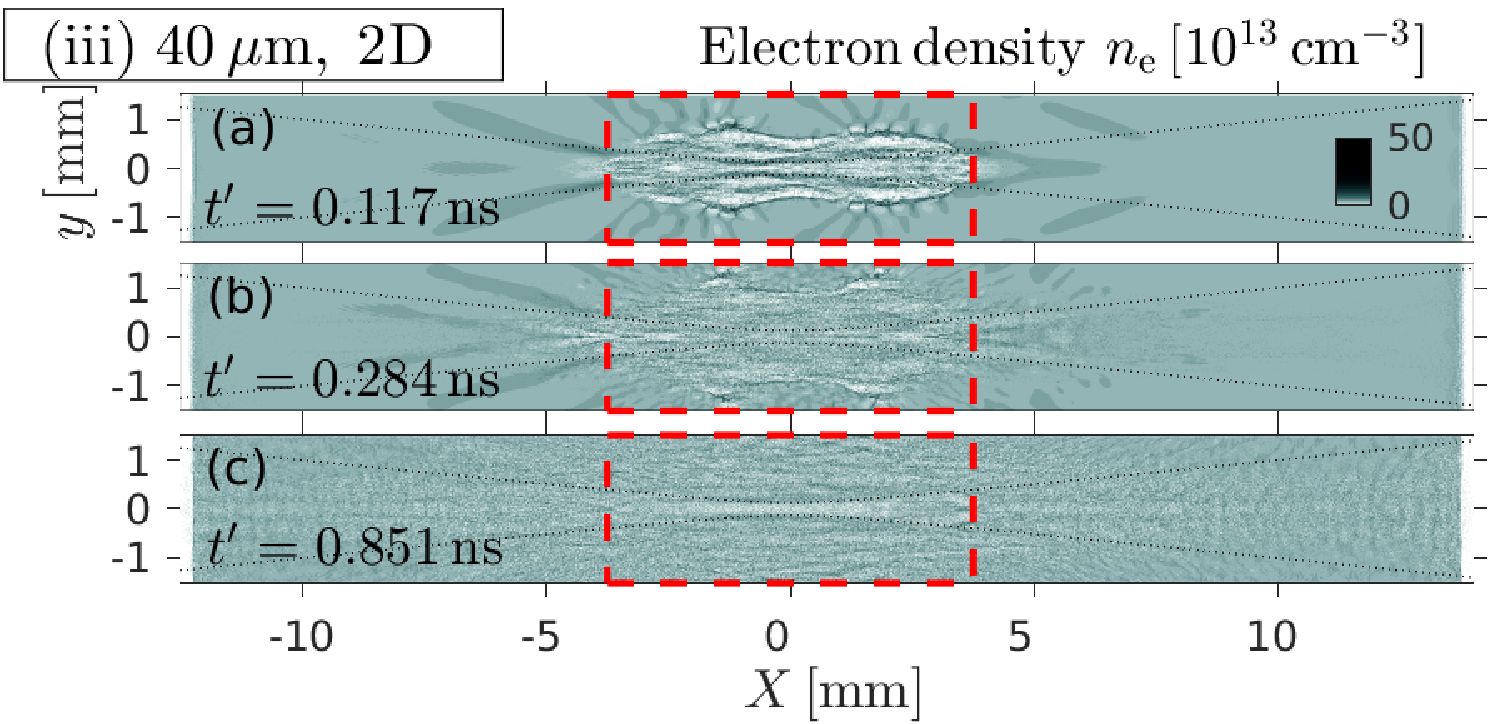} \includegraphics[width=0.48\textwidth]{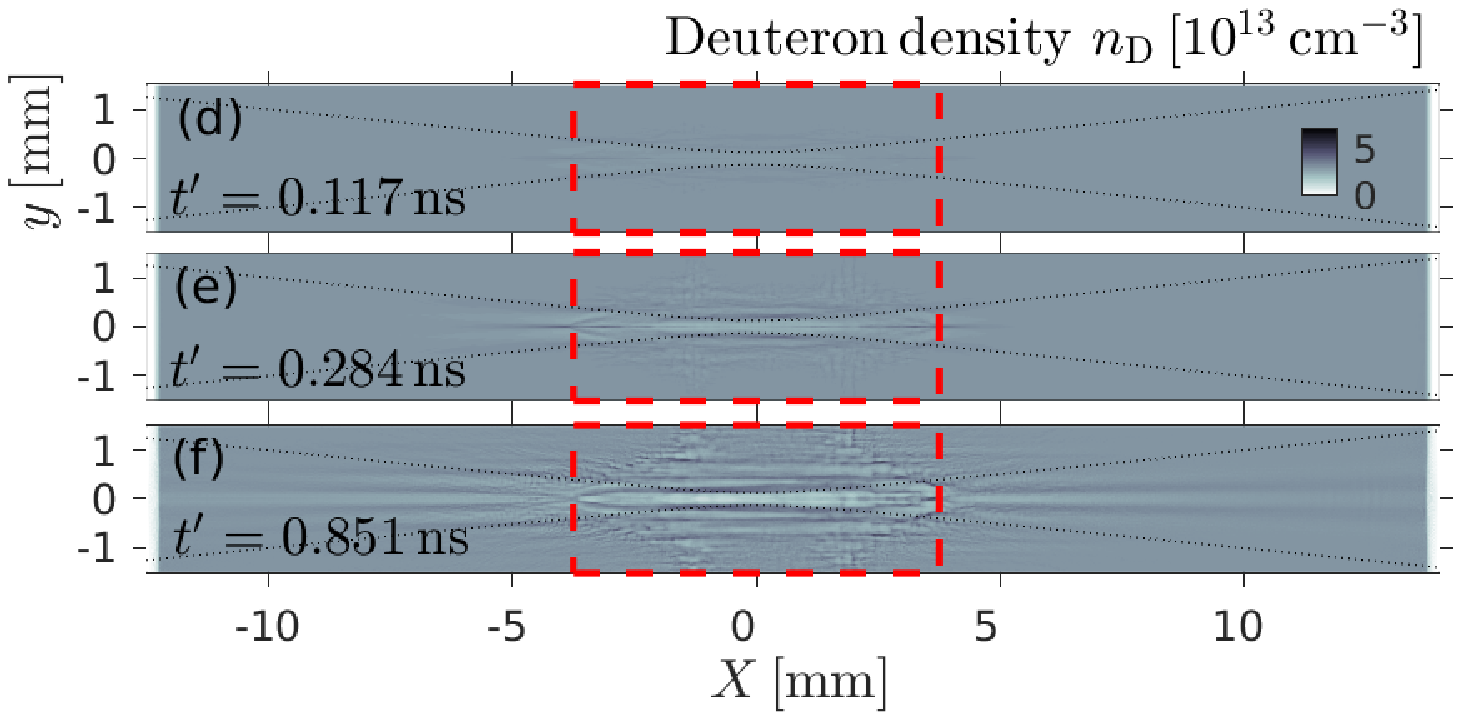}
\caption{Long-time evolution ($t' = t - t_{\rm foc} \sim 0.1$...$0.9\,{\rm ns}$) of the electron density $n_{\rm e}(X,y)$ (left) and the deuteron density $n_{\rm D}(X,y)$ (right) in the three cases (i)--(iii) shown in Fig.~\protect\ref{fig:04_fix_2d_10-40um_ne} above.}
\label{fig:05_fix_2d_10-40um_nei_late}%
\end{figure*}

Thus, in the following, we will consider only cases with an axial magnetic field as in Fig.~\ref{fig:02_win10mm_10um_b0-bx2-bz2_ne}(b) with $B_x = 2\,{\rm T}$, which will serve us as a reference case. The structures seen in Fig.~\ref{fig:02_win10mm_10um_b0-bx2-bz2_ne}(b) may be viewed as a distorted planar projection of the 3D dynamics, where the charged particles gyrate around the $x$-axis in a helical manner. The 2D code evolves the particle momenta in all three directions, but the motion along $z$ is not executed, and we have $\delta B_x = \delta B_y = \delta J_z = 0$, where $\delta{\bm B} \equiv {\bm B} - {\bm B}_{\rm ext}$ is the fluctuating part of the magnetic field. In any case, regardless of where ${\bm B}_{\rm ext}$ is pointing, the results of 2D simulations must always be interpreted with great care.

\subsection{Length of the electron-free cavity}
\label{sec:channel_len}

For the same case as in Fig.~\ref{fig:02_win10mm_10um_b0-bx2-bz2_ne}(b), the series of snapshots in Fig.~\ref{fig:03_win10mm_10um_bx2-evol_ne} shows how the structure of the magnetized wake behind the laser pulse changes as the laser focuses and defocuses. In snapshots (a,b,g,h,i), where the laser intensity is low, one can clearly see the theoretically predicted $4\,{\rm mm}$ wavelength of the wake given by Eq.~(\ref{eq:lWake}). In the other snapshots (c--f), where the laser intensity is relativistic, the wave period is obscured by multi-flow structures as we already saw in Fig.~\ref{fig:02_win10mm_10um_b0-bx2-bz2_ne}.

\begin{figure*}
[tb]
\centering
\includegraphics[width=0.96\textwidth]{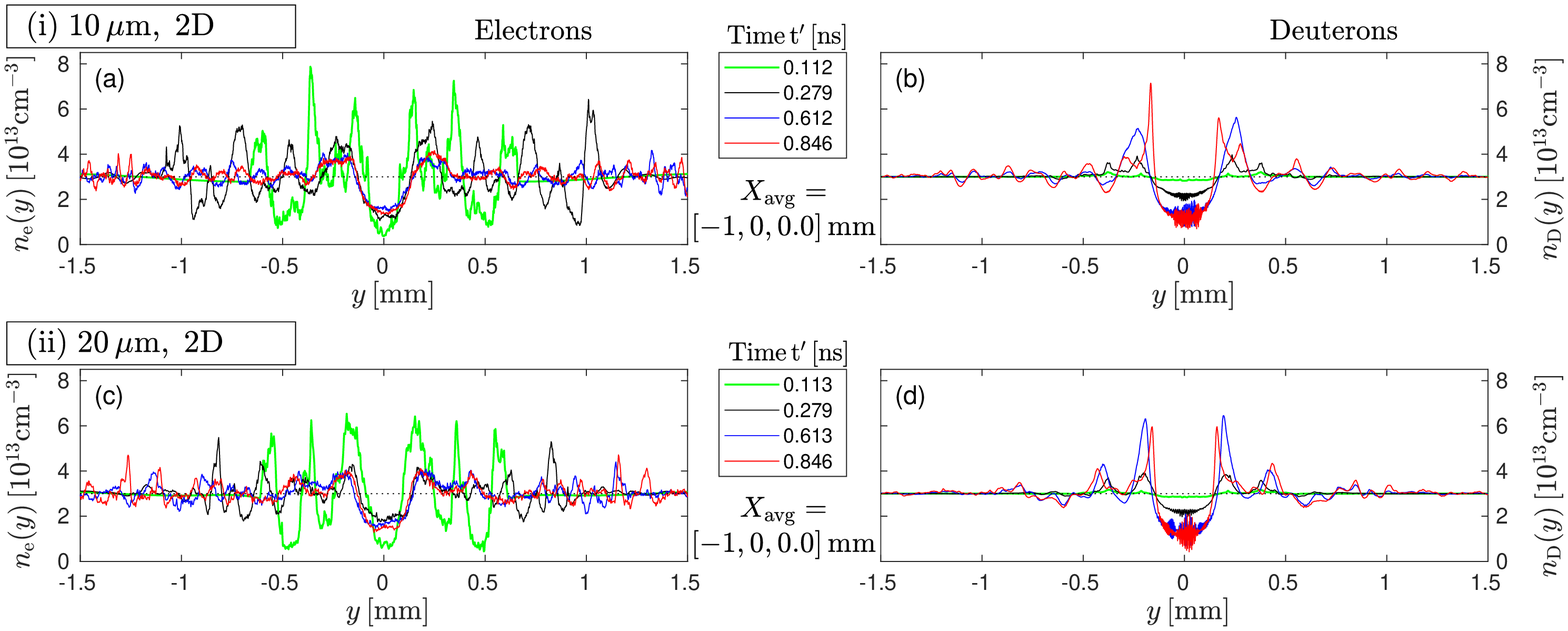} \\
\includegraphics[width=0.96\textwidth]{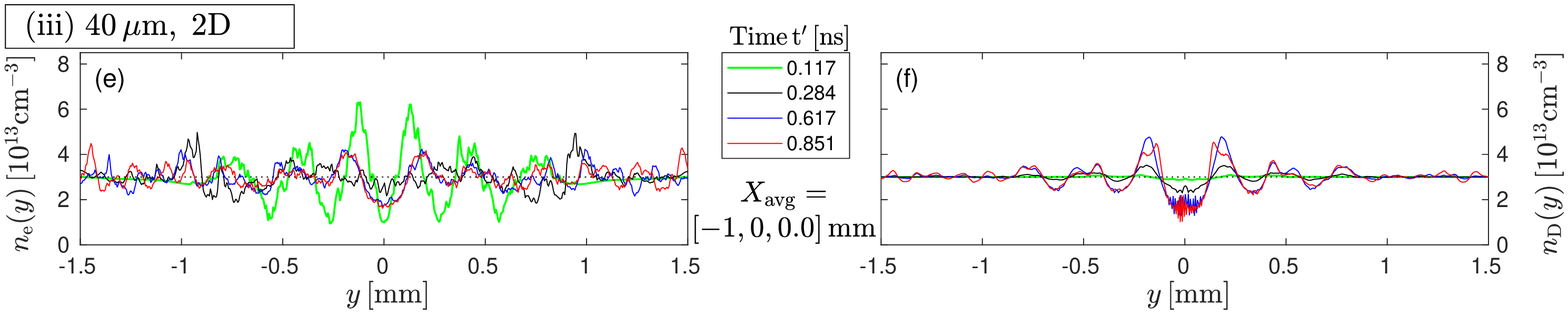}
\caption{Long-time evolution of the transverse profiles of the electron density $n_{\rm e}(y)$ (left) and the deuteron density $n_{\rm D}(y)$ (right) at the focal point ($X = 0$) in each of the three cases (i)--(iii) shown in Fig.~\protect\ref{fig:05_fix_2d_10-40um_nei_late}. For noise reduction, the profiles were spatially averaged over a distance of $1\,{\rm mm}$.}
\label{fig:06_fix_2d_scan-L_n-prof-evol}%
\end{figure*}

Figure~\ref{fig:03_win10mm_10um_bx2-evol_ne} shows that the laser is able to carve out an electron-free cavity in a region extending approximately $15\,{\rm mm}$ on both sides of the focal point $X = 0$. According to the predicted evolution of the 10 micron laser pulse in Fig.~\ref{fig:01_paraxial} above, this corresponds to the region where $a_0 \gtrsim \sqrt{1/3}$ in 2D ($N=2$) or $a_0 \gtrsim 1/3$ in 3D ($N=3$). Based on this observation, the length $L_{\rm cav}$ of the domain where a cavity forms may be estimated by substituting the condition
\begin{equation}
a_0(|X| = L_{\rm cav}/2) \approx (1/3)^{\frac{N-1}{2}}.
\label{eq:def_len_ch}
\end{equation}

\noindent into Eq.~(\ref{eq:paraxial_E}), which yields
\begin{equation}
\frac{a(|X| = L_{\rm cav}/2)}{a_{\rm 0,foc}} = \left(1 + \frac{L_{\rm cav}^2}{4 L_{\rm R}^2}\right)^{-\frac{N-1}{4}} \approx \frac{(1/3)^{\frac{N-1}{2}}}{a_{\rm 0,foc}},
\end{equation}

\noindent or
\begin{equation}
L_{\rm cav} \approx 2 L_{\rm R} \sqrt{(1/3)^{-2} a_{\rm 0,foc}^{\frac{4}{N-1}} - 1} \propto \lambda_{\rm las}^{-1},
\label{eq:len_cav}
\end{equation}

\noindent for $a_{\rm 0,foc} \geq (1/3)^{\frac{N-1}{2}}$. We will refer to this quantity $L_{\rm cav}$ as the ``cavitation length''. The exponent $N-1$ in our definition of $L_{\rm cav}$ in Eq.~(\ref{eq:def_len_ch}) implies that the cavitation length is the same in 2D and 3D when $a_{\rm 0,foc} = 1$. This empirical choice seems to be valid at least for the scenario studied here, as will be confirmed later by the 3D simulation reported in Section~\ref{sec:3d}. The empirical value $1/3$ on the right-hand side of Eq.~(\ref{eq:def_len_ch}) is somewhat arbitrary and may be replaced by another value of similar magnitude. In our reference case with $a_{\rm 0,foc} = 1$ and $\lambda_{\rm las} = 10\,\mu{\rm m}$, where $L_{\rm R} \approx 5\,{\rm mm}$, the formula in Eq.~(\ref{eq:len_cav}) yields the cavitation length
\begin{equation}
L_{\rm cav}(a_{\rm 0,foc}=1) \approx 6\times L_{\rm R}(\lambda_{\rm las}=10\mu{\rm m}) \approx 30\,{\rm mm},
\label{eq:len_cav_ref}
\end{equation}

\noindent which is indicated in Fig.~\ref{fig:03_win10mm_10um_bx2-evol_ne} by a red dashed rectangle with boundaries at $X = \pm L_{\rm cav}/2$, which is only partially visible due to its large size.

Note that the length of the cavity can be shorter than the cavitation length $L_{\rm cav}$, as in the unmagnetized case shown in Fig.~\ref{fig:02_win10mm_10um_b0-bx2-bz2_ne}(a). Furthermore, note that the length of the plasma channel evolves in time and its boundaries will deviate from our estimate $X = \pm L_{\rm cav}/2$ as particles enter the cavity by streaming along the magnetic field.

\subsection{Effect of the laser wavelength}
\label{sec:channel_Lscan}

It is clear that the cavitation length $L_{\rm cav}$ varies with the laser wavelength $\lambda_{\rm las}$ via the Rayleigh length $L_{\rm R}(\lambda_{\rm las})$, which appears as a factor in Eq.~(\ref{eq:len_cav}). In the following, we compare the magnetized wake dynamics induced by relativistic laser pulses with different wavelengths. All other parameters are the same as in Table~\ref{tab:parm_laser}.

For each of the three cases (i)--(iii) in Table~\ref{tab:parm_num_fix}, Fig.~\ref{fig:04_fix_2d_10-40um_ne} shows three snapshots (a)--(c) of the electron density $n_{\rm e}$ during the first $12\,{\rm ps}$ after the laser pulse has passed its focal point. The laser's $1/{\rm e}$ amplitude radius is indicated by a yellow dotted ellipse. While the contour plots provide an image of the overall structure of the density field, more quantitative information about the density perturbations can be gleaned from panel (d) in Fig.~\ref{fig:04_fix_2d_10-40um_ne}, which shows the transverse profile $n_{\rm e}(y)$ of the electron density at $X = 0$ for snapshot (a) taken at $t' = 4\,{\rm ps}$ in each case.

The results in Fig.~\ref{fig:04_fix_2d_10-40um_ne} show that laser pulses with different wavelength, here $10,\,20,\,40\,\mu{\rm m}$ ($\sigma = 1,\,2,\,4$), produce similar wake channels. One can also see that the starting point of the strongly perturbed region is consistent with the cavitation length $L_{\rm cav} \approx 30\,{\rm mm}/\sigma$ (red dashed rectangles) estimated by Eq.~(\ref{eq:len_cav}).

About $100$-$200\,{\rm ps}$ after the formation of the electron-free cavity, the ion response becomes visible in the contour plots of $n_{\rm D}(X,y)$ shown in Fig.~\ref{fig:05_fix_2d_10-40um_nei_late}(right) and in the corresponding transverse profiles $n_{\rm D}(y)$ in Fig.~\ref{fig:06_fix_2d_scan-L_n-prof-evol}(right). Owing to their large inertia, the ion motion becomes the dominant factor determining the further evolution of the plasma channel thereafter. During the $\lesssim 1\,{\rm ns}$ interval shown in Figs.~\ref{fig:05_fix_2d_10-40um_nei_late} and \ref{fig:06_fix_2d_scan-L_n-prof-evol}, the ion density also develops a pronounced cavity in the original cavitation region of the electron density, $-L_{\rm cav}/2 \lesssim X \lesssim L_{\rm cav}/2$ (red dashed rectangle). At this stage, the electrons in Fig.~\ref{fig:05_fix_2d_10-40um_nei_late}(left) seem to mostly follow the relatively slow ion motion.

The plasma channel grows in length along the $X$ axis, far beyond the initial cavity boundaries, first towards the left ($X < 0$) and with a small delay also towards the right ($X > 0$). The diameter of the elongated channel remains similar to the diameter of the primary cavity near the focal point, which is a clear indication of the fact that the axial elongation is due to the particles flowing along the magnetic field.\footnote{If the axial elongation of the channel was a direct but retarded effect of the laser pulse, we would expect it to become wider with increasing distance from the focal point, but this is not the case here.}

\section{Limited similarity of magnetized wakes in 2D}
\label{sec:limsim}

The results in Figs.~\ref{fig:04_fix_2d_10-40um_ne}--\ref{fig:06_fix_2d_scan-L_n-prof-evol} indicate that the overall structure and temporal evolution of the magnetized plasma wake channel in the parameter regime considered here is largely independent of the laser wavelength $\lambda_{\rm las}$. We interpret this as a manifestation of so-called {\it limited similarity} or {\it limited scaling} (see Ref.~\cite{Esirkepov12} for a review of these concepts). In this section, we examine in more detail to what extent the 2D dynamics are similar.

\begin{figure}
[tb]
\centering
\includegraphics[width=0.48\textwidth]{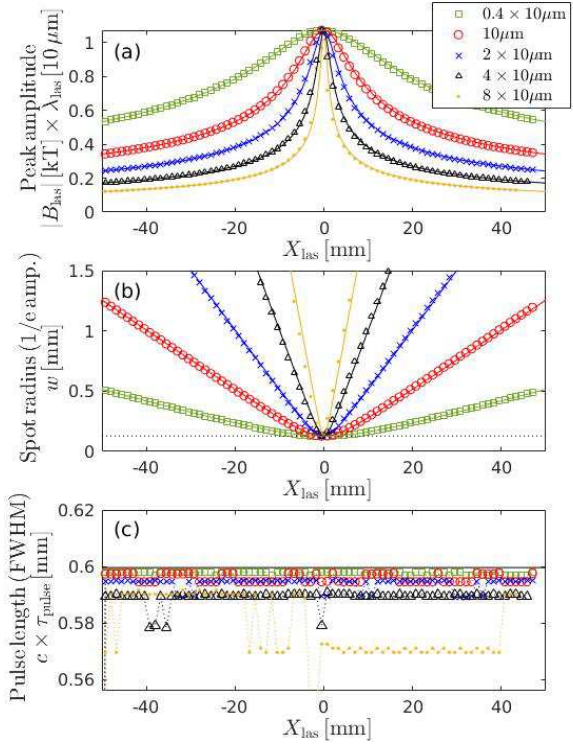}
\caption{Evolution of the laser pulse in 2D with wavelengths $\lambda_{\rm las} = \sigma \times 10\,\mu{\rm m}$ scaled by factors $\sigma = 0.4$, $1$, $2$, $4$ and $8$. (a): Peak amplitude of the laser's magnetic field multiplied by $\sigma$, which is proportional to the normalized amplitude $a_0$ in Eq.~(\protect\ref{eq:anrm}). (b): $1/{\rm e}$ spot radius. (c): FWHM pulse length. All curves are plotted as functions of the position $X_{\rm las}(t) = c(t - t_{\rm D}) - x_{\rm foc}$ of the Gaussian pulse center. The symbols are the simulation results (2D EPOCH with moving window) and the solid lines are the analytical solutions (\protect\ref{eq:paraxial_E}) and (\protect\ref{eq:paraxial_w}) of the paraxial wave equation, with $|B_{\rm las}|\times\lambda_{\rm las} \approx 2\pi a_0 m_{\rm e} c/e$ according to Eq.~(\protect\ref{eq:anrm}).}
\label{fig:07_win10mm_evol-pulse_L04-80}%
\end{figure}

\subsection{Preliminary considerations}
\label{sec:limsim_prep}

Two systems are guaranteed to behave identically if their dynamics are governed by the same equations and if all dimensionless parameters have the same values. In contrast, limited scaling requires only that the same phenomena dominate in the simulation ``and in nature, i.e. dimensionless quantities in nature which are small compared to unity should be small in the model, but not necessarily by the same order of magnitude'' \cite{Block67}. In our case, the small key parameter is the ratio of the initial electron density to the critical density,
\begin{equation}
n_{\rm e}/n_{\rm crit} = \omega_{\rm pe}^2/\omega_{\rm las}^2 \ll 1,
\label{eq:small}
\end{equation}

\noindent where
\begin{equation}
n_{\rm crit} = \frac{\epsilon_0 m_{\rm e}}{e^2} \omega_{\rm las}^2 = \frac{4\pi^2 m_{\rm e}}{\mu_0 e^2} \frac{1}{\lambda_{\rm las}^2} \approx 10^{21} {\rm cm}^{-3} \left(\frac{[\mu{\rm m}]}{\lambda_{\rm las}}\right)^2
\label{eq:ncrit}
\end{equation}

\noindent is the density for which the laser frequency $\omega_{\rm las} = 2\pi c/\lambda_{\rm las}$ equals the electron plasma frequency $\omega_{\rm pe}$ given by Eq.~(\ref{eq:wpe}).

\begin{figure*}
[tb]
\centering
\includegraphics[width=0.48\textwidth]{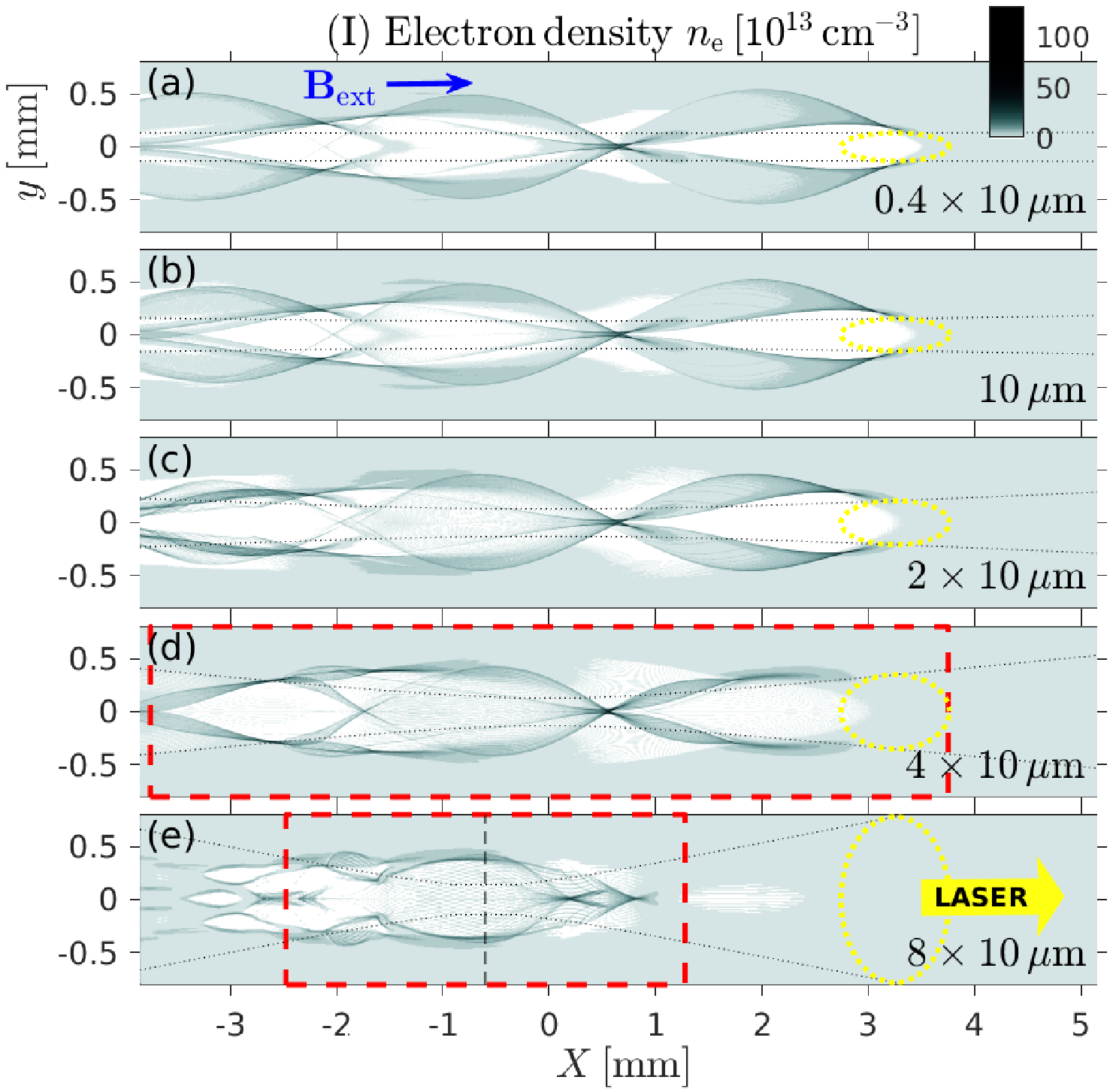}
\includegraphics[width=0.48\textwidth]{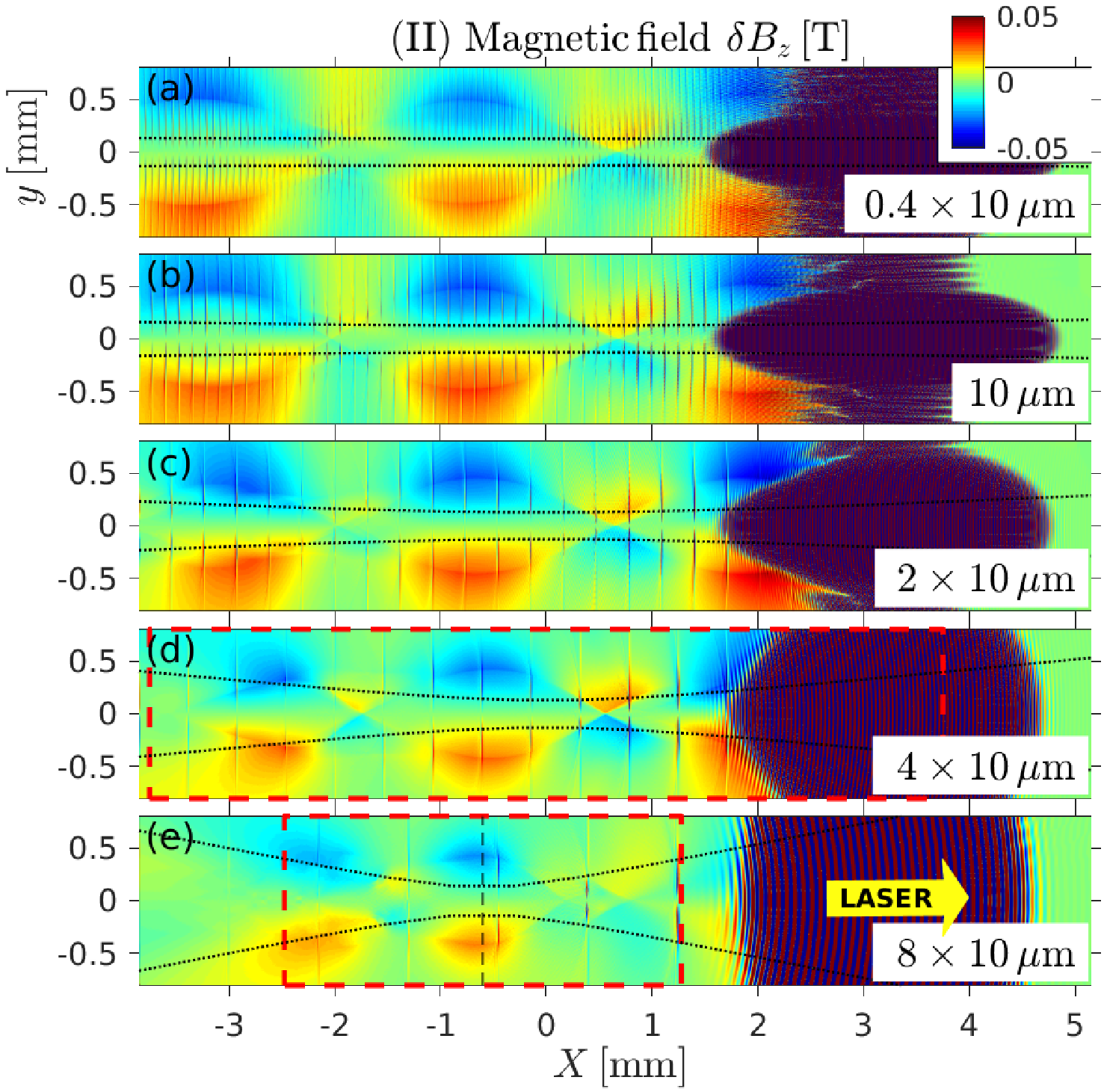}
\vspace{-0.25cm}
\caption{Contour plots of (I) the electron density field $n_{\rm e}(X,y)$ (left) and (II) the out-of-plane magnetic field fluctuations $\delta B_z(X,y)$ (right) in 2D simulations with axial magnetic field $B_x = 2\,{\rm T}$ and different wavelengths $\lambda_{\rm las} = 4,\,10,\,20,\,40,\,80\,\mu{\rm m}$ (a)--(e). The laser pulse has entered from the left-hand side and all snapshots are taken at the time where the pulse center is located at $X_{\rm las} \approx 3.25\,{\rm mm}$ (yellow ellipses), a few millimeters beyond its focal point ($X = 0$). The black dotted curves indicate the local spot diameter $2 w(X)$ of the laser (cf.~Fig.~\protect\ref{fig:07_win10mm_evol-pulse_L04-80}(b)). In (d) and (e), the boundaries of the cavitation domain, $X = \pm L_{\rm cav}/2$ (cf.~Eq.~(\protect\ref{eq:len_cav_ref})), lie within the plotted range and are indicated by a dashed red rectangle. In (e), that rectangle was shifted by $\Delta x_{\rm foc} = -0.6\,{\rm mm}$ to match the actual location of the focal point inferred from Fig.~\protect\ref{fig:07_win10mm_evol-pulse_L04-80}(b). The magnetic fluctuation amplitude $\delta B_z$ was clipped at $\pm 50\,{\rm mT}$, so the magnetic field in the vicinity of the laser appears as a saturated dark area in the contour plots on the right-hand side. The vertical stripes in the $\delta B_z$ contours seem to be harmless post-processing artifacts of the EPOCH code and coincide with the MPI domain boundaries along $X$. (2D EPOCH simulations with moving window and immobile ions using the parameters in Tables~\protect\ref{tab:parm_laser} and \protect\ref{tab:parm_num}.)}
\label{fig:08_win10mm_scan-L_ne-bz}%
\end{figure*}

As explained in Sections~\ref{sec:intro} and \ref{sec:setup} above, we intend to apply our simulations to magnetically confined fusion plasmas, in particular tokamaks. Such plasmas tend to have a relatively flat density profile and are constrained by an empirical limit known as the ``Greenwald density limit'' \cite{Greenwald02}, which implies that the electron density near the plasma surface cannot be much higher than $10^{14}\,{\rm cm}^{-3}$. The reference density of $n_{\rm e} = 3\times 10^{13}\,{\rm cm}^{-3}$ in our simulations was chosen to lie somewhat below this limit. From Eq.~(\ref{eq:ncrit}) one can readily see that this is at least 5 orders of magnitude smaller than the critical density for typical high power lasers, whose wavelengths $\lambda_{\rm las}$ lie between $0.8\,\mu{\rm m}$ (for solid-state Ti:sapphire lasers) and $10\,\mu{\rm m}$ (for ${\rm CO}_2$ gas lasers). This means that the condition $n_{\rm e}/n_{\rm crit} \ll 1$ in Eq.~(\ref{eq:small}) will be satisfied even if we increase the laser wavelength in the simulation far above the original reference value $\lambda_{\rm ref}$ by a large scaling factor $1 \ll \sigma \ll (n_{\rm crit}/n_{\rm e})^{1/2}$ as
\begin{equation}
\lambda_{\rm las} = \sigma \lambda_{\rm ref}.
\label{eq:fscale}
\end{equation}

In the following subsections \ref{sec:limsim_demo}--\ref{sec:limsim_discuss}, we compare in detail the magnetized plasma dynamics induced by relativistic laser pulses with wavelength $\lambda_{\rm las} = \sigma \times 10\,\mu{\rm m}$ and scaling factors
\begin{equation}
\sigma = 0.4,\, 1,\, 2,\, 4\, 8.
\end{equation}

\noindent All other parameters are the same as in Table~\ref{tab:parm_laser}. For these values, the chosen pulse length $\tau_{\rm pulse} = 2\,{\rm ps}$ is sufficient to accommodate a reasonable number of wave cycles, which is $N_\lambda = c\tau_{\rm pulse}/\lambda_{\rm las} = 7.5$ for $\sigma = 8$.

\begin{figure*}
[tb]
\centering\includegraphics[width=0.48\textwidth]{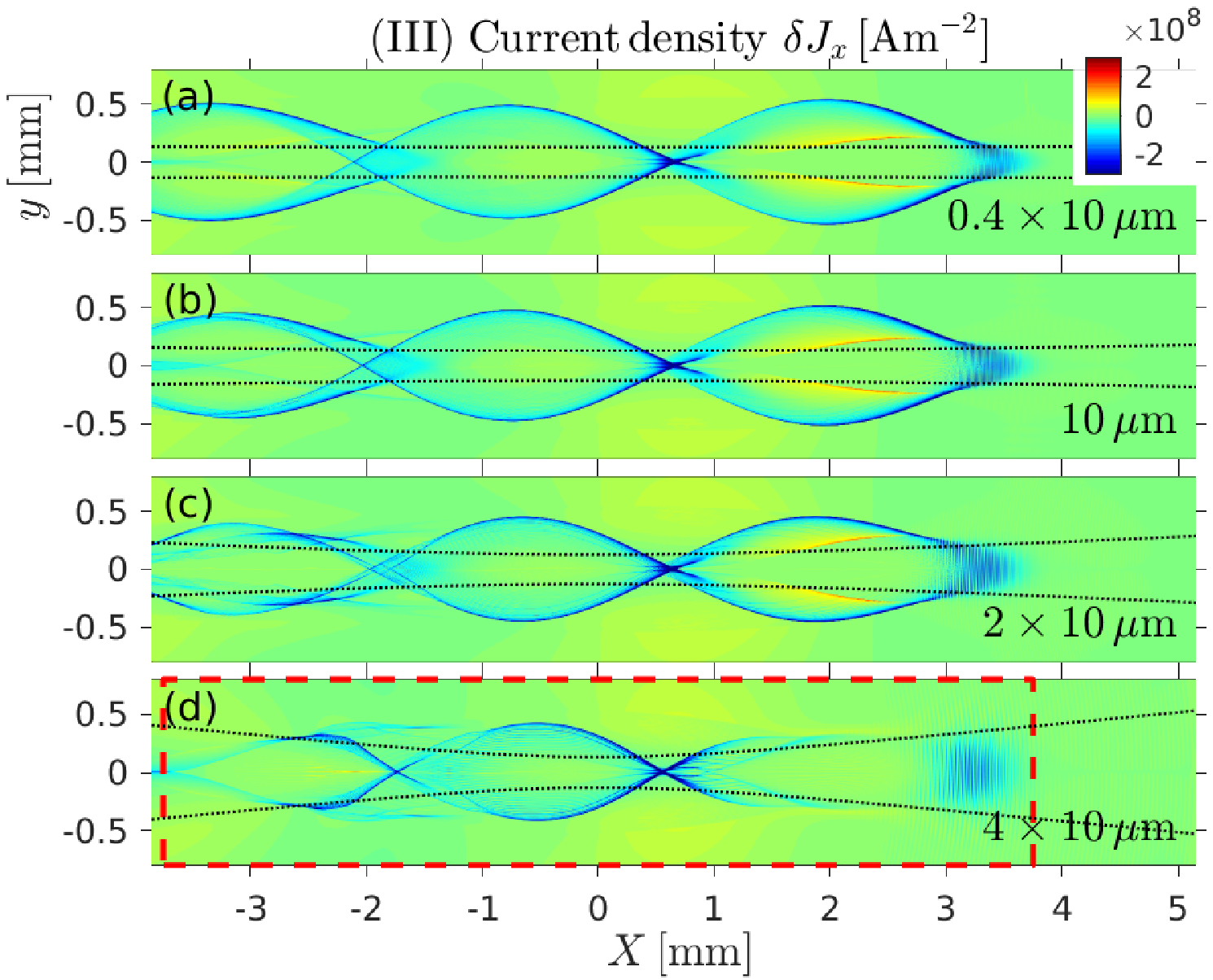}
\includegraphics[width=0.48\textwidth]{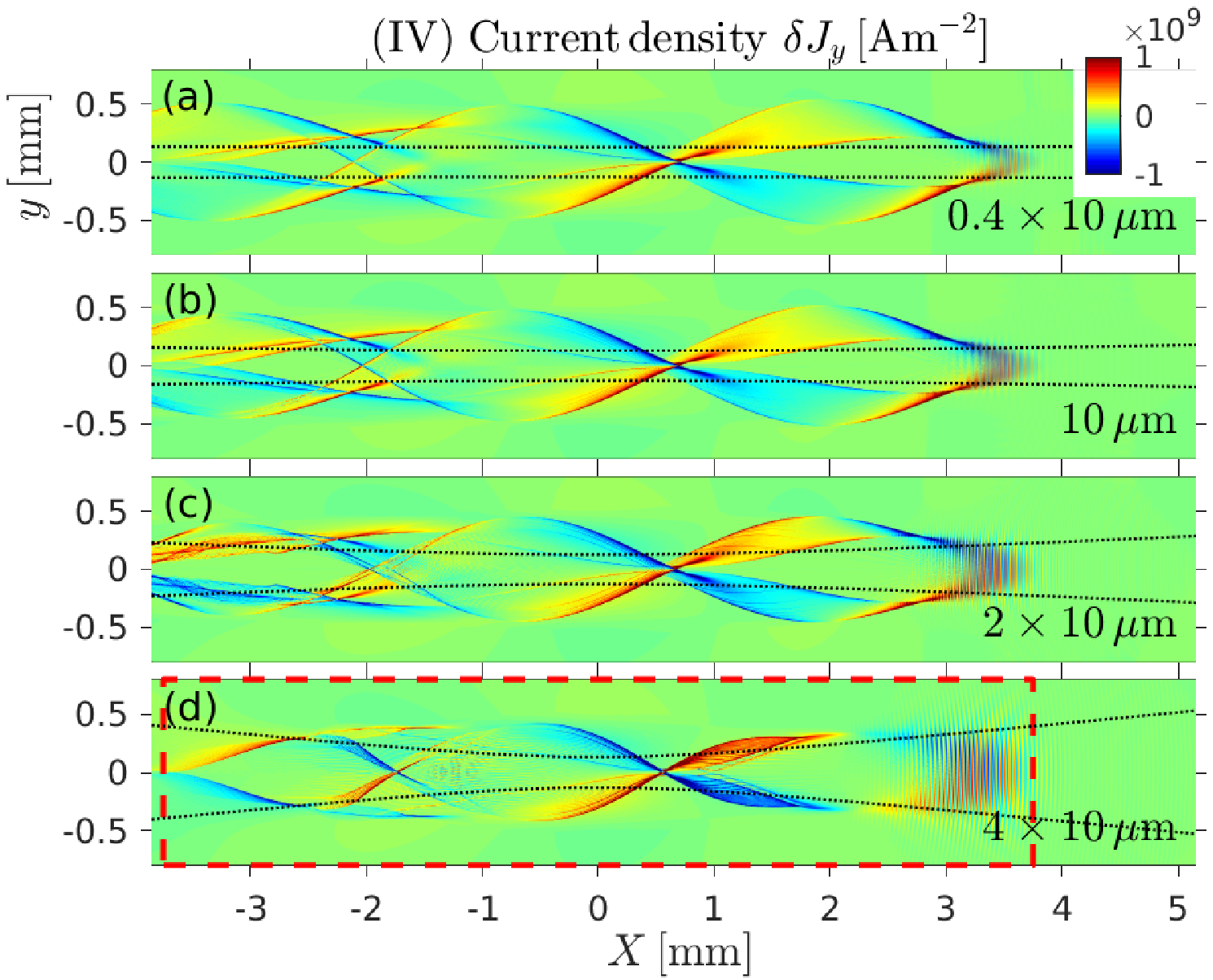} \\
\centering\includegraphics[width=0.48\textwidth]{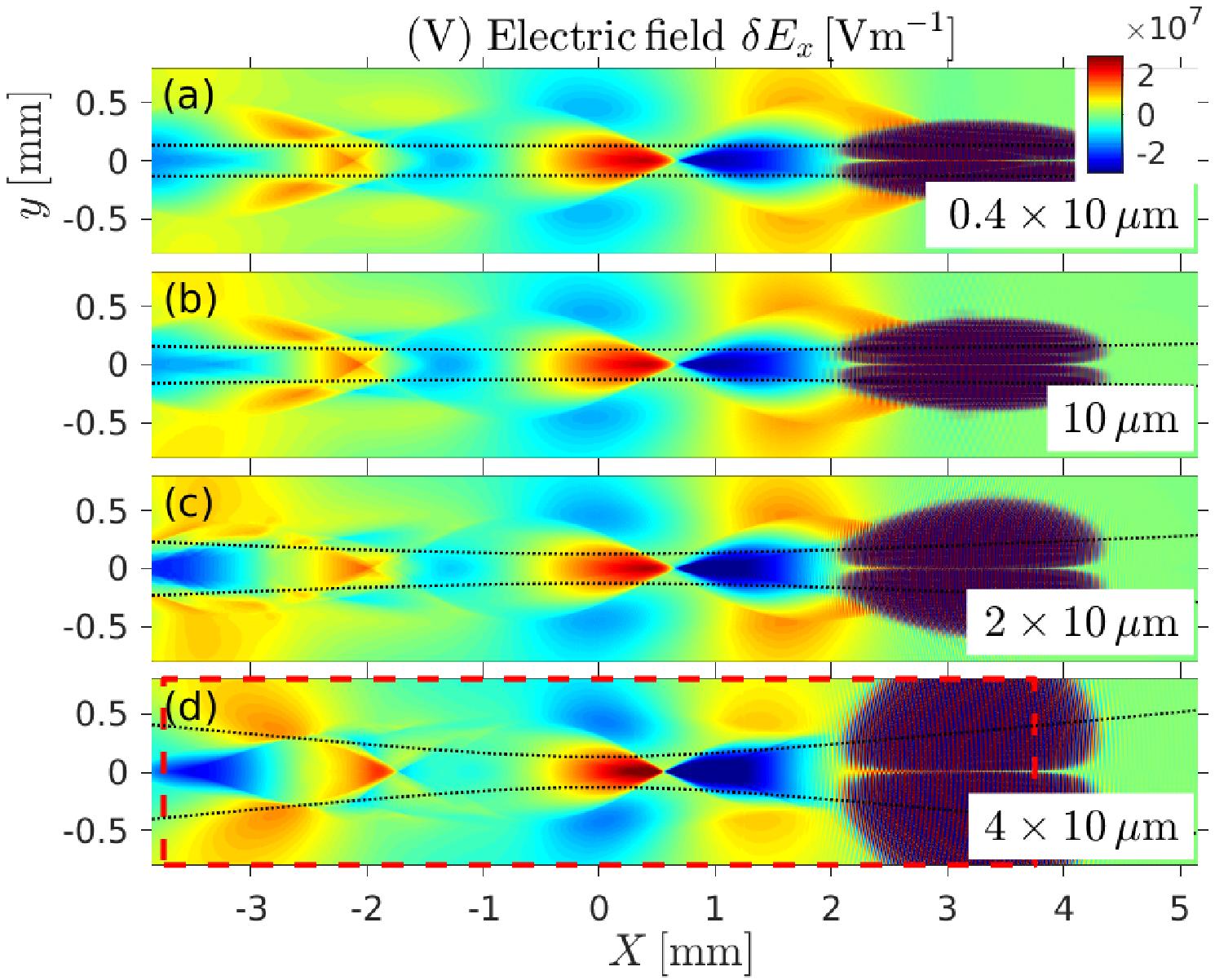}
\includegraphics[width=0.48\textwidth]{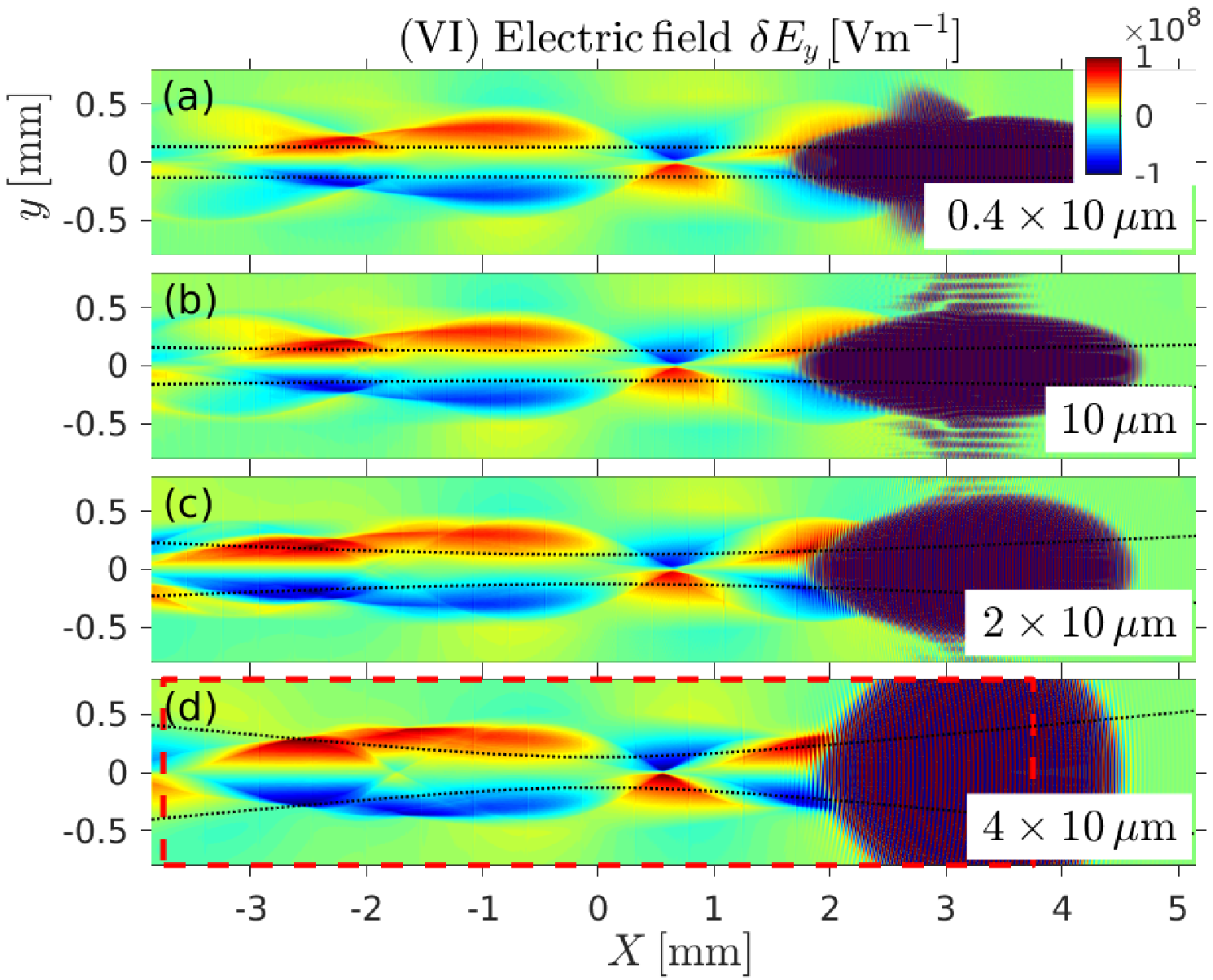}
\caption{Contour plots of the $x$ and $y$ components of the induced current density $\delta{\bm J}$ (top) and the electric field strength $\delta{\bm E}$ (bottom) in the cases with laser wavelengths $\lambda_{\rm las} = 4,\,10,\,20,\,40\,\mu{\rm m}$ from Fig.~\protect\ref{fig:08_win10mm_scan-L_ne-bz}. (Results of 2D EPOCH simulations with moving window and immobile ions.)}
\label{fig:09_win10mm_scan-L_Jxy-Exy}%
\end{figure*}

Before examining the plasma dynamics, it is useful to take a glance at the evolution of the laser pulse in each case, which is shown in Fig.~\ref{fig:07_win10mm_evol-pulse_L04-80} as a function of the position $X_{\rm las}(t) = c(t - t_{\rm D}) - x_{\rm foc}$ of the Gaussian pulse center. Figure~\ref{fig:07_win10mm_evol-pulse_L04-80}(a) shows the peak amplitude of the laser's magnetic field scaled by $\lambda_{\rm las}$, which is proportional to the normalized amplitude $a_0$ in Eq.~(\protect\ref{eq:anrm}). Figure~\ref{fig:07_win10mm_evol-pulse_L04-80}(b) shows the spot radius $w$ (defined in terms of the $1/{\rm e}$ amplitude), which drops to $w_{\rm foc} = 0.13\,{\rm mm}$ at the focal point $X = 0$ in all cases. The simulation results (symbols) are close to the respective analytical solutions (solid lines) of the paraxial wave equation for a Gaussian laser pulse propagating through vacuum; namely, Eqs.~(\ref{eq:paraxial_E}) and (\ref{eq:paraxial_w}). There is, however, a small but noticeable difference between the theoretically estimated focal point (defined to be at $X=0$) and the simulated one: in the simulation, the laser reaches its focus slightly before $X=0$, and that difference seems to increase with increasing wavelength $\lambda_{\rm las}$.

Figure~\ref{fig:07_win10mm_evol-pulse_L04-80}(c) serves as an additional test for numerical accuracy as it shows the constancy of the pulse duration $\tau_{\rm pulse}$ throughout the simulation. Note that the pulse length $c\times\tau_{\rm pulse}$ plotted in Fig.~\ref{fig:07_win10mm_evol-pulse_L04-80}(c) was simply measured from the profile of the oscillatory field magnitude, $|B_{\rm las}|$, so it tends to be underestimated and fluctuates by some value of magnitude $\lesssim \lambda_{\rm las}$.

\subsection{Demonstration of limited similarity in 2D}
\label{sec:limsim_demo}

We consider the time slice where the laser has traveled about $3.5\,{\rm mm}$ past the focal point located at $X = 0$. Figure~\ref{fig:08_win10mm_scan-L_ne-bz} shows color-shaded contour plots of the spatial structure of (I) the electron density $n_{\rm e}$ in the left column and (II) the fluctuating component of the out-of-plane magnetic field $\delta B_z = ({\bm B} - {\bm B}_{\rm ext})\cdot\hat{\bm e}_z$ in the right column (here ${\bm B}_{\rm ext}\cdot\hat{\bm e}_z = 0$ since ${\bm B}_{\rm ext}$ is directed along $x$). One can see that the structure of the electron density fluctuations and of the induced magnetic perturbations is similar in the cases with $\lambda_{\rm las} \leq 40\,\mu{\rm m}$ ($\sigma \leq 4$). The same can be said about the structure of the induced current density $\delta{\bm J}$ and electric field $\delta{\bm E}$, whose $x$ and $y$ components are shown in Fig.~\ref{fig:09_win10mm_scan-L_Jxy-Exy}.

The $40\,\mu{\rm m}$ case in row (d) of Figs.~\ref{fig:08_win10mm_scan-L_ne-bz} and \ref{fig:09_win10mm_scan-L_Jxy-Exy} is marginal because the cavitation length --- which is $L_{\rm cav}(40\,\mu{\rm m}) \approx 7.5\,{\rm mm}$ according to Eq.~(\ref{eq:len_cav_ref}) and indicated by a dashed red rectangle in Figs.~\ref{fig:08_win10mm_scan-L_ne-bz}(d) and \ref{fig:09_win10mm_scan-L_Jxy-Exy}(d) --- can still accommodate nearly two cycles of the magnetized wake, whose characteristic length is approximately $\lambda_{\rm wake} \approx 4\,{\rm mm}$ (Eq.~(\ref{eq:lWake})). In contrast, in the $80\,\mu{\rm m}$ case in Figs.~\ref{fig:08_win10mm_scan-L_ne-bz}(e) and \ref{fig:09_win10mm_scan-L_Jxy-Exy}(e), the cavitation length has shrunk to roughly the same size as the wake length or shorter, $L_{\rm cav}(80\,\mu{\rm m}) \approx 3.3\,{\rm mm} \lesssim \lambda_{\rm wake}$, so that similarity is lost for the present choice of parameters.\footnote{As is clear from Eqs.~(\protect\ref{eq:wUH}) and (\protect\ref{eq:lWake}), the wake length $\lambda_{\rm wake}$ may be reduced by increasing the strength of the external field $B_{\rm ext}$ or increasing the density $n_{\rm e0}$.}

Note that the data shown in rows (a)--(c) of Figs.~\ref{fig:08_win10mm_scan-L_ne-bz} and \ref{fig:09_win10mm_scan-L_Jxy-Exy} have been filtered and down-sampled to the same resolution as in the $40\,\mu{\rm m}$ case in row (d). Originally, the electron density spikes and cusps in the $4\,\mu{\rm m}$ case are much narrower and about 5 times higher (${\rm max}\{n_{\rm e}\} \approx 240 \times 10^{13}\,{\rm cm}^{-3}$) than in the $40\,\mu{\rm m}$ case ($50 \times 10^{13}\,{\rm cm}^{-3}$).

\begin{figure}
[tbp]
\centering
\includegraphics[width=0.48\textwidth]{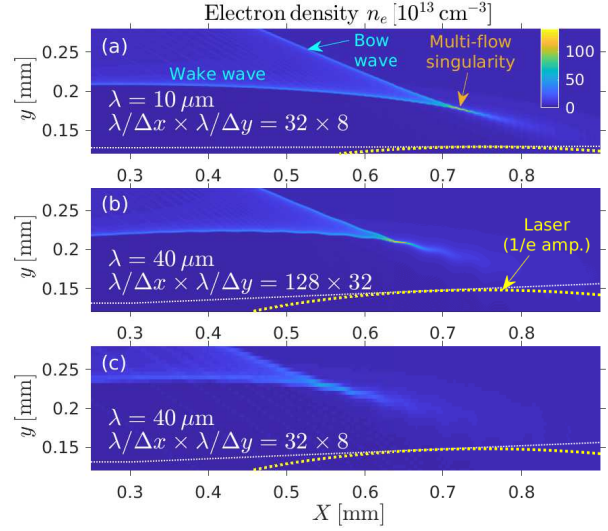}
\caption{Zoomed-up view of the density singularity near the laser pulse. Panels (a) and (c) show data from the same simulations as Figs.~\ref{fig:08_win10mm_scan-L_ne-bz}(b) and \ref{fig:08_win10mm_scan-L_ne-bz}(d) for the $10$ and $40\,\mu{\rm m}$ cases, respectively, albeit at a slightly earlier time where the laser is closer to its focal point ($X_{\rm las} \approx 0.75\,{\rm mm}$). Panel (b) shows the results of the $40\,\mu{\rm m}$ case simulated with 4 times higher resolution in both $x$ and $y$; i.e., the same resolution as was used by default in the $10\,\mu{\rm m}$ case. The white dotted curves indicate the local spot radius $w(X)$ of the laser (cf.~Fig.~\protect\ref{fig:07_win10mm_evol-pulse_L04-80}(b)) and the partially visible dotted orange ellipse is the laser's $1/{\rm e}$ amplitude radius. (Results of 2D EPOCH simulations with moving window and immobile ions.)}
\label{fig:10_win10mm_singularity}%
\end{figure}

Figure~\ref{fig:10_win10mm_singularity} shows a zoom-up of the density spike located near the laser pulse, where the bow wave detaches from the wake. Recently, this structure has attracted attention since it may serve as a practically useful source of coherent energetic photons via a mechanism dubbed Burst Intensification by Singularity Emitting Radiation (BISER) \cite{Pirozhkov17b, Pirozhkov12, Pirozhkov14}. This structure is a true (albeit integrable) singularity, which means that its width in the simulation varies with the spatial resolution used. This is demonstrated in Fig.~\ref{fig:10_win10mm_singularity}, where panels (a) and (c) show the singularities for the $10\,\mu{\rm m}$ and $40\,\mu{\rm m}$ cases with the default resolution of ($\lambda_{\rm las}/\Delta x \times \lambda_{\rm las}/\Delta y = 32 \times 8$ points per wavelength), and panel (b) shows the $40\,\mu{\rm m}$ case simulated with 4 times higher resolution ($128 \times 32$); i.e., with the same grid spacing as used in the $10\,\mu{\rm m}$ case. One can see that the singularity in (b) is sharper than that in (c) and it is located a little closer to the laser, whose $1/{\rm e}$ amplitude radius is indicated by the yellow dotted curve in the bottom right portion of each panel.

Note that there are differences between the equally resolved singularities in panels (a) and (b). For instance, one can observe a modulation of both the wake and the bow wave on the scale of the laser wavelength, and the amplitude of this modulation is larger in Fig.~\ref{fig:10_win10mm_singularity}(b) than in (a). This observation implies that singularity-related phenomena such as harmonic generation by the BISER mechanism \cite{Pirozhkov17b, Pirozhkov12, Pirozhkov14} depend at least quantitatively on the laser wavelength $\lambda_{\rm las}$. Meanwhile, the required resolution and, hence, the computational effort required to study these phenomena is not determined by the laser wavelength but by the singular structures, which should be resolved in as much detail as possible.

\begin{figure}
[tbp]
\centering
\includegraphics[width=0.48\textwidth]{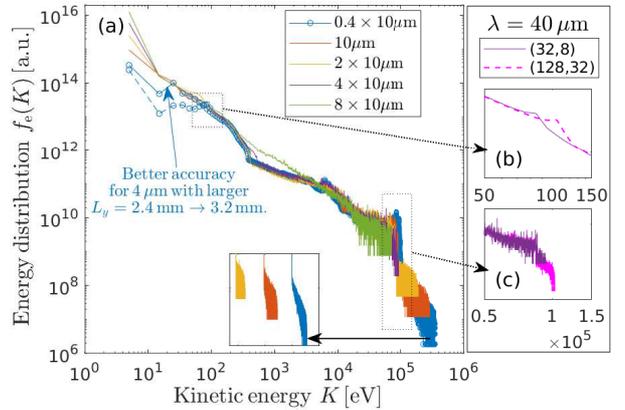}
\caption{Electron energy distribution $f_{\rm e}(K)$ in arbitrary units (a.u.). Panel (a) shows the energy distributions for the cases with $\lambda_{\rm las} = 4,\,10,\,20,\,40,\,80\,\mu{\rm m}$ at the same snapshot time as in Fig.~\protect\ref{fig:08_win10mm_scan-L_ne-bz}. For $\lambda_{\rm las} = 4\,\mu{\rm m}$ (blue circles) results are shown for two different heights of the simulation box: $L_y = 2.4\,{\rm mm}$ (dashed) and $3.2\, {\rm mm}$ (solid). The inset at the bottom of panel (a) shows the structure of the (noisy) high-energy tails, $K \gtrsim 10^5\,{\rm eV}$. Panels (b) and (c) show zoom-ups of the energy distribution in the $\lambda_{\rm las} = 40\,\mu{\rm m}$ case simulated with the default $(\lambda_{\rm las}/\Delta x,\lambda_{\rm las}/\Delta y) = (32,8)$ grid points per wavelength (solid) and with 4 times higher resolution, $(128,32)$ (dashed) as in Fig.~\protect\ref{fig:10_win10mm_singularity}. Notable resolution-dependent differences in $f_{\rm e}(K)$ were found only near $10^2\,{\rm eV}$ and $10^5\,{\rm eV}$. (Results of 2D EPOCH simulations with moving window and immobile ions.)}
\label{fig:11_win10mm_scan-L_enrDistr-e}%
\end{figure}

Next, let us inspect the electron energy distribution $f_{\rm e}(K)$, where $K = (\gamma - 1)m_{\rm e} c^2$ is the kinetic energy of an electron with rest mass $m_{\rm e}$ and Lorentz factor $\gamma = (1 - v^2/c^2)^{-1/2}$. Figure~\ref{fig:11_win10mm_scan-L_enrDistr-e}(a) shows the spatially averaged energy distribution $f_{\rm e}(K)$ for the same snapshot time as in Fig.~\ref{fig:08_win10mm_scan-L_ne-bz}. One can see that the form and amplitude of $f_{\rm e}(K)$ is similar for wavelengths $\lambda_{\rm las} \leq 40\,\mu{\rm m}$ and energies in the range $10^2\,{\rm eV} \lesssim K \lesssim 10^5\,{\rm eV}$.

The differences at energies below $100\,{\rm eV}$ may be explained as follows. The different values of the first sample ($K = 5\,{\rm eV}$) in Fig.~\ref{fig:11_win10mm_scan-L_enrDistr-e}(a) can be attributed to the fact that, for a fixed focal spot size $d_{\rm foc} = 0.15\,{\rm mm}$, lasers with longer wavelengths perturb a larger volume of the gas in regions away from the focal point, where their spot size is larger (cf.~Fig.~\ref{fig:07_win10mm_evol-pulse_L04-80}(b)). The perturbation in those regions is, of course, very weak, which is why the resulting differences in $f_{\rm e}(K)$ are seen only at the lowest energies. To some extent, the differences at low energies are also due to the fact that the height of the simulation box and, thus, the number of weakly perturbed low-energy electrons is varied in proportion to the laser wavelength as $L_y = \sigma\times 6\,{\rm mm}$ while keeping $L_{\rm x} = {\rm const}$.\ (cf.~Table~\ref{tab:parm_num}). Here, this has a noticeable effect in the case with $\lambda_{\rm las} = 4\,\mu{\rm m}$ for energies $K \lesssim 100\,{\rm eV}$. As shown in Fig.~\ref{fig:11_win10mm_scan-L_enrDistr-e}(a), the differences are reduced when the box height is increased from $L_y = 2.4\,{\rm mm}$ to $3.2\, {\rm mm}$.\footnote{As explained in \ref{apdx:resource}, a larger transverse box size may be needed to reduce the accumulation of artifacts in the electromagnetic fields; in particular, for shorter wavelengths $\lambda_{\rm las}$, because the laser converges only slowly due to its longer focal length, so that its outer portions interact with the artificial boundaries for a longer time. While this is not critical for the simulations with moving window discussed in the present section, more care is required when choosing the transverse size of the simulation box in long-time simulations with stationary window.}

Since we know from Fig.~\ref{fig:10_win10mm_singularity} that there are underresolved singular structures in the present simulations, we have also checked whether the spatial resolution affects the energy distribution in Fig.~\ref{fig:11_win10mm_scan-L_enrDistr-e} in the $40\,\mu{\rm m}$ case. Noteworthy differences were found only near $K \approx 10^2\,{\rm eV}$ and $10^5\,{\rm eV}$, and zoom-ups around these points are shown in panels (b) and (c) of Fig.~\ref{fig:11_win10mm_scan-L_enrDistr-e}: with 4 times higher resolution (dashed line), the ``knee'' near $K \approx 10^2\,{\rm eV}$ in (b) is slightly shifted towards higher energies, and in (c) one can see that the high-energy ``tail'' around $10^5\,{\rm eV}$ is somewhat longer. Since these resolution-dependent differences are small, we have not tried to identify their reasons. They may be related to the structural differences at small scales, such as those seen in Fig.~\ref{fig:10_win10mm_singularity}(b) and (c).

Finally, we discuss the high-energy tail. At $K \approx 100\,{\rm keV}$ ($v/c \approx 0.55$), the energy distribution $f_{\rm e}(K)$ in Fig.~\ref{fig:11_win10mm_scan-L_enrDistr-e}(a) drops abruptly by about one order of magnitude. This cut-off lies well below the theoretical maximum for the acceleration of a single electron in the 1D plane wave limit, which is
\begin{align}
K_{\rm max}^{\rm 1D} &\sim \left(\sqrt{1 + a_0^2 + a_0^4/4} - 1\right) m_{\rm e} c^2 \nonumber
\\
&\approx a_{\rm 0,foc}^2 m_{\rm e} c^2/2 \approx a_{\rm foc}^2\times 255\,{\rm keV},
\label{eq:emax}
\end{align}

\noindent for acceleration due to the laser field (e.g., see p.~327 in Ref.~\cite{Mourou06} and p.~231 in Ref.~\cite{BulanovSV_Book01}). Nevertheless, the fact that the present $100\,{\rm keV}$ cut-off in Fig.~\ref{fig:11_win10mm_scan-L_enrDistr-e} does not depend on the wavelength of the laser suggests that this is, in fact, the effective maximal electron energy that can be achieved with the present paraxially focused Gaussian pulse with FWHM spot diameter $d_{\rm foc} = 0.15\,{\rm mm}$ and pulse length $\tau_{\rm pulse} = 2\,{\rm ps}$. Indeed, it is reasonable to assume that some electrons can be accelerated to full speed in all our cases, because even with $\lambda_{\rm las} = 80\,\mu{\rm m}$ the Rayleigh length $L_{\rm R}(80\,\mu{\rm m}) \approx 0.7\,{\rm mm}$ is an order of magnitude larger than the theoretically predicted distance over which full acceleration takes place, which is given by (again in the 1D limit)\footnote{Equations~(\protect\ref{eq:smax_perp}) and (\protect\ref{eq:smax_par}) determine the distance that an electron travels along the laser path until it reaches its maximal perpendicular and parallel momentum, $p_\perp = a_0 m_{\rm e}c$ and $p_\parallel = a_0^2 m_{\rm e}c/2$, respectively, assuming that this acceleration takes place while the electron experiences half a period of the laser field.}
\begin{align}
s_{{\rm max}\perp}^{\rm 1D} &\sim \lambda_{\rm las} a_0,
\label{eq:smax_perp}
\\
s_{{\rm max}\parallel}^{\rm 1D} &\sim 2\lambda_{\rm las} a_0^2 / (2 + a_0^2).
\label{eq:smax_par}
\end{align}

\noindent and becomes $s_{\rm max}^{\rm 1D} \lesssim \lambda_{\rm las} \ll L_{\rm R}$ for $a_0 \leq 1$.

In other words, our $K_{\rm max} \approx 100\,{\rm keV}$ in Fig.~\ref{fig:11_win10mm_scan-L_enrDistr-e} is independent of $\lambda_{\rm las}$ because electrons can be accelerated to full speed well within the time interval (or distance), where the laser pulse is maximally focused. We consider this to be one of the necessary conditions for limited similarity, which are discussed in Section~\ref{sec:limsim_cond} below.

A clear dependence of $f_{\rm e}(K)$ on $\lambda_{\rm las}$ can be seen in Fig.~\ref{fig:11_win10mm_scan-L_enrDistr-e}(a) only beyond the $100\,{\rm keV}$ cut-off, where a noisy high-energy tail extends to higher energies, reaching nearly $400\,{\rm keV}$ in the $\lambda_{\rm las} = 0.4\,\mu{\rm m}$ case. The inset panel in Fig.~\ref{fig:11_win10mm_scan-L_enrDistr-e}(a) shows the structure of these noisy tails for three cases with $\sigma = 0.4,\,1,\,2$. We have not investigated this further, because the number of electrons in the noisy tail is very small, so we do not expect them to have any significant influence on the overall dynamics of the plasma wake channel.

\subsection{Conditions for insensitivity w.r.t.\ laser wavelength}
\label{sec:limsim_cond}

The results in Figs.~\ref{fig:08_win10mm_scan-L_ne-bz}, \ref{fig:09_win10mm_scan-L_Jxy-Exy} and \ref{fig:11_win10mm_scan-L_enrDistr-e} discussed in the previous section show that the overall structure and dynamics of the magnetized plasma wakes in the vicinity of the laser's focal point are fairly independent of the laser wavelength for $4\,\mu{\rm m} \lesssim \lambda_{\rm las} \lesssim 40\,\mu{\rm m}$. It is also reasonable to assume that this limited similarity holds for shorter wavelengths below $4\,\mu{\rm m}$. Of course, the parameter regime where such similarity can be found has some constraints, besides the key condition $n_{\rm e}/n_{\rm crit} \ll 1$ in Eq.~(\ref{eq:small}) that was already pointed out in Section~\ref{sec:limsim_prep}.

One constraint that limits the range of values for the wavelength scaling factor $\sigma$ in Eq.~(\ref{eq:fscale}) arises from the fact that, according to Eq.~(\ref{eq:len_cav}), the cavitation length scales with the wavelength as $L_{\rm cav} \propto \lambda_{\rm las}^{-1} \propto \sigma^{-1}$, so that the initial length of the plasma channel becomes shorter with increasing $\sigma$. However, the channel should not be shorter than the typical size $\lambda_{\rm wake}$ of the structures that form in the magnetized plasma wake of the laser (cf.~Eq.~(\ref{eq:lWake})). In our case, this condition was violated for $\lambda_{\rm las} > 40\,\mu{\rm m}$.

In order to ensure similarity of the bow waves as well as the electron energy distributions $f_{\rm e}(K)$, we require that some electrons must be accelerated to the maximally possible kinetic energy $K_{\rm max}$ (Eq.~(\ref{eq:emax}) in 1D) while the pulse is fully focused. This means that the scaling factor $\sigma$ should be sufficiently small, so that the Rayleigh length $L_{\rm R} = \pi w_{\rm foc}^2 / \lambda_{\rm las} \propto 1/(\lambda_{\rm ref}\sigma)$ is long compared to the acceleration distance $s_{\rm max}$ (Eqs.~(\ref{eq:smax_perp}) and (\ref{eq:smax_par}) in 1D).

Another constraint on the wavelength scaling factor $\sigma$ arises from the fact that we were able to demonstrate limited similarity only for fixed pulse length, $\tau_{\rm pulse} = {\rm const}$. Since the pulse length should be sufficiently long to contain several wave cycles, this implies that the scaling factor is constrained by the condition $c\tau_{\rm pulse} \gg \lambda_{\rm las} = \sigma\lambda_{\rm ref}$.

The requirement that $\tau_{\rm pulse} = {\rm const}$.\ for limited similarity at fixed amplitude $a_{\rm 0,foc}$ is related to the fact the overall amount of energy transferred from the laser to the particles should be fixed. Although the details can be complicated (see \ref{apdx:tpulse}), one simple argument is that the absorption efficiency of the medium increases with increasing wavelength as $n_{\rm e}/n_{\rm crit} \propto \lambda_{\rm las}^2$, and that this increase of the absorption efficiency is compensated by the reduction of the energy
\begin{equation}
W_{\rm las} \propto \tau_{\rm pulse} \left(a_{\rm 0,foc} d_{\rm foc}/\lambda_{\rm las}\right)^2
\end{equation}

\noindent that is carried by an electromagnetic pulse with longer wavelength, when its amplitude $a_{\rm 0,foc}$, spot size $d_{\rm foc}$ and duration $\tau_{\rm pulse}$ are fixed.

Finally, our definition of limited similarity requires that the spot size $d_{\rm foc}$ must not change as it controls the width of the plasma channel. This is because we require here that lasers of different wavelengths produce structures of both similar shape and similar size.\footnote{Of course, it may be possible to find other manifestations of limited similarity, where one scales the size of the entire system and of all the structures it contains (as is done in laboratory astrophysics) in proportion to, say, the laser wavelength. However, this is not our intention here.}

In summary, the size and structure of the magnetized plasma wake with characteristic wavelength $\lambda_{\rm wake}$ is insensitive with respect to the laser wavelength $\lambda_{\rm las}$ if the following conditions are satisfied:
\begin{align}
n_{\rm e} / n_{\rm crit}(\lambda_{\rm las}) \ll 1 & \quad ({\rm highly\, subcritical}),
\label{eq:limsim_subcrit}
\\
L_{\rm cav}(\lambda_{\rm las})/\lambda_{\rm wake} > 1 & \quad ({\rm \#\, of\, wake\, cycles}),
\label{eq:limsim_size}
\\
c\tau_{\rm pulse}/\lambda_{\rm las} \gg 1 & \quad ({\rm \#\, of\, laser\, cycles}),
\label{eq:limsim_cycle}
\\
L_{\rm R}(\lambda_{\rm las})/s_{\rm max}(\lambda_{\rm las}) \gg 1 & \quad ({\rm max.\, kin.\, energy}),
\label{eq:limsim_emax}
\\
d_{\rm foc} = {\rm const}. & \quad ({\rm cavity\, diameter}).
\label{eq:limsim_cavity}
\end{align}

\noindent In addition, the amplitude and pulse length must either both be constant,
\begin{align}
a_{\rm 0,foc}, \tau_{\rm pulse}\omega_{\rm pe} &  = {\rm const}. & ({\rm pulse\, amp.\, \&\, length})
\label{eq:limsim_laser}
\end{align}

\noindent (which is the simplest case and is realized here), or co-varied in a certain way as discussed in \ref{apdx:tpulse}. The relation between $a_0$ and $\tau_{\rm pulse}\omega_{\rm pe}$ is not a straightforward one. It is determined by the requirement that the spatial structure and amplitude of the electric field induced by the laser pulse remains similar. If one uses the optimal pulse length that maximized the electric field, the relation between $a_0$ and $\tau_{\rm pulse}$ may be inferred from the theory in Ref.~\cite{BulanovSV16}. As was mentioned in Section~\ref{sec:setup_param}, our pulse length was not optimized.

\begin{figure}
[tb]
\centering
\includegraphics[width=0.48\textwidth]{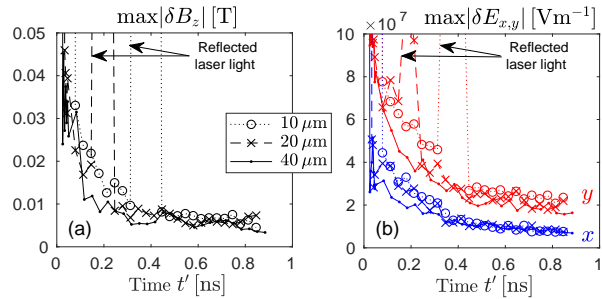}
\caption{Evolution of the global maxima of the electromagnetic field components (a) $\delta B_z$, (b) $\delta E_x$ (blue) and $\delta E_y$ (red) in the domain of the cavity ($|X| \leq L_{\rm cav}/2 \approx 3.8\,{\rm mm}$, $|y| \leq 10\times d_{\rm foc} = 1.5\,{\rm mm}$). Results are shown for three laser wavelengths $\lambda_{\rm las} = 10$, $20$ and $40\,\mu{\rm m}$. The simulations were performed using EPOCH 2D with stationary window and imperfect ``open'' boundaries, which reflect about $0.1\%$ of the laser light. When the reflected light returns to the cavity, its amplitude is still one order of magnitude larger than the wake field in the present low density plasma in the 10 and 20 micron cases, which obscures the results for $\delta B_z$ and $\delta E_y$ around $t'\approx 0.35$ and $0.2$ (arrows). This reflection has no significant effect on the electron dynamics.}
\label{fig:12_fix_evol-maxEB_1ns}%
\end{figure}

Note that the range of wavelengths for which limited similarity holds depends on the values of $n_{\rm e}$ and/or ${\bm B}_{\rm ext}$, because these quantities determine the characteristic wavelength $\lambda_{\rm wake}$ via Eqs.~(\ref{eq:wUH}) and (\ref{eq:lWake}), which enters condition (\ref{eq:limsim_size}).

\begin{figure*}
[tb]
\centering
\includegraphics[width=0.96\textwidth]{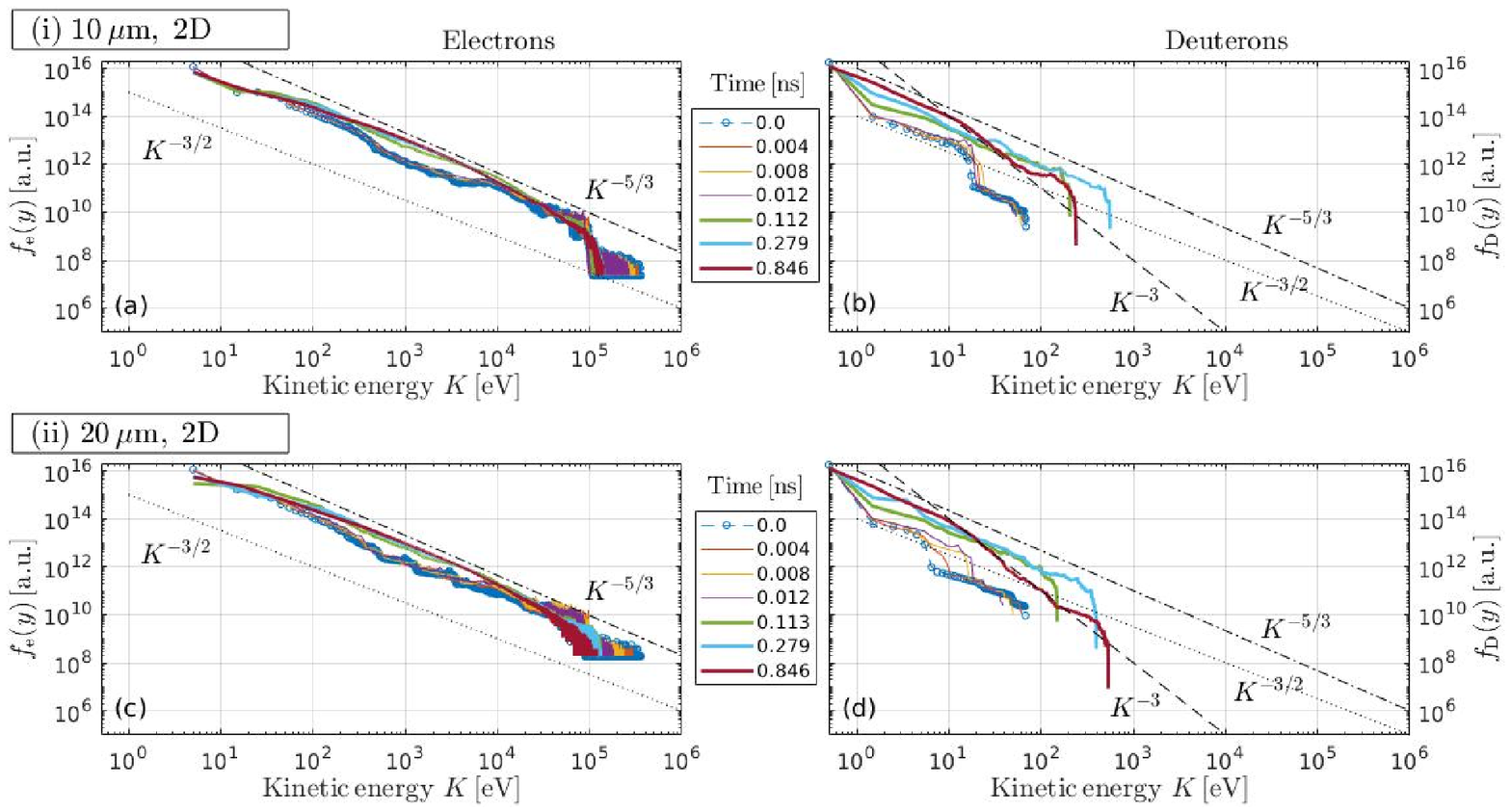} \\
\includegraphics[width=0.96\textwidth]{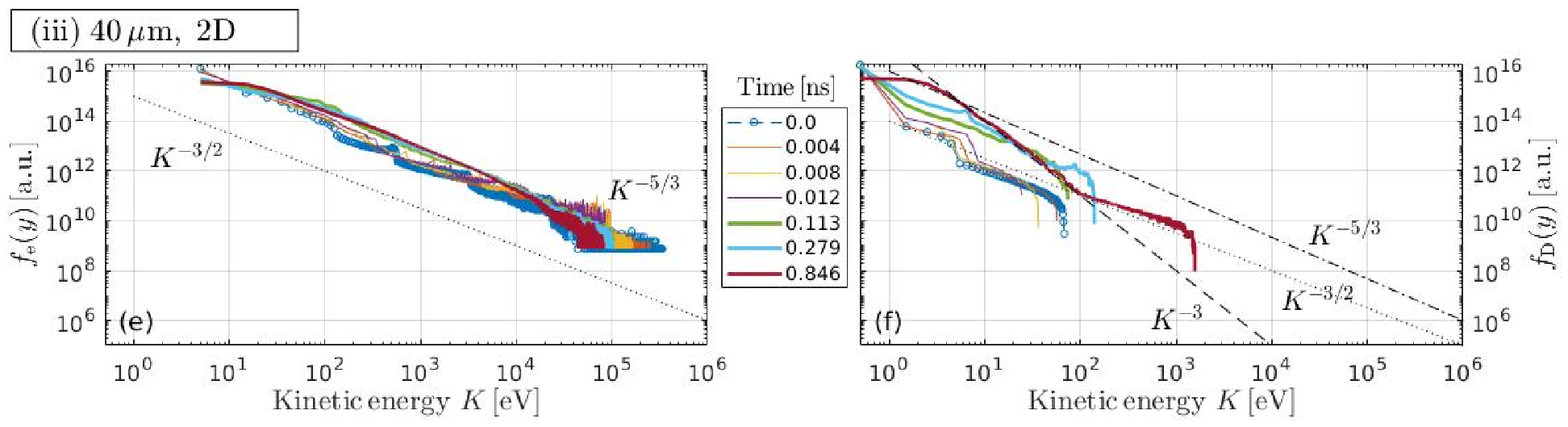}
\caption{Evolution of the energy distributions of electrons $f_{\rm e}(K)$ (left column) and deuterons $f_{\rm D}(K)$ (right column) in each of the three cases (i)--(iii) shown in Figs.~\protect\ref{fig:04_fix_2d_10-40um_ne}--\protect\ref{fig:06_fix_2d_scan-L_n-prof-evol} and \protect\ref{fig:12_fix_evol-maxEB_1ns}. These REMP 2D data were multiplied by the respective wavelength scaling factors ($\sigma = 1,2,4$) in order to account for the different heights $L_y$ of the respective stationary simulation windows (cf.~Table~\ref{tab:parm_num_fix}). A few power laws, $K^{-3/2}$, $K^{-5/3}$, $K^{-3}$, are plotted for ease of comparison and orientation (without claiming physical relevance, although it might exist).}
\label{fig:13_fix_2d_scan-L_enrDistr-e-D}%
\end{figure*}

Although the results are not presented here, we have also performed scans of the electron density in order to obtain information about how small the ratio $n_{\rm e}/n_{\rm crit}$ needs to be. For instance, it is found that for $n_{\rm e0} = 10^{16}\,{\rm cm}^{-3}$ and higher, absorption becomes significant for the $40\,\mu{\rm m}$ laser, whose critical density is $n_{\rm crit} \approx 6\times 10^{17}\,{\rm cm}^{-3}$. On this basis, the condition for subcriticality in Eq.~(\ref{eq:limsim_subcrit}) may be written more concretely as
\begin{equation}
n_{\rm e} / n_{\rm crit}(\lambda_{\rm las}) \ll 10^{-2} \quad ({\rm for\, parms.\, in\, Table~\protect\ref{tab:parm_laser}}).
\label{eq:limsim_density}
\end{equation}

\noindent Note that this is the result for the present choice of parameters; in particular, for normalized laser amplitude $a_{\rm 0,foc} = 1$. The upper limit in Eq.~(\ref{eq:limsim_density}) may be different for other parameter settings.

\subsection{Discussion on the validity of limited similarity on longer time scales}
\label{sec:limsim_discuss}

The fact that the electron density, the current density, the electric and magnetic fields as well as the electron energy distributions are all similar at the time of the snapshot shown in Figs.~\ref{fig:08_win10mm_scan-L_ne-bz}, \ref{fig:09_win10mm_scan-L_Jxy-Exy} and \ref{fig:11_win10mm_scan-L_enrDistr-e} suggests that the subsequent evolution of the plasma wake will continue to be similar, at least for a certain amount of time. This assertion is supported by the long-time simulation results reported in Section~\ref{sec:channel_Lscan} above; namely, Figs.~\ref{fig:05_fix_2d_10-40um_nei_late} and \ref{fig:06_fix_2d_scan-L_n-prof-evol}. It is difficult and perhaps unnecessary to ascertain the long-time persistence of limited similarity in detail for the entire global structure of the perturbed density and electromagnetic fields as we have done in Section~\ref{sec:limsim_demo}, because the structures may be similar at slightly different times. Therefore, we will provide further evidence only on an average or statistical level.

Figure~\ref{fig:12_fix_evol-maxEB_1ns} shows the time traces of the global maxima of the electromagnetic field components $\delta B_z$, $\delta E_x$ and $\delta E_y$ in the region of the cavity. Even without low-pass filtering --- which could be used to smooth out differences at singular structures as seen in Figs.~\ref{fig:05_fix_2d_10-40um_nei_late} and \ref{fig:10_win10mm_singularity} and eliminate spurious reflections of the laser pulse --- the results are quantitatively similar for $t' \gtrsim 0.3\,{\rm ns}$. This suggests that wavelength-dependent differences, which exist initially at small scales, gradually vanish, so that the limited similarity of the expanding plasma channels is enhanced or maintained; at least, until the internal dynamics of the channels begin to be affected by the different initial cavitation lengths $L_{\rm cav}(\lambda_{\rm las})$, or the longitudinal expansion hits the simulation domain boundaries.

Figure~\ref{fig:13_fix_2d_scan-L_enrDistr-e-D} summarizes the energy distributions $f_{\rm e}(K)$ and $f_{\rm D}(K)$ of electrons and deuterons in 2D simulations with $\lambda_{\rm las} = 10,\,20,\,40\,\mu{\rm m}$ for a series of snapshots, including those that were shown in Figs.~\ref{fig:04_fix_2d_10-40um_ne}--\ref{fig:06_fix_2d_scan-L_n-prof-evol} above. For all three wavelengths, the electron energy distributions in the left column of Fig.~\ref{fig:13_fix_2d_scan-L_enrDistr-e-D} are similar below $100\, {\rm keV}$ and exhibit relatively little variation in time for $t' > 0.1\,{\rm ns}$. Differences in the noisy high-energy tails beyond $100\,{\rm keV}$ were already discussed in Section~\ref{sec:limsim_demo}, and variations at earlier times can be expected from the different cavitation lengths $L_{\rm cav}(\lambda_{\rm las})$ (Eq.~(\ref{eq:len_cav})).

On the one hand, the similarity and relative constancy of $f_{\rm e}(K)$ suggests that the electrons have more or less equilibrated through their interactions with the electromagnetic fluctuations of the plasma wake channel, whose evolution is henceforth determined primarily by the spatial distribution and relatively slow motion of the much heavier deuterons. On the other hand, the observed similarity and relative constancy of $f_{\rm e}(K)$ also means that the electron energy distribution is insensitive to that of the deuterons, because the snapshots of $f_{\rm D}(K)$ in the right column of Fig.~\ref{fig:13_fix_2d_scan-L_enrDistr-e-D} do vary significantly, both in time and between cases with different wavelengths.

It should be noted that the ends of the plasma channels are approaching the boundaries of the simulation domain as they expand along $X$, following the magnetic field. Indeed at the time of the last snapshot ($t' \approx 0.85\,{\rm ns}$) shown in Figs.~\ref{fig:05_fix_2d_10-40um_nei_late}, \ref{fig:06_fix_2d_scan-L_n-prof-evol} and \ref{fig:13_fix_2d_scan-L_enrDistr-e-D}, the ends of the channel in the $40\,\mu{\rm m}$ case (which has the shortest simulation box with $L_{\rm x} = 26.5\,{\rm mm}$) have already reached the boundaries, as one can see in Fig.~\ref{fig:05_fix_2d_10-40um_nei_late}(iii). This may explain why the deuteron distribution in the $40\,\mu{\rm m}$ case in Fig.~\ref{fig:13_fix_2d_scan-L_enrDistr-e-D}(f) changes more drastically than in the cases with $10\,\mu{\rm m}$ and $20\,\mu{\rm m}$ in panels (b) and (d).

\section{Long-time 3D simulation with scaled wavelength $\lambda_{\rm las} = 40\,\mu{\rm m}$ and comparison with 2D results}
\label{sec:3d}

Our estimates for the computational expenses in \ref{apdx:resource} suggest that a long-time 3D simulation would currently be infeasible in our parameter regime of interest, if we were to use actually available wavelengths of high power lasers, which lie in the range from $0.8\,\mu{\rm m}$ (Ti:sapphire) to $10\,\mu{\rm m}$ (${\rm CO}_2$). This problem can be circumvented if we assume that the limited similarity demonstrated in Section~\ref{sec:limsim} for 2D does also hold in 3D. That is, we assume that 3D simulations with an up-scaled laser wavelength, such as $\lambda_{\rm las} = 40\,\mu{\rm m}$, will capture the salient features of the laser-induced plasma wake channels that would be produced by conventional high power lasers with $\lambda_{\rm las} \lesssim 10\,\mu{\rm m}$.

Based on this assumption, we have successfully simulated the after-glow dynamics of the laser-induced plasma wake channel in 3D on a relatively long time of about $1\,{\rm ns}$ using an up-scaled wavelength $\lambda_{\rm las} = 40\,\mu{\rm m}$, and first results obtained with the REMP code are presented in this section to demonstrate the numerical feasibility. Another purpose of this section is to compare the 3D results qualitatively with 2D results (case (iii) in Table~\ref{tab:parm_num_fix}) for the same parameters, except that the 3D simulation (case (iv) in Table~\ref{tab:parm_num_fix}) used only half the number of grid points along the $x$ axis. This is sufficient for our present purposes and reduces the computational expenses by a factor 1/4. With this setup, the 3D simulation of $1\,{\rm ns}$ of physical time took about 13 days using 6,400 cores on the supercomputer JFRS-1 \cite{jfrs1}. The 2D results presented in this section were obtained with REMP, so one may compare them with the EPOCH results in Figs.~\ref{fig:04_fix_2d_10-40um_ne}--\ref{fig:06_fix_2d_scan-L_n-prof-evol} above to ascertain that both codes yield very similar results in 2D for at least $1\,{\rm ns}$.

\begin{figure}
[tb]
\centering
\includegraphics[width=0.48\textwidth]{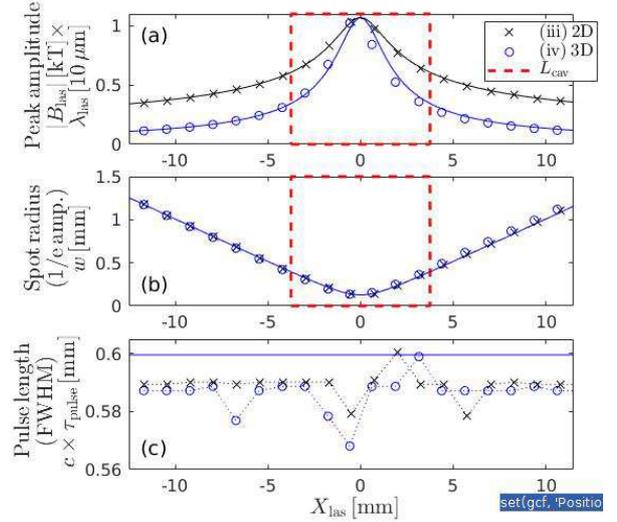}
\caption{Evolution of the $40\,\mu{\rm m}$ laser pulse in 2D (crosses) and 3D (circles) REMP simulations with a stationary window, which are identified as cases (iii) and (iv) in Table~\protect\ref{tab:parm_num_fix}. The solid lines indicate the analytical solutions given by (a) Eq.~(\ref{eq:paraxial_E}), (b) Eq.~(\ref{eq:paraxial_w}), and (c) the expected pulse length $c\times \tau_{\rm pulse} \approx 0.6\,{\rm mm}$. The red dashed rectangle in (a) and (b) indicates the cavitation length $L_{\rm cav} \approx 7.5\,{\rm mm}$ based on Eq.~(\protect\ref{eq:len_cav}). Otherwise arranged as Fig.~\protect\ref{fig:07_win10mm_evol-pulse_L04-80}.}
\label{fig:14_fix_evol-pulse_L40_bx2_2d-3d}%
\end{figure}

\begin{figure*}
[tb]
\centering
\includegraphics[width=0.48\textwidth]{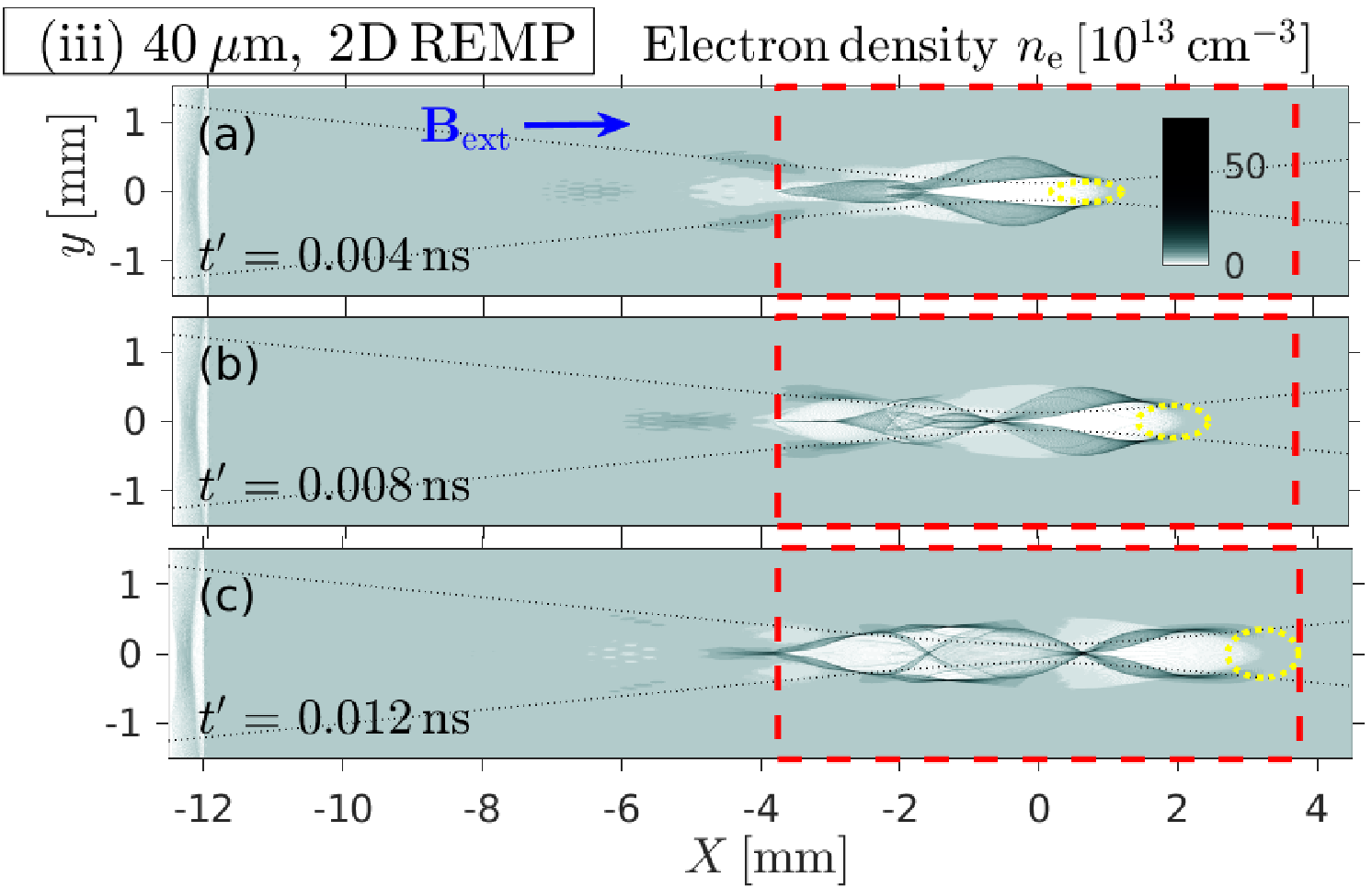} \includegraphics[width=0.48\textwidth]{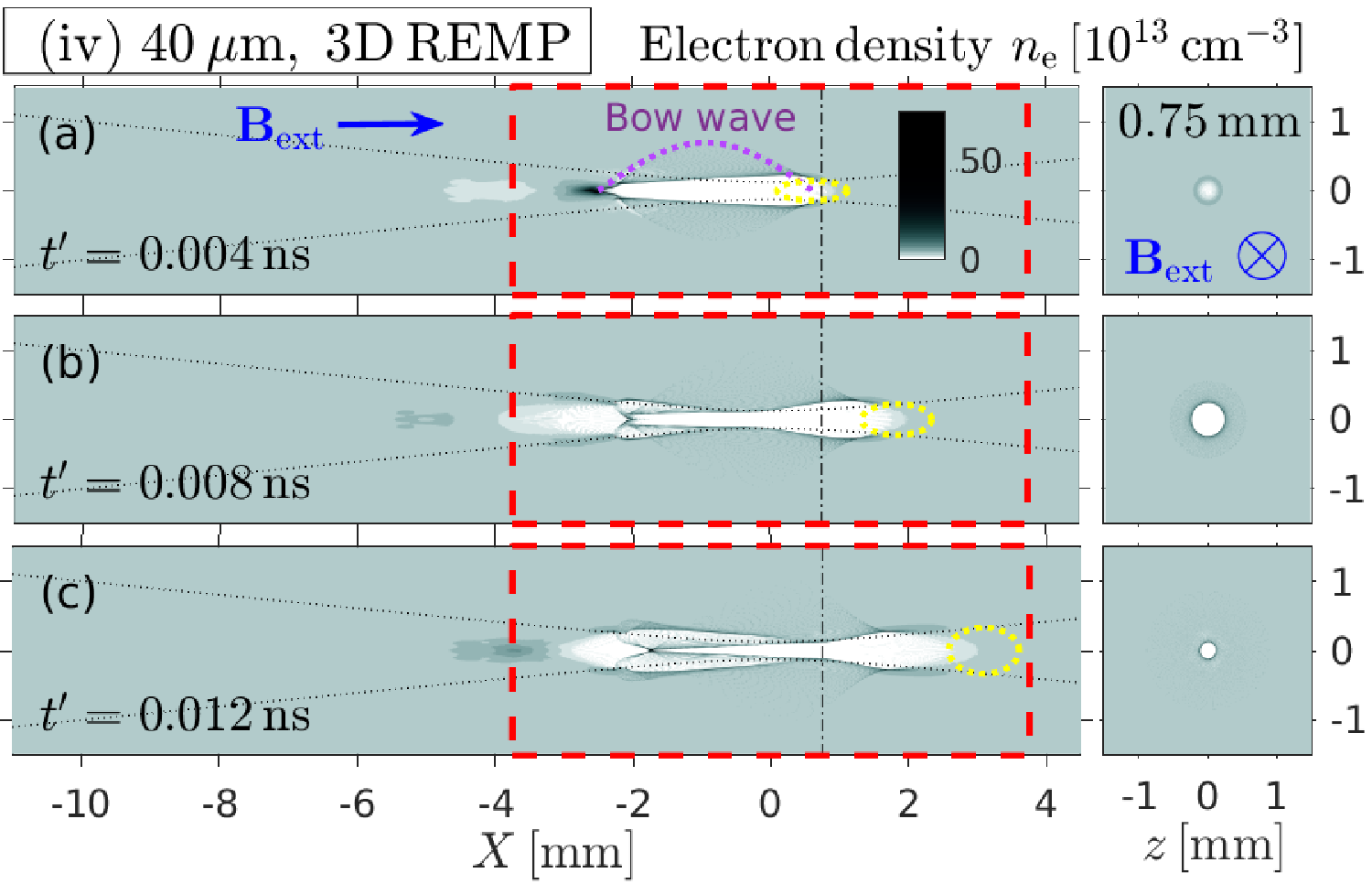}
\caption{Early phase ($t' = t - t_{\rm foc} \lesssim 12\,{\rm ps}$) of the evolution of the magnetized plasma wake in (iii) 2D and (iv) 3D REMP simulations. Three snapshots (a)--(c) of the electron density are shown. The contour plots of $n_{\rm e}(X,y)$ are arranged as in Fig.~\protect\ref{fig:04_fix_2d_10-40um_ne}. The vertical dash-dotted line in column (iv) indicates the location $X = 0.75\,{\rm mm}$, where we have measured the transverse cross-section of the electron density $n_{\rm e}(y,z)$ in the 3D case, whose contours are shown on the right-hand side of this figure. (The simulations were performed with a stationary window and including the ion response, using the parameters in Tables~\protect\ref{tab:parm_laser} and \protect\ref{tab:parm_num_fix}.)}
\label{fig:15_fix_2d-3d_40um_ne_early}%
\end{figure*}

\subsection{Evolution of the laser pulse}
\label{sec:3d_laser}

Figure~\ref{fig:14_fix_evol-pulse_L40_bx2_2d-3d} shows the evolution of the Gaussian $40\,\mu{\rm m}$ laser pulse as it propagates through the simulation box of length $L_x = 26.5\,{\rm mm}$ in 2D (crosses) and 3D (circles). The measured amplitudes, which are plotted as symbols in Fig.~\ref{fig:14_fix_evol-pulse_L40_bx2_2d-3d}(a), vary in accordance with the 2D ($N=2$) and 3D ($N=3$) solutions of the paraxial wave equation in Eq.~(\ref{eq:paraxial_E}), which are plotted as solid lines and are a good approximation for the present case with highly subcritical electron density. The same is true for the evolution of the spot radius $w$ (at $1/{\rm e}$ amplitude) shown in Fig.~\ref{fig:14_fix_evol-pulse_L40_bx2_2d-3d}(b), which follows closely the form of $w(X)$ from Eq.~(\ref{eq:paraxial_w}) in both cases.

Of course, the analytical formulas in Eqs.~(\ref{eq:paraxial_E}) and (\ref{eq:paraxial_w}) are approximations themselves, so that small deviations can be expected. In fact, we think that this is the reason for why the theoretically predicted focal point is close to but not identical to the measured one seen in Fig.~\ref{fig:14_fix_evol-pulse_L40_bx2_2d-3d} (the same is true for Fig.~\ref{fig:07_win10mm_evol-pulse_L04-80} above). The small difference between $w(X)$ in the 2D and 3D simulations seems to be due to the lower resolution that was used in the 3D case ($\lambda_{\rm las}/\Delta x = 16$ instead of $32$; cf.~Table~\ref{tab:parm_num_fix}).

Finally, Fig.~\ref{fig:14_fix_evol-pulse_L40_bx2_2d-3d}(c) shows that the deviation of the measured pulse length from its expected value $c\times\tau_{\rm pulse} = 0.6\,{\rm mm}$ lies within the systematic measurement error of order $\lesssim \lambda_{\rm las} = 0.04\,{\rm mm}$.

\begin{figure}
[tb]
\centering
\includegraphics[width=0.48\textwidth]{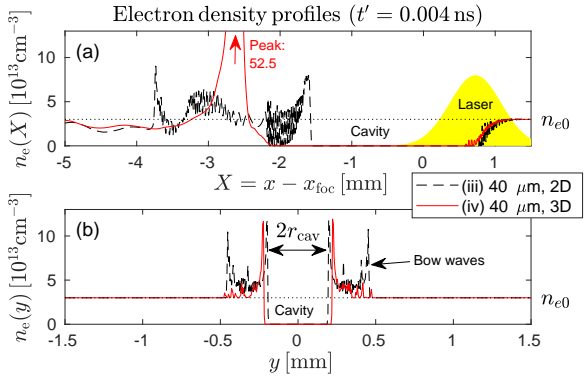}
\caption{Profiles of the electron cavity and bow waves in 2D (dashed) and 3D (solid) for the same snapshot as in Fig.~\protect\ref{fig:15_fix_2d-3d_40um_ne_early}(a). (a): Axial profiles $n_{\rm e}(X)$ evaluated at $y = z = 0$. The location and shape of the laser pulse is indicated by a shaded Gaussian (arbitrary amplitude). (b): Transverse profiles $n_{\rm e}(y)$ evaluated in the focal plane at $X = z = 0$}
\label{fig:16_fix_2d-3d_n-prof_early}%
\end{figure}

\begin{figure*}
[tb]
\centering
\includegraphics[width=0.48\textwidth]{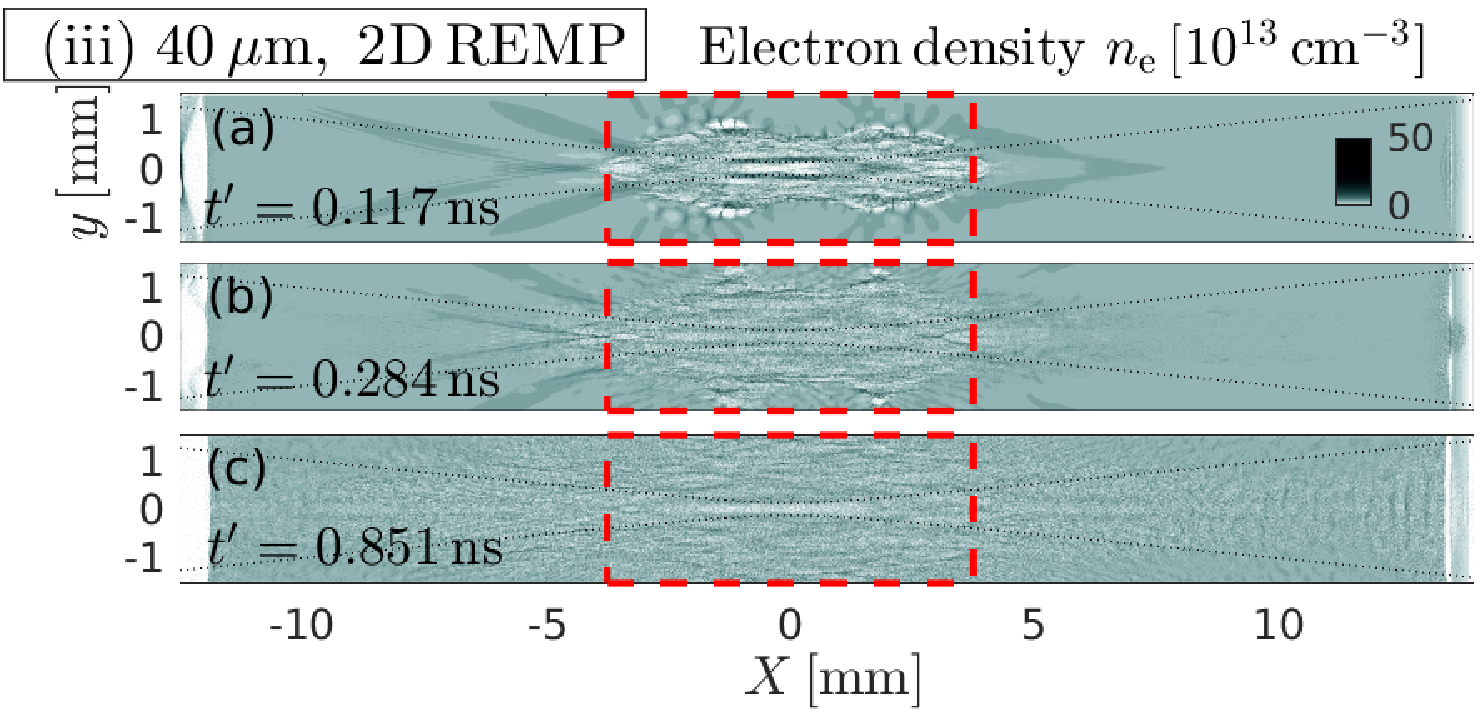} \includegraphics[width=0.48\textwidth]{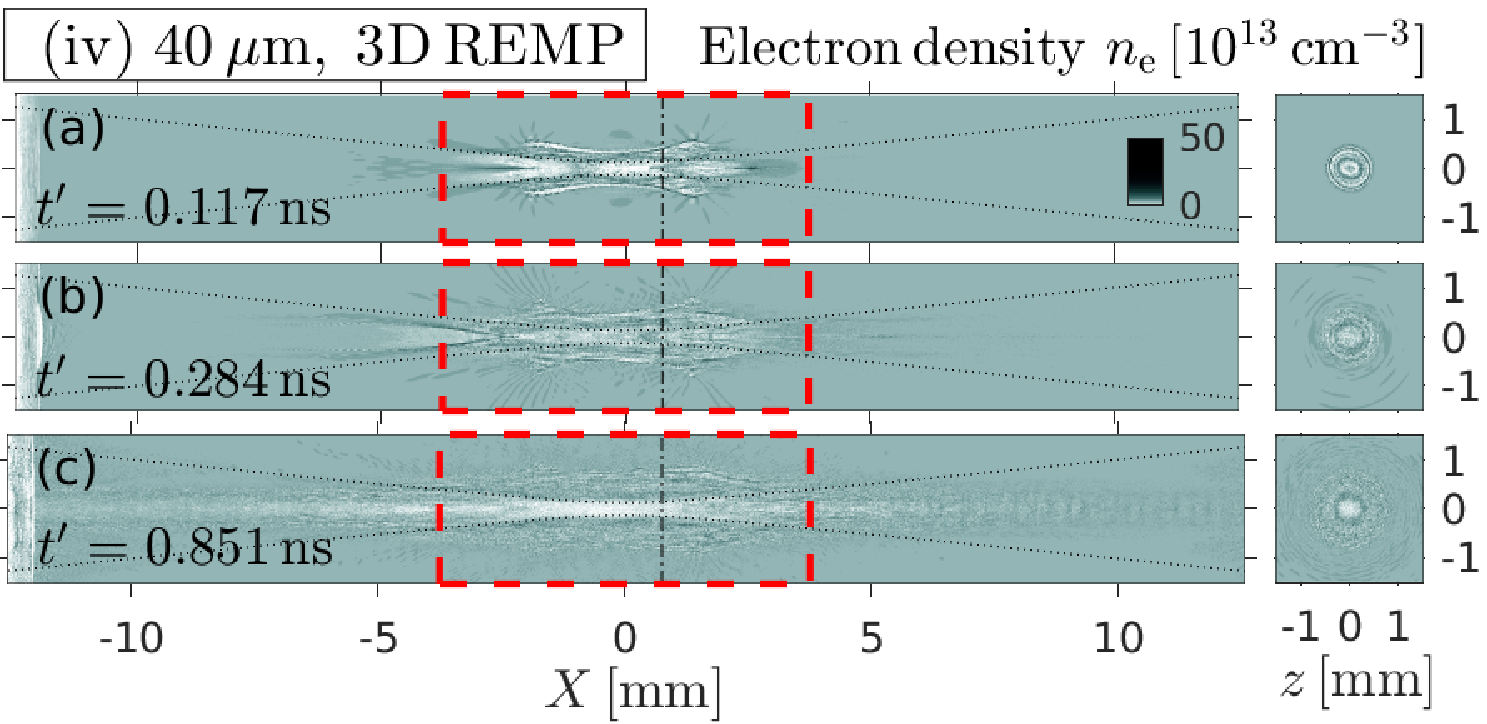} \\
\includegraphics[width=0.48\textwidth]{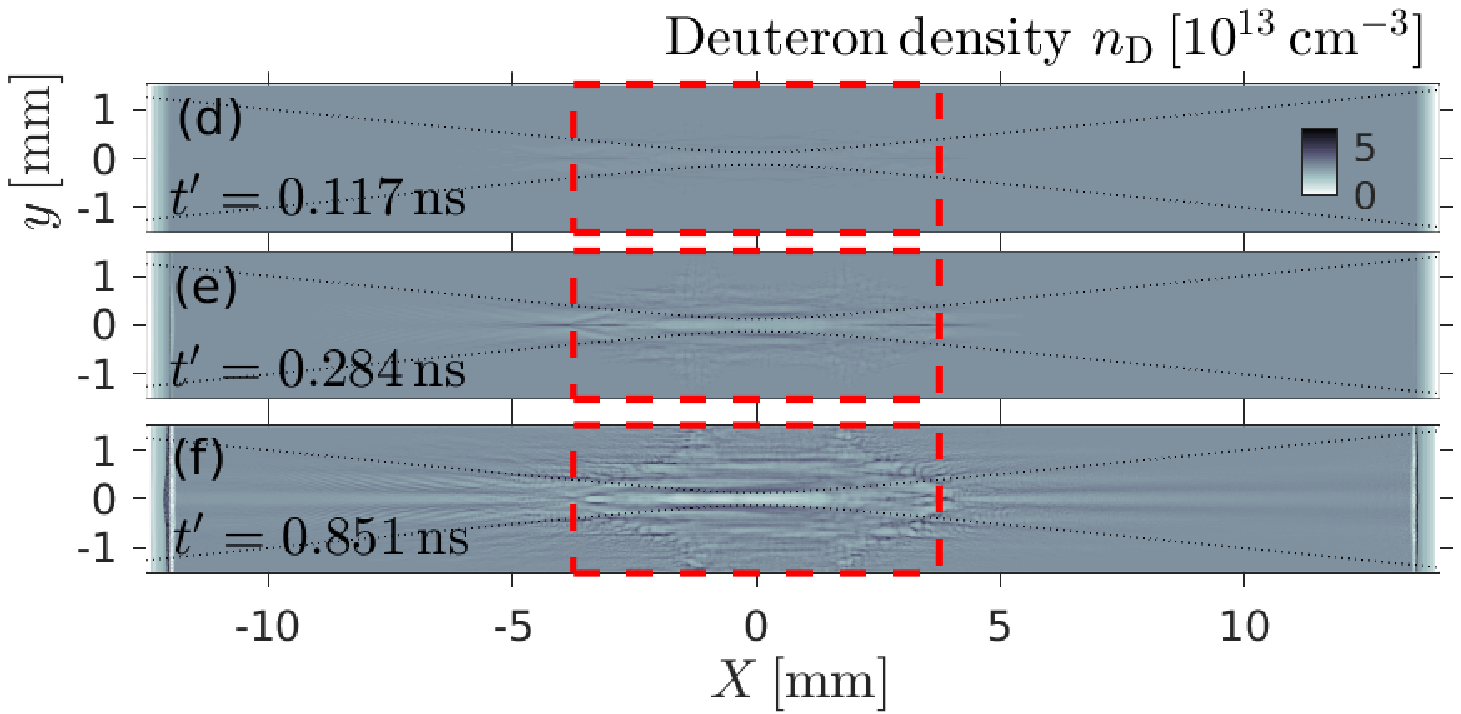} \includegraphics[width=0.48\textwidth]{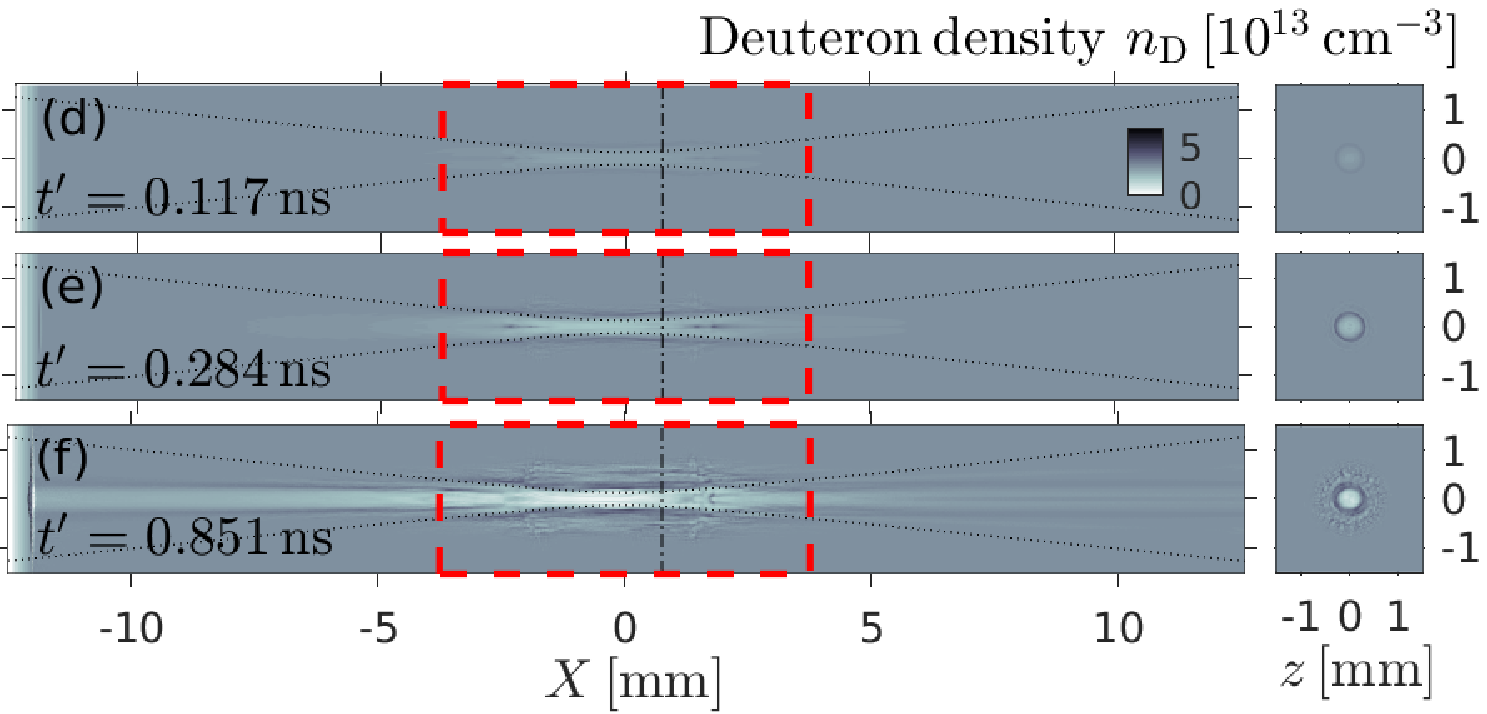}
\caption{Late phase ($t' = t - t_{\rm foc} \sim 0.1$...$0.9\,{\rm ns}$) of the evolution of the magnetized plasma wake channels in (iii) 2D and (iv) 3D REMP simulations. Continuation of and arranged as Fig.~\protect\ref{fig:15_fix_2d-3d_40um_ne_early} for electrons (top) and deuterons (bottom).}
\label{fig:17_fix_2d-3d_40um_ne-ni-late}%
\end{figure*}

\subsection{Early evolution on the picosecond scale}
\label{sec:3d_early}

Figures~\ref{fig:15_fix_2d-3d_40um_ne_early} and \ref{fig:16_fix_2d-3d_n-prof_early} summarize the results for the evolution of the electron cavity in the vicinity of the laser pulse, shortly after it has passed its focal point ($X=0$). At first glance, one can see many differences between the 2D and 3D wake channels, for instance in the longitudinal structure of the electron density $n_{\rm e}(X)$ at $y=z=0$ that is shown in Fig.~\ref{fig:16_fix_2d-3d_n-prof_early}(a). However, there are also some important similarities. For instance, the transverse density profile $n_{\rm e}(y)$ in Fig.~\ref{fig:16_fix_2d-3d_n-prof_early}(b) shows that the initial radius of the cavity wall in the focal plane is similar in 2D and 3D, where it is measured to be around $r_{\rm cav}^{\rm 3D} \approx 0.21\,{\rm mm}$.

As expected from theoretical analyses \cite{BulanovSV13}, the contour plots in Fig.~\ref{fig:15_fix_2d-3d_40um_ne_early}(iv) show clearly that the external magnetic field $B_x = 2\,{\rm T}$ along the laser axis prevents the cavity wall in the 3D simulation from fully collapsing at a distance $\lambda_{\rm wake}/2 \approx 2\,{\rm mm}$ (cf.,~Eq.~(\ref{eq:lWake})) behind the laser pulse. Substituting $r_{\rm cav}^{\rm 3D} \approx 0.21\,{\rm mm}$ and all other simulation parameters into Eq.~(48) of Ref.~\cite{BulanovSV13}, the theoretically estimated minimal radius of the magnetized cavity is $r_{\rm min}^{\rm theory} \approx 0.75\times r_{\rm cav} \approx 0.16\,{\rm mm}$. The value measured in the 3D simulation at $X = 0$ in Fig.~\ref{fig:15_fix_2d-3d_40um_ne_early}(iv,b) is $r_{\rm min}^{\rm 3D} \approx 0.11\,{\rm mm}$; i.e., about half of the initial cavity radius and smaller than the theoretical prediction. A difference on that order may originate from the approximations made in the theoretical analysis.

Although the cavity wall may seem to collapse in the 2D case shown in Fig.~\ref{fig:15_fix_2d-3d_40um_ne_early}(iii), a zoom-up reveals that the cavity remains open, albeit with a much smaller radius than in 3D; here about $r_{\rm min}^{\rm 2D} \sim 0.02\,{\rm mm}$ (cf., Fig.~\ref{fig:21_win_2d_10um_scan-tp-a_ne} in \ref{apdx:tpulse}).

In this context, it should also be noted that the minimal radius given in Ref.~\cite{BulanovSV13} was derived for the cavity wall and that the magnetized bow waves may be subject to different conditions. Indeed, some portions of the bow waves pass through the $x$ axis when they undergo transverse wave breaking in our simulations. The resulting multi-flow singularities can be clearly seen in the contour plots of the electron density in Fig.~\ref{fig:15_fix_2d-3d_40um_ne_early}, both in 2D (iii) and 3D (iv). The density spikes are surrounded by cusps and loops that are reminiscent of those known from transverse wave breaks in the unmagnetized case (cf.~Fig.~1 in Ref.~\cite{BulanovSV97}).

The longitudinal profile of one such density spike can be partially seen in Fig.~\ref{fig:16_fix_2d-3d_n-prof_early}(a) around $X \approx -2.6\,{\rm mm}$, but it should be noted that this structure is affected here by the finite cavitation length $L_{\rm cav}$ (red dashed rectangle in Fig.~\ref{fig:15_fix_2d-3d_40um_ne_early}). This density spike appears to propagate at about 30\% of the speed of light, reaching $X \approx -1.8\,{\rm mm}$ at $t' = 12\,{\rm ps}$ in Fig.~\ref{fig:15_fix_2d-3d_40um_ne_early}(iv,c).

While the magnetized bow wave fronts are clearly visible in the 2D case (Figs.~\ref{fig:15_fix_2d-3d_40um_ne_early}(iii) and \ref{fig:16_fix_2d-3d_n-prof_early}(b)), they are very faint in 3D (Figs.~\ref{fig:15_fix_2d-3d_40um_ne_early}(iii) and \ref{fig:16_fix_2d-3d_n-prof_early}(b)) because the finite number of simulation particles is more rapidly diluted with increasing distance from the channel in 3D than in 2D. We have indicated the upper border of the 3D bow wave in Fig.~\ref{fig:15_fix_2d-3d_40um_ne_early}(iv,a) by an arched curve of purple dots, whose maximal radius from the channel axis is about $r_{\rm bow}^{\rm 3D} \approx 0.7\,{\rm mm}$. There are some interesting aspects concerning the 3D bow waves in the present scenario, which will be discussed in a separate physics-oriented paper.

\begin{figure*}
[tb]
\centering
\includegraphics[width=0.96\textwidth]{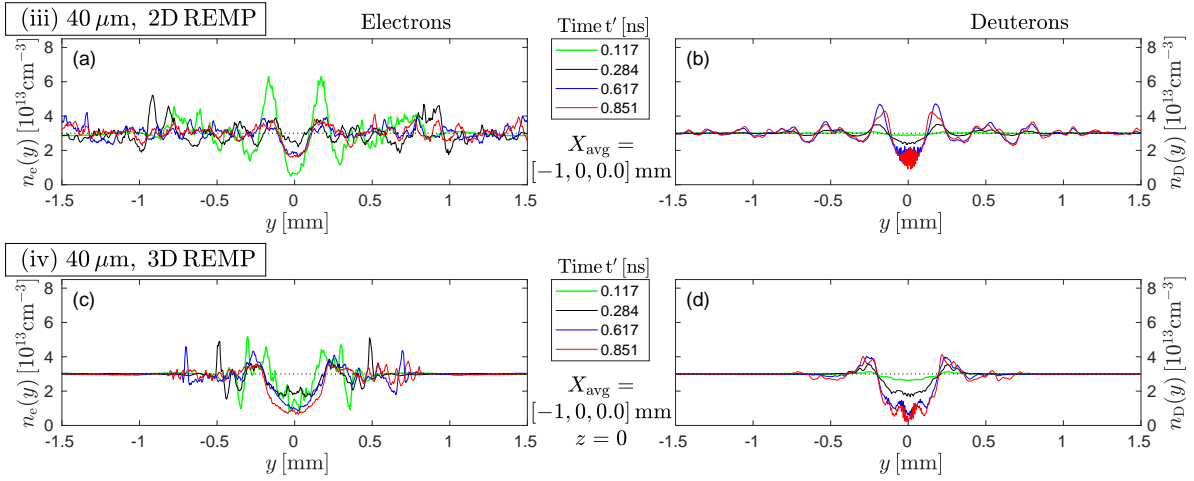}
\caption{Long-time evolution of the transverse profiles evaluated at $X = z = 0$ for (a,c) electrons $n_{\rm e}(y)$ and (b,d) deuterons $n_{\rm D}(y)$ in (iii) 2D and (iv) 3D REMP simulations. Four snapshots are shown, three of which correspond to those in Fig.~\protect\ref{fig:17_fix_2d-3d_40um_ne-ni-late}. For noise reduction, the data were averaged over the interval $-1\,{\rm mm} \leq X \leq 0$. Arranged in the same way as the 2D EPOCH results in Fig.~\ref{fig:06_fix_2d_scan-L_n-prof-evol}.}
\label{fig:18_fix_2d-3d_scan-L_n-profY-evol}%
\end{figure*}

\subsection{Long-time evolution up to the nanosecond scale}
\label{sec:3d_late}

\begin{figure}
[tb]
\centering
\includegraphics[width=0.48\textwidth]{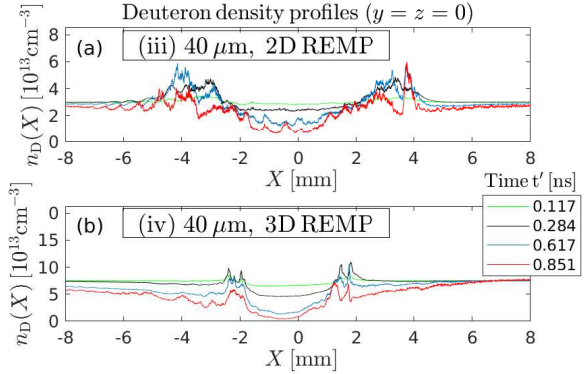}
\caption{Long-time evolution of the axial profiles of the deuteron density $n_{\rm D}(X)$ at $y = z = 0$ in (a) 2D and (b) 3D REMP simulations for the same snapshot times as in Fig.~\protect\ref{fig:18_fix_2d-3d_scan-L_n-profY-evol}. Unlike in Fig.~\protect\ref{fig:18_fix_2d-3d_scan-L_n-profY-evol}, the raw data are plotted here, without spatial averaging.}
\label{fig:19_fix_2d-3d_scan-L_nD-profX-evol}%
\end{figure}

Figures~\ref{fig:17_fix_2d-3d_40um_ne-ni-late}--\ref{fig:19_fix_2d-3d_scan-L_nD-profX-evol} show the long-time evolution of the electron and deuteron densities during the interval $0.1\,{\rm ns} \lesssim t' \lesssim 0.9\,{\rm ns}$. From the contour plots in Fig.~\ref{fig:17_fix_2d-3d_40um_ne-ni-late} one can see that the 2D simulation (iii) predict fairly well the time scales of the various stages of the plasma channel dynamics that are observed in the 3D case (iv):
\begin{itemize}
\item  beginning with the formation of the primary electron cavity that is maintained by magnetized plasma oscillations in snapshot (a) at $t' = 0.117\,{\rm ns}$ (and in preceding snapshots in Fig.~\ref{fig:15_fix_2d-3d_40um_ne_early}),
\item  followed by the ion expansion that becomes clearly visible in snapshot (b) at $t' = 0.284\,{\rm ns}$,
\item  all the way to the axial expansion of the channel along ${\bm B}_{\rm ext}$ as seen in snapshot (c) at $t' = 0.851\,{\rm ns}$.
\end{itemize}

\noindent For the duration of this simulation, the 2D channels preserve most of their up-down symmetry in the $X$-$y$ plane, and the 3D channel preserves its high degree of rotational symmetry around the $X$ axis.

However, there are also several differences between the 2D and 3D channels. For instance, the 3D channel has a larger diameter and it maintains a deeper density cavity with sharper boundaries for a longer time than in 2D. This is visible more clearly in Fig.~\ref{fig:18_fix_2d-3d_scan-L_n-profY-evol}, which shows snapshots of the transverse density profiles of electrons and deuterons, $n_{\rm e}(y)$ and $n_{\rm D}(y)$, averaged over the range $-1\,{\rm mm} \leq X \leq 0$ near the focal plane. Figure~\ref{fig:19_fix_2d-3d_scan-L_nD-profX-evol} shows that the depth of the 3D channel is also larger along its axis, and that its expansion is strongly asymmetric along $X$, proceeding more rapidly and more strongly in the negative $X$ direction.

Moreover, the 2D deuteron density profiles in Figs.~\ref{fig:17_fix_2d-3d_40um_ne-ni-late}(f-iii) and \ref{fig:18_fix_2d-3d_scan-L_n-profY-evol}(b) exhibit laminar quasi-harmonic ripples that extend a relatively long distance of about $1\,{\rm mm}$ from the channel center. In the 3D case, such fluctuations are found only very close to the channel edge. Further investigations are required to explain these observations physically.

\section{Conclusion}
\label{sec:conclusion}

In the present paper, we have demonstrated that a relativistically intense laser pulse ($a_0 \sim 1$) shot into in a sparse cold deuterium gas ($n_{\rm e} \sim 10^{13}\,{\rm cm}^{-3}$) that is permeated by an external magnetic field of moderate strength (few tesla, $\omega_{\rm Be}/\omega_{\rm pe} \sim 1$) leaves behind a plasma wake channel whose structure is largely independent of the laser wavelength $\lambda_{\rm las}$ (Section~\ref{sec:limsim_demo}) when certain conditions are satisfied (Section~\ref{sec:limsim_cond}). Most importantly, the electron density should be highly subcritical (here: $n_{\rm e}/n_{\rm crit} \ll 10^{-2}$). This manifestation of so-called limited similarity \cite{Block67, Esirkepov12} has practically useful implications, such as the freedom to choose the most convenient type of laser for an experiment and the possibility to reduce computational expenses by performing scaled simulations with increased wavelengths, while retaining comparability between the results of the scaled simulations and experimental conditions.

The prospect of reducing the computational expenses by scaling the laser wavelength was the main theme of the present paper. This was motivated by our future goal to simulate the long-term dynamics (or after-glow) of the magnetized plasma wake channel that would be produced by a high power laser with relativistic intensity in a magnetically confined fusion (MCF) plasma with parameters that are typical for tokamak and stellarator experiments. Since the relatively inexpensive 2D simulations cannot capture the full motion of charged particles in a magnetic field, it is important to have the ability to perform 3D simulations, and the present wavelength scaling method justified by limited similarity makes meaningful long-time 3D simulations possible with existing supercomputer systems.

As a proof of principle, we have reported first results of a long-time 3D simulation (Section~\ref{sec:3d}), which covered about $1\,{\rm ns}$ of physical time and was performed with an up-scaled laser wavelength of $\lambda_{\rm las} = 40\,\mu{\rm m}$. Such a 3D simulation would not be feasible on presently available supercomputers if it was to be performed with the actual wavelengths ($0.8...10\,\mu{\rm m}$) of presently available high power lasers that can achieve relativistic intensities (\ref{apdx:resource}).

However, it may not always be easy to find a well-balanced compromise between computational efficiency and realism. In our case, the choice $\lambda_{\rm las} = 40\,\mu{\rm m}$ reduced the computational expenses to a comfortable level, but due to its relatively short Rayleigh length $L_{\rm R} \propto \lambda_{\rm las}^{-1}$, it also placed us at the limit for a reasonable cavitation length $L_{\rm cav}$ (Eq.~(\ref{eq:len_cav})), which is the length of the electron-free cavity that is carved out by the laser when its amplitude is sufficiently high; here $a_0 \gtrsim 0.3$ in 3D and $a_0 \gtrsim \sqrt{0.3}$ in 2D (Eq.~(\ref{eq:def_len_ch})). In the scenario studied here, the cavitation length is $L_{\rm cav}(40\,\mu{\rm m}) \approx 7.5\,{\rm cm}$, which barely covers two wavelength $\lambda_{\rm UH} = 2\pi c(\omega_{\rm pe}^2 + \omega_{\rm Be}^2)^{-1/2} \approx 4\,{\rm mm}$ of the upper-hybrid wave that determines the characteristic size of the magnetized laser wake \cite{BulanovSV13}. At later times, this may affect the rate at which the channel is expanding and refilled via charged particle motion along the magnetic field. Indeed, we believe that the wavelength dependence of the cavitation length may be responsible for the differences seen in the deuteron energy distribution $f_{\rm D}(K)$ and its continuing evolution (cf.~Fig.~\ref{fig:13_fix_2d_scan-L_enrDistr-e-D}(right)). For further research, a somewhat shorter wavelength may be more appropriate, such as $\lambda_{\rm las} = 20\,\mu{\rm m}$, which would be up to $2^5 = 32$ times more expensive computationally (cf.~\ref{apdx:resource}, Eq.~(\ref{eq:flop_lambda})).

Limited similarity with respect to varying $\lambda_{\rm las}$ has been demonstrated here only in 2D. We do not see any obvious reason why limited similarity under similar conditions should not hold in 3D and have assumed that it does, but this remains to be demonstrated.

There are several other points to be considered before applying the wavelength scaling method described in this paper. While the overall trajectory of the bow wave front is similar, its cross-section is broader and shallower, and small-scale features such as undulations may be lost or depend on the wavelength used (Fig.~\ref{fig:10_win10mm_singularity}). This means that our scaling method cannot be used to study effects that depend on the microscopic features of singular structures, such as harmonic generation via the BISER mechanism \cite{Pirozhkov17b, Pirozhkov12, Pirozhkov14}.\footnote{3D simulations of such processes can be performed with realistic wavelengths as long as the structures of interest are located near the laser pulse, since one has to simulate only a small window that follows the laser pulse for a short time. There are, however, conditions where this is not possible; for instance when the counter-propagating stimulated Raman scattering (SRS) wave cannot be neglected.}

Another point to be considered is the required number of particles per cell (PPC), $N_{\rm PPC}$. If we fix $N_{\rm PPC}$ as in the present study, the limited similarity assumption helps us to reduce the total number of simulation particles without changing the noise level. This reduces the computational cost, but it may also dilute certain features to such a degree that they become difficult to detect. Here, this happened to the radially expanding 3D bow wave in Fig.~\ref{fig:15_fix_2d-3d_40um_ne_early}(iv). The value of $N_{\rm PPC}$ may have to be increased substantially in order to study this phenomenon and its effects in detail.

At later times, our 3D results (in particular, Figs.~\ref{fig:17_fix_2d-3d_40um_ne-ni-late}--\ref{fig:19_fix_2d-3d_scan-L_nD-profX-evol}) show that the ions are continuously expanding, and they take the electrons with them as they evacuate the channel. The resulting vacuum bubble is likely to persist well beyond the end of the simulation. Thus, our results reliably demonstrate that the laser-induced plasma channel survives for $1\,{\rm ns}$ and perhaps longer.

Although this result was obtained for a cold gas, the arguments provided in \ref{apdx:thermal} give us reasons to assume that this result is also valid under MCF plasma conditions with typical thermal energies on the multi-keV level. In essence, our arguments in \ref{apdx:thermal} imply that the thermal electron motion is not the primary factor. Instead, the life time of the laser-induced plasma wake channel beyond the $1\,{\rm ns}$ simulated here is determined by collisionless thermal mixing on the time scale of ion gyration. This process will be studied in the future, using the methods described in the this work.

When such long-time simulations with finite temperature are performed with scaled laser wavelengths and reduced resolution (taking advantage of the limited similarity argument), it must be noted that the Debye length $\lambda_{\rm D} \approx 74\,\mu{\rm m}$ in Eq.~(\ref{eq:debye}) also sets an upper limit on the allowed grid size in a PIC simulation. If we require that the grid spacing satisfies $\lambda_{\rm D}/\Delta x > \lambda_{\rm D}/\Delta y > 10$, a simulation with $\lambda_{\rm las}/\Delta y = 8$ such as ours can only be performed with scaled wavelengths satisfying $\lambda_{\rm las} < 60\,\mu{\rm m}$ (for the plasma parameters assumed in Eq.~(\ref{eq:debye})).

This upper limit on the resolution also has an impact on the possibility of continuing the simulation at reduced resolution (and, perhaps, with a larger simulation box) after the laser pulse has left. In our example with $\lambda_{\rm las} = 40\,\mu{\rm m}$, where $2\Delta x = \Delta y = 5\mu{\rm m}$, the above condition $\Delta x \leq \Delta y < \lambda_{\rm D}/10 \approx 7.4\,\mu{\rm m}$ implies that the resolution may be reduced by, say, a factor $3$ along $x$, and a factor $1.5$ along $y$ and $z$. This is not much but sufficient to justify the effort of down-sampling the fields, because it would allow us to extend the length $L_x$ of the simulation domain by a factor 2 or so, without increasing the computational cost too much. The simulation domain should be as long as possible in order to accommodate the continuous stretching of the channel and the inflow of particles along the magnetic field.
 
The insights that can be won with such simulations would be useful for advancing the physical understanding of the long-term after-glow dynamics of magnetized low-density plasmas irradiated by relativistically intense laser pulses and to explore possible applications.

\section*{Acknowledgements}

We thank S.V.~Bulanov for fruitful discussions and comments. This work was supported by the QST Director Fund for Creative Research (Project No.~16), JSPS KAKENHI Grant Number JP19H00669, and by the Japanese government MEXT as ``Priority Issue on Post-K computer'' (Accelerated Development of Innovative Clean Energy Systems). The simulations were carried out using the JFRS-1 supercomputer system at Computational Simulation Centre of International Fusion Energy Research Centre (IFERC-CSC) in Rokkasho Fusion Institute of QST (Aomori, Japan) (Project ID: LASART) and the ITO supercomputer at Kyushu University in (Fukuoka, Japan) (Project group ID: gr180494). The development of the EPOCH code employed in this work was in part funded by the UK EPSRC grants EP/G054950/1, EP/G056803/1, EP/G055165/1 and EP/M022463/1.

\appendix

\section{Estimation of computational resources required for 3D simulations}
\label{apdx:resource}

In order to simulate the dynamics of the plasma wake channel, the length $L_x$ of the simulation box must be significantly larger than the initial length $L_{\rm cav}$ of the electron-free cavity defined by Eq.~(\ref{eq:len_cav}). In particular, this is necessary for our intended long-time simulations with axial magnetic field, because the plasma channel may expand along the $x$-direction, so we add buffer zones of length $L_{\rm buf}$ on each side: $L_x = L_{\rm cav} + 2L_{\rm buf}$. In order to accommodate the unfocussed laser pulse at $x=0$ and $x=L_x$, the lateral cross-section $L_yL_z$ must also be increased in proportion to $w^2(X=L_x/2)$ given by Eq.~(\ref{eq:paraxial_w}). Consequently, the volume $V = V_x V_y V_z$ of the simulation box scales as
\begin{equation}
V \propto \underbrace{(2L_{\rm buf} + L_{\rm cav})}\limits_{L_x} \underbrace{\left(1 + \frac{(2L_{\rm buf} + L_{\rm cav})^2}{4 L_{\rm R}^2}\right)}\limits_{w^2(X=L_x/2) \; \propto \; L_y L_z}.
\label{eq:scale_v}
\end{equation}

Here, we work with a buffer zone of size $L_{\rm buf} \approx L_{\rm cav}$ on each side. Substituting Eq.~(\ref{eq:len_cav_ref}) for $L_{\rm cav}$, our box length for a simulation with $\lambda_{\rm las} = 10\,\mu{\rm m}$ becomes $L_x \approx 3\times 6 L_{\rm R}(10\,\mu{\rm m}) \approx 100\,{\rm mm}$. The $1/{\rm e}$ radius of the unfocused laser amplitude at a distance $L_x/2 \approx 50\,{\rm mm}$ from the focal point is $w(X=\pm L_x/2) = w_{\rm foc}\sqrt{1 + (3\times 6 L_{\rm R}/2)^2/L_{\rm R}^2} \approx 9 w_{\rm foc} \approx 1.2\,{\rm mm}$ and we make the box about 5 times wider than that: $L_y = L_z = 6\,{\rm mm} \approx 5\times w(L_x/2)$. With this, the volume of our simulation box becomes $V = 100 \times 6 \times 6\,{\rm mm}^3$. The numerical parameters in Table~\ref{tab:parm_num} and the location of the focal point $x_{\rm foc} = 50\,{\rm mm} \approx L_x/2$ for $\lambda_{\rm las} = 10\,\mu{\rm m}$ were chosen on this basis.

The number of mesh points in the simulation box is given by
\begin{equation}
N_{\rm mesh} = \frac{L_x L_y L_z}{\Delta x \Delta y \Delta z}.
\label{eq:nmesh}
\end{equation}

\noindent Substituting the values from Table~\ref{tab:parm_num} gives
\begin{equation}
\frac{100\,{\rm mm}}{10\,\mu{\rm m}/32} \times \left(\frac{6\,{\rm mm}}{10\,\mu{\rm m}/8}\right)^2 = 320{,}000\times 4{,}800^2 \sim 7\times 10^{12}
\end{equation}

\noindent mesh points. The required memory for 9 fields (${\bm E}$, ${\bm B}$, ${\bm J}$) and 6 particle coordinates (${\bm x}$, ${\bm v}$) stored in double precision is then $7\times 10^{12} \times 15 \times 8\,{\rm bytes} = 840\,{\rm Tbyte}$ for a PIC simulation initialized with only 1 particle per cell (PPC). If $5\,{\rm Gbyte}$ per core are available, the simulation would require at least 168,000 cores, which is 3 times the total number of cores available on our 4.2 Petaflop/s supercomputer JFRS-1 \cite{jfrs1}. Even the memory requirements alone show clearly that such a 3D simulation is hardly feasible on presently available supercomputers.

A similarly discouraging result is obtained if one estimates the computational effort in terms of the floating-point operations (flop) necessary for the simulation:
\begin{equation}
N_{\rm flop} = N_{\rm mesh} N_{\rm step} N_{\rm flop/unit},
\label{eq:nflop}
\end{equation}

\noindent where $N_{\rm flop/unit}$ denotes the flop number per mesh point and time step. The number of time steps $\Delta t$ during a simulation of length $T_{\rm sim}$ is
\begin{equation}
N_{\rm step} = \frac{T_{\rm sim}}{\Delta t} = \frac{cT_{\rm sim}}{\Delta x} \times \frac{\Delta x}{c\Delta t}.
\label{eq:nstep}
\end{equation}

\noindent The factor $\Delta x/(c\Delta t)$ is fixed at the maximum value permitted by the Courant-Friedrichs-Levy (CFL) condition,
\begin{equation}
\frac{\Delta x}{c\Delta t} = {\rm const.} \gtrsim \left\{\begin{array}{lcl}
1 & : & {\rm 1D}, \\
\left(1 + \frac{\Delta x^2}{\Delta y^2}\right)^{1/2} & : & {\rm 2D}, \\
\left(1 + \frac{\Delta x^2}{\Delta y^2} + \frac{\Delta x^2}{\Delta z^2}\right)^{1/2} & : & {\rm 3D}.
\end{array}\right.
\label{eq:cfl}
\end{equation} 

\noindent With $\lambda_{\rm las}/\Delta (y,z) = 8$ and $\lambda_{\rm las}/\Delta x = 32$, the condition $\Delta x/(c\Delta t) \approx 1$ yields a CFL time step of order $\Delta t \sim 1\,{\rm fs}$, so that one has to perform $N_{\rm step} \sim 10^6$ steps in order to simulate $T_{\rm sim} = 1\,{\rm ns}$. Assuming that each step consists of about $100$ floating-point operations per mesh point and variable\footnote{100 flop per mesh point and variable is an empirical estimate. The actual number of operations (particle pushing, field solver, mapping between particle and mesh positions) is difficult to determine; in part, because it depends on how the compiler optimizes the source code.},
we have $N_{\rm flop/unit} \sim (9+6)\times 100$ for our 9 field variables and 6 particle variables. Hence, this simulation would require
\begin{equation}
N_{\rm flop} \sim \underbrace{7\times 10^{12}}\limits_{N_{\rm mesh}} \times \underbrace{10^6}\limits_{N_{\rm step}} \times \underbrace{15\times 100}\limits_{N_{\rm flop/unit}} \approx 10^{22}\,{\rm flop},
\end{equation}

\noindent or $10{,}000$ Exaflop. Assuming 5\% efficiency\footnote{Computational efficiency is determined by factors such as communication overhead, cache misses, and so on. For the type of simulations performed here, an efficiency of 5\% can be considered outstanding.}, such a simulation would fully occupy a high performance computer system with nominal 4.2 Petaflop/s like JFRS-1 for 1.5 years (provided that there is sufficient memory). Even a single simulation (not to speak of parameter scans and convergence tests) is infeasible under these conditions.

However, the computational effort depends strongly on the laser wavelength $\lambda_{\rm las}$ if one is only interested in dynamics on larger spatial scales. For a given simulation time $T_{\rm sim} = {\rm const}$., and with fixed ratios $\lambda_{\rm las}/\Delta (x,y,z) = {\rm const}$.\ and $c\Delta t/\Delta x = {\rm const}$., the flop number given by Eqs.~(\ref{eq:nflop}) and (\ref{eq:nstep}) scales as
\begin{equation}
N_{\rm flop} \propto \underbrace{\frac{V}{\Delta x \Delta y \Delta z}}\limits_{N_{\rm mesh}} \times \underbrace{\frac{1}{\Delta x}}\limits_{N_{\rm step}} \propto \lambda_{\rm las}^{-4} V(\lambda_{\rm las}),
\end{equation}

\noindent with $V(\lambda_{\rm las})$ given by Eq.~(\ref{eq:scale_v}). Recalling that $L_{\rm cav} \propto L_{\rm R} \propto \lambda_{\rm las}^{-1}$, we obtain
\begin{equation}
N_{\rm flop} \propto \lambda_{\rm las}^{-4} \times \left\{\begin{array}{lcl}
\lambda_{\rm las}^{-1} & : & L_{\rm buf} \ll L_{\rm cav}, \\
\lambda_{\rm las}^{2} & : & L_{\rm buf} \gg L_{\rm cav}.
\end{array}\right.
\label{eq:flop_lambda}
\end{equation}

\noindent If the size $L_{\rm buf}$ of the buffer zone is much smaller than the size of the cavity, the volume of the simulation box scales favorably as $V \propto L_{\rm R} \propto \lambda_{\rm las}^{-1}$ and enhances the wavelength-dependence of the computational effort to $N_{\rm flop} \propto \lambda_{\rm las}^{-5}$. However, this limit may be useful only for simulations of a relatively short time window, where the inflow of particles and associated expansion of the channel along the $x$ axis is still small. In the other limit, $L_{\rm buf} \gg L_{\rm cav}$, the volume scales unfavorably as $V \propto L_{\rm R}^{-2} \propto \lambda_{\rm las}^2$, so that the computational effort depends more weakly on the wavelength: $N_{\rm flop} \propto \lambda_{\rm las}^{-2}$. Note, however, that the unfavorable scaling $V \propto L_{\rm R} \propto \lambda_{\rm las}^2$ is actually somewhat overpessimistic, because it holds only for laser wavelengths with a relatively short focus. According to our experience, the underlying scaling $L_yL_z \propto w^2(L_x/2) \propto \lambda_{\rm las}^2$ for the transverse cross-sectional area of the simulation box breaks down as one goes towards shorter wavelengths, because a slowly focusing laser tends to accumulate artifacts near the lateral boundaries of the simulation box, which is most safely avoided by increasing $L_y L_z$ (e.g., see Fig.~\ref{fig:11_win10mm_scan-L_enrDistr-e} below). Thus, it is clear that the scaling for actual simulation tasks will probably lie somewhere between the two extremes, $\lambda_{\rm las}^{-5...-2}$, that we have estimated here in Eq.~(\ref{eq:flop_lambda}).

In any case, the fact that the computational effort decreases rapidly with increasing laser wavelength (if only dynamics on larger scales are of interest) motivates the use of artificially increased wavelengths in order to make 3D simulations feasible. The limited similarity of the laser-induced plasma wake channel that is demonstrated in the present paper offers a justification for such wavelength scaling.

\section{Role of the pulse length and amplitude}
\label{apdx:tpulse}

\begin{figure}
[tb]
\centering
\includegraphics[width=0.48\textwidth]{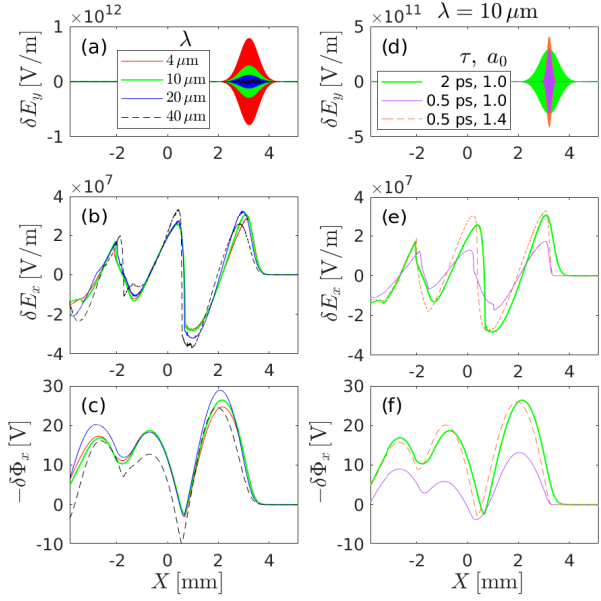}
\caption{Profiles of the electric field and its potential along the $x$ axis ($y = 0$) in the vicinity of the laser pulse for the same snapshot time as in Figs.~\protect\ref{fig:08_win10mm_scan-L_ne-bz}, \protect\ref{fig:09_win10mm_scan-L_Jxy-Exy} and \protect\ref{fig:11_win10mm_scan-L_enrDistr-e} above. The left columns shows results for different wavelengths $\lambda_{\rm las} = 4$...$40\,\mu{\rm m}$, while fixing the pulse length $\tau_{\rm pulse} = 2\,{\rm ps}$ and the normalized amplitude $a_{\rm 0,foc} = 1$. For the $10\,\mu{\rm m}$ case, the right column shows results for different pulse lengths and amplitudes. Panels (a) and (d) show the vertical electric field component $\delta E_y$, which is mainly due to the laser itself. Panels (b) and (e) show the axial electric field component $\delta E_x$, and panels (c) and (f) show the associated potential $\delta\Phi_x$ given by Eq.~(\protect\ref{eq:epotx}).}
\label{fig:20_win_2d_scan-L-tp_E-Phi}%
\end{figure}

The requirement $\tau_{\rm pulse} = {\rm const}$.\ discussed in Section~\ref{sec:limsim_cond} (Eq.~(\ref{eq:limsim_laser})) seems to be connected with the strength of the electric field $\delta {\bm E}$ that is induced by the laser displacing electrons with respect to the ions and by the ensuing plasma oscillations. To show this, we plot in Fig.~\ref{fig:20_win_2d_scan-L-tp_E-Phi} the profiles of the axial component of the fluctuating electric field $\delta E_x(x)$ and its associated potential
\begin{equation}
\delta\Phi_x \equiv -\int_{\infty}^x{\rm d}x'\, \delta E_x = \delta\Phi(x) - \delta\Phi(\infty) - \int_{\infty}^x{\rm d}x'\,  \frac{\partial \delta A_x}{\partial t},
\label{eq:epotx}
\end{equation}

\noindent measured along the $x$ axis ($y=0$) for different laser wavelengths $\lambda_{\rm las} = 4$...$40\,\mu{\rm m}$, while fixing the values of $\tau_{\rm pulse} = 2\,{\rm ps}$ and $a_{\rm 0,foc} = 1$. One can see that the amplitudes of $\delta E_x$ and $\delta\Phi_x$ are similar in all cases shown, in spite of the fact that the amplitude of the laser's electric field $\delta E_y$ in Fig.~\ref{fig:20_win_2d_scan-L-tp_E-Phi}(a) varies in proportion to $\lambda_{\rm las}$.

\begin{figure}
[tb]
\centering
\includegraphics[width=0.48\textwidth]{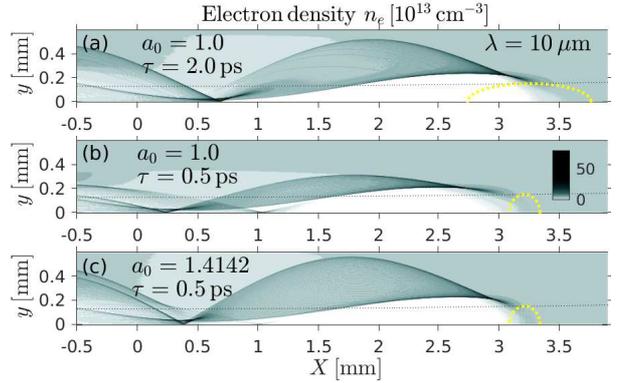}
\caption{Electron density contours for the $10\,\mu{\rm m}$ case with different pulse lengths and amplitudes as in the right column of Fig.~\protect\ref{fig:20_win_2d_scan-L-tp_E-Phi}.}
\label{fig:21_win_2d_10um_scan-tp-a_ne}%
\end{figure}

As a counter-example, the panels in the right column of Fig.~\ref{fig:20_win_2d_scan-L-tp_E-Phi} show how the amplitudes of $\delta E_x$ and $\delta \Phi_x$ decrease when the pulse length in the $10\,\mu{\rm m}$ case is reduced by a factor 4 from $2\,{\rm ps}$ to $0.5\,{\rm ps}$ (so that the number of wave cycles per pulse, $c\tau_{\rm pulse}/\lambda_{\rm las}$, is the same as in the $40\,\mu{\rm m}$ case). The contour plots of the electron density $n_{\rm e}(x,y)$ for these two cases are shown in Fig.~\ref{fig:21_win_2d_10um_scan-tp-a_ne}(a) and (b), where one can see that the transverse excursion of the bow wave is significantly reduced when the pulse length is shortened.

The original magnitudes of $\delta E_x$ and $\delta\Phi_x$ and the transverse excursion of the bow wave as seen in the default case with $\tau_{\rm pulse} = 2\,{\rm ps}$ and $a_{\rm 0,foc} = 1.0$ can be obtained in the short pulse case $\tau_{\rm pulse} = 0.5\,{\rm ps}$ by increasing the laser amplitude. As one can see in Fig.~\ref{fig:20_win_2d_scan-L-tp_E-Phi}(d)--(f) and by comparing the density contours in Figs.~\ref{fig:21_win_2d_10um_scan-tp-a_ne}(a) and (c), a similar result was obtained with $a_{\rm 0,foc} = \sqrt{2} \approx 1.4$.

Note that these parameter values had to be obtained by trial-and-error since the relation between $\tau_{\rm pulse}$ and $a_{\rm 0,foc}$ is generally complicated, as we have briefly discussed in Sections~\ref{sec:setup_param} and \ref{sec:limsim_cond} on the basis of Ref.~\cite{BulanovSV16}.

\section{Discussion concerning thermal motion}
\label{apdx:thermal}

Thermal electrons in a $3\,{\rm keV}$ tokamak plasma travel about $30\,{\rm mm}$ per nanosecond on average. According to Eqs.~(\ref{eq:len_cav}) and (\ref{eq:len_cav_ref}), this corresponds to $10\%$ of the initial length $L_{\rm cav}(1\,\mu{\rm m}) \approx 300\,{\rm mm}$ of the cavity as produced by a pulse from a 0.8 micron Ti:sapphire laser and is comparable to $L_{\rm cav}(10\,\mu{\rm m}) \approx 30\,{\rm mm}$ in the case of a 10 micron ${\rm CO}_2$ gas laser, both with a spot diameter of $d_{\rm foc} = 0.15\,{\rm mm}$ (Table~\ref{tab:parm_laser}). In simulations performed with up-scaled wavelength, the thermal velocity of electrons may be down-scaled accordingly in order to represent the actual situation.

However, the thermal electrons are not free to move through the channel at their initial speed, even along the magnetic field ${\bm B}_{\rm ext}$. During the early stages of the evolution of the plasma wake channel, where the ion displacement is still small ($t' \lesssim 0.1\,{\rm ns}$), electrons approaching the laser-induced cavity at thermal speeds are likely to encounter repulsive forces from accumulations of electrons that are formed by the wake dynamics at the boundaries of the cavity (cf.~Figs.~\ref{fig:18_fix_2d-3d_scan-L_n-profY-evol} and \ref{fig:19_fix_2d-3d_scan-L_nD-profX-evol}). The scale of these barriers is larger than the Debye length for $3\,{\rm keV}$ electrons with a density of $n_{\rm e} = 3\times 10^{13}\,{\rm cm}^{-3}$,
\begin{equation}
\lambda_{\rm D} = \left(\epsilon_0 K_{\rm e}/(n_{\rm e} e^2)\right)^{1/2} \approx 74\,\mu{\rm m},
\label{eq:debye}
\end{equation}

\noindent so that the attractive force of the positively charged cavity is effectively shielded.

The penetration of thermal electrons across the magnetic field is even less of a threat because their thermal Larmor radii are on the order of
\begin{equation}
\rho_{\rm Be} = p_\perp/(e B_{\rm ext}) = \gamma m_{\rm e} v_\perp/(e B_{\rm ext}) \lesssim 0.1\,{\rm mm}
\label{eq:rhoBe}
\end{equation}

\noindent which is much smaller than the cavity diameter in our case ($2r_{\rm cav}^{\rm 3D} \approx 0.42$ in Fig.~\ref{fig:16_fix_2d-3d_n-prof_early}(b)), and because the strong radial electric field would deflect them ($\delta{\bm E}\times{\bm B}_{\rm ext}$ drift), if they did manage to enter the positively charged region.

Consequently, we may assume that the life time of the channel is primarily determined by the motion of thermal ions, which move on a much slower time scale ($\gtrsim 0.1\,{\rm ns}$). For instance, a $3\,{\rm keV}$ deuteron with Larmor radius $\rho_{\rm BD} \lesssim 3\,{\rm mm}$ travels about $0.5\,{\rm mm}$ per nanosecond, which corresponds to the diameter of the plasma wake channel in the final stages of our simulation (Fig.~\ref{fig:18_fix_2d-3d_scan-L_n-profY-evol}).

\bigskip

\noindent {\bf Declarations of interest:} none.

\bibliographystyle{unsrt}
\bibliography{references}

\end{document}